\documentclass[aps,prl,twocolumn,superscriptaddress,nofootinbib,floatfix]{revtex4-1}

\usepackage[utf8]{inputenc}
\usepackage{graphicx}
\usepackage[dvipsnames]{xcolor}
\usepackage{amsthm, amsfonts, amsmath, mathrsfs, bm}
\usepackage{bbm}
\usepackage{slashed}
\usepackage{physics}
\usepackage{xspace}
\usepackage[caption=false]{subfig}

\usepackage{orcidlink}
\usepackage{comment}
\usepackage{glossaries}
\usepackage{quantikz}
\usepackage{braket}
\setacronymstyle{long-short}
\glsdisablehyper

\usepackage{hyperref}
\hypersetup{colorlinks=true,
    linkcolor=magenta,
    citecolor=blue,
    urlcolor=cyan,
    pdfpagemode=FullScreen,}

\usepackage{fullpage}
\usepackage{physics2}
\usephysicsmodule{ab}
\usephysicsmodule{braket}
\usepackage{bm}
\usepackage{amsmath}
\usepackage{amssymb}
\usepackage{faktor}

\usepackage{soul,color}
\definecolor{chromeyellow}{rgb}{1.0, 0.65, 0.0}
\definecolor{DodgeBlue}{rgb}{0.118, 0.565,1.000}
\definecolor{asparagus}{rgb}{0.53, 0.66, 0.42}
\definecolor{cadmiumgreen}{rgb}{0.0, 0.42, 0.24}

\definecolor{jlab_red}{RGB}{192,39,45}
\definecolor{jlab_orange}{RGB}{249,102,0}
\definecolor{jlab_blue}{RGB}{47,122,121}
\definecolor{jlab_green}{RGB}{65,125,10}
\definecolor{bobcat_green}{RGB}{2,66,48}
\definecolor{lbl_burgandy}{RGB}{103,46,69}
\definecolor{lbl_red}{RGB}{224,78,57}

\usepackage{mfirstuc} 
\newcommand{\addReviewer}[2]{
  \expandafter\newcommand\csname #1\endcsname[1]{{\sf \color{#2} {#1}:\,##1}}
  \expandafter\newcommand\csname #1cor\endcsname[2]{{\color{#2} {#1}:\,\st{##1}{\sf ##2}}}
  \expandafter\newcommand\csname #1color\endcsname{#2}
}

\addReviewer{ib}{jlab_green}
\addReviewer{rb}{jlab_blue}
\addReviewer{ac}{jlab_orange}
\addReviewer{er}{jlab_red}
\addReviewer{trr}{ForestGreen}
\addReviewer{awl}{lbl_red}

\newacronym{pionless}{EFT$_\slashed \pi$}{pionless effective field theory}
\newacronym{eft}{EFT}{effective field theory}
\newacronym{lec}{LEC}{low energy coefficient}
\newacronym{qed}{QED}{quantum electrodynamics}
\newacronym{qcd}{QCD}{quantum chromodynamics}
\newacronym{nrqed}{NRQED}{nonrelativistic QED}
\newacronym{vnrqed}{vNRQED}{velocity NRQED}
\newacronym{nrqcd}{NRQCD}{nonrelativistic QCD}
\newacronym{ChiPT}{\ensuremath{\chi}PT}{chiral perturbation theory}
\newacronym{ChiEFT}{\ensuremath{\chi}EFT}{chiral effective field theory}
\newacronym{BSM}{BSM}{Beyond the Standard Model}
\newacronym{NN}{\ensuremath{N\!N}\xspace}{nucleon-nucleon}
\newacronym{vRG}{vRG}{velocity renormalization group}
\newacronym{lo}{LO\ensuremath{_\slashed \pi}}{leading order}
\newacronym{nlo}{\ensuremath{\text{NLO}_{\slashed \pi}}}{next-to-leading order}
\newacronym{n2lo}{\ensuremath{\text{N}^2\text{LO}_{\slashed \pi}}}{next-to-next-to-leading order}
\newacronym{n3lo}{\ensuremath{\text{N}^3\text{LO}_{\slashed \pi}}}{next-to-next-to-next-to-leading order}
\newacronym{n4lo}{\ensuremath{\text{N}^4\text{LO}}}{next-to-next-to-next-to-next-to-leading order}
\newacronym{dimreg}{DimReg}{dimensional regularization}
\newacronym{av18}{AV18}{Argonne \ensuremath{v18}}
\newacronym{LLalpha}{\ensuremath{\text{LL}_\alpha}}{leading-logarithm-in \ensuremath{\alpha}}
\newacronym{NLLalpha}{\ensuremath{\text{NLL}_\alpha}}{next-to-leading-logarithm-in \ensuremath{\alpha}}
\newacronym{PDS}{PDS}{Power Divergence Subtraction}
\newacronym{BBN}{BBN}{big bang nucleosynthesis}

\usepackage{ulem}
\newcommand\redsout{\bgroup\markoverwith{\textcolor{red}{\rule[0.5ex]{2pt}{1.4pt}}}\ULon}

\renewcommand{\emph}[1]{\textit{#1}}

\begin{document}

\newcommand{\nersc}{NERSC, Lawrence Berkeley National Laboratory, Berkeley, CA 94720, USA}
\newcommand{\ucb}{Department of Physics, University of California, Berkeley, CA 94720, USA}
\newcommand{\lbnl}{Nuclear Science Division, Lawrence Berkeley National Laboratory, Berkeley, CA 94720, USA}
\newcommand{\lbnlCS}{Applied Math and Computational Research Division, Lawrence Berkeley National Laboratory, Berkeley, CA 94720, USA}
\newcommand{\umd}{Department of Physics, Maryland Center for Fundamental Physics, and Joint Center for Quantum Information and Computer Science, University of Maryland, College Park, MD 20742, USA}
\newcommand{\lbnlphys}{Physics Division,
Lawrence Berkeley National Laboratory, Berkeley,
CA 94720, USA}

\title{Scattering amplitudes from quantum hardware \`a la RESOs}

\author{Ra\'ul A. Brice\~no\,\orcidlink{0000-0003-1109-1473}}
\email[e-mail: ]{rbriceno@berkeley.edu}
\affiliation{\ucb}
\affiliation{\lbnl}

\author{Ivan M. Burbano \,\orcidlink{0000-0002-3792-1773}}
\email{iburbano@umd.edu}
\affiliation{\umd}
\affiliation{\ucb}
\affiliation{\lbnl}
\affiliation{\lbnlphys}

\author{Anthony N. Ciavarella\,\orcidlink{0000-0003-3918-4110}}
\email{anciavarella@lbl.gov}
\affiliation{\lbnlCS}

\author{Ermal Rrapaj\,\orcidlink{0000-0002-3222-7010}}
\email{ermalrrapaj@lbl.gov}
\affiliation{\nersc}

\author{Thomas R.~Richardson\,\orcidlink{0000-0001-6314-7518}}
\email{thomas.richardson@berkeley.edu}
\affiliation{\ucb}
\affiliation{\lbnl}

\author{Andr\'{e}~Walker-Loud\,\orcidlink{0000-0002-4686-3667}}\email{walkloud@lbl.gov}
\affiliation{\lbnl}
\affiliation{\ucb}

\date{\today}

\begin{abstract}
We report the first quantum-hardware implementation of Real-time Estimators for Scattering
Observables (RESOs). The physical model is a one-dimensional lattice theory of non-relativistic spin-1/2 fermions with a single contact interaction. 
The lattice sizes studied are up to $N_x=54$ lattice sites and up to $20$ time steps. 
Spacetime correlation functions were computed on IBM's quantum computers, using $109$ qubits with $10,806$ CZ gates applied, reaching a two-qubit gate depth of $246$. 
We demonstrate how to study bound states and extract scattering amplitudes from these correlation functions.
\end{abstract}

\maketitle

\paragraph*{Introduction.} Scattering amplitudes are central to the nuclear and particle physics program,
encoding the hadron spectrum~\cite{Briceno:2017max,Lebed:2016hpi},
hadron structure~\cite{Accardi:2012qut}, and processes that test
the Standard Model~\cite{AbdulKhalek:2022hcn,DUNE:2016hlj,DUNE:2020fgq,LEGEND:2021bnm,CUPID:2019imh,KamLAND-Zen:2024eml,GERDA:2020xhi}.
Presently, the only pathway towards theoretically computing scattering amplitudes for strongly interacting systems directly from the Standard Model is paved by Euclidean lattice field theories,  e.g., lattice QCD, which leverage the power of Monte Carlo sampling of the Euclidean path integral. 

Modern lattice QCD calculations can access an increasingly broad range of two- and three-body reactions~\cite{Wilson:2014cna,Dudek:2016cru,Woss:2020ayi,Boyle:2024hvv,PitangaLachini:2026lyd,Mai:2018djl,Hansen:2020otl,Dawid:2025zxc,Yan:2025mdm,Yan:2024gwp} by constructing nonperturbative mappings between finite-volume Euclidean observables and infinite-volume amplitudes~\cite{Luscher:1985dn,Luscher:1986pf,Luscher:1990ux,Kim:2005gf,Briceno:2014oea,Hansen:2014eka,Hansen:2015zga,Mai:2017bge,Doring:2018xxx, Jackura:2022gib,Raposo:2025dkb,Briceno:2017max,Hansen:2019nir,Mai:2021lwb}. The development and implementation of these mappings may be the largest impediment for modern and future lattice QCD in studying increasingly rich and interesting reactions.

Quantum computers offer a complementary route to study multi-particle dynamics as they allow for direct real-time unitary evolution. Rapid development in quantum hardware in recent years~\cite{PRXQuantum.4.027001,osti_2588210,doi:10.1126/science.adz8659} has motivated the field to explore the feasibility of using these platforms for the simulation of nuclear and high-energy systems~\cite{Jordan:2012xnu,Jordan:2011ci,Jordan:2014tma,Bennewitz:2025nhz,Davoudi:2025rdv,Davoudi:2024wyv,Guo:2026nuc,Guo:2026qkx,Guo:2025vgk,Gustafson:2019mpk,Davoudi:2021ney,Bauer:2021gek,Bauer:2022hpo,Davoudi:2022bnl,Catterall:2022wjq,DiMeglio:2023nsa,Halimeh:2025vvp,Davoudi:2025kxb,Bauer:2025nzf,Raychowdhury:2019iki,Klco:2018zqz,Ciavarella:2021nmj,Farrell:2023fgd,Farrell:2024fit,Farrell:2025nkx,Schuhmacher:2025ehh,Bauer:2025oqz,Jha:2023ecu,Thompson:2021eze,Marshall:2015mna,Ale:2025sxz,deJong:2021wsd,Choi:2020pdg,
Ciavarella:2022zhe,Ciavarella:2022qdx,Ciavarella:2023mfc,Ciavarella:2024fzw,Ciavarella:2025bsg,Ciavarella:2025tdl,Modi:2026syn,Balaji:2025afl,Balaji:2025yua,Draper:2026bcj,Hidalgo:2026zsz,
Raychowdhury:2018tfj,Raychowdhury:2018osk,Kadam:2022ipf,Davoudi:2022xmb,Kadam:2024ifg,Burbano:2024uvn,Kadam:2025trs,Ilcic:2025gel,Das:2025utp,Gupta:2026tcg,
Chandrasekharan:1996ih,Brower:1997ha,Brower:2003vy,Zache:2021ggw,Halimeh:2021ufh,Osborne:2023rzx,Joshi:2025pgv,Cao:2026qky,Gandon:2026das,Joshi:2026hfe,Rule:2026brk}.  Presently, there are two classes of proposals for studying reactions using quantum computers.
The first uses wavepacket states to simulate scattering processes~\cite{Jordan:2012xnu,Jordan:2011ci,Jordan:2014tma}.
If these wavepackets are asymptotically free, the Fourier transform of this time evolution can be related to the S-matrix of the system. Preliminary simulations of elastic scattering events in one dimensional systems have been demonstrated~\cite{Gustafson:2019mpk,Gustafson:2021imb,Parks:2022kdb,Belyansky:2023rgh,Chai:2023qpq,Yusf:2024igb,Jha:2024jan,Zemlevskiy:2024vxt,Abel:2025zxb,Davoudi:2025rdv,Chai:2025qhf,Joshi:2025rha,Chai:2025kbi,Schuhmacher:2025ehh,Zemlevskiy:2026kpc} and extensions of this formalism for inelastic two-particle scattering~\cite{Belyansky:2023rgh,Jha:2024jan,Bennewitz:2025nhz,Farrell:2025nkx,Ingoldby:2025bdb,Schuhmacher:2025ehh,Davoudi:2025rdv,Chai:2025qhf,Wang:2025ocn,Artiaco:2025qqq,Zemlevskiy:2026kpc,Surace:2026rtg}, and electroweak processes~\cite{Farrell:2022vyh,Chernyshev:2025lil} have been initiated.

The second class of proposals for studying reactions are correlation function-based methods~\cite{Briceno:2020rar,Briceno:2023xcm,Carrillo:2024chu,Burbano:2025pef, Ciavarella:2020vqm, Guo:2025vgk, Guo:2026nuc, Guo:2026qkx}. Real-time Estimators for Scattering Observables (RESOs)~\cite{Briceno:2020rar,Briceno:2023xcm,Carrillo:2024chu,Burbano:2025pef} is unique among these methods because it provides a pathway to access any scattering observable, including those where external electroweak probes can be inserted. RESOs was originally designed to circumvent an obstacle that is shared by classical and quantum calculations, namely that a finite spatial volume has no asymptotic scattering states. RESOs addresses this problem by first constructing regulated energy-dependent quantities from finite-volume real-time correlators which have been proven to converge to the desired infinite-volume observable exponentially quickly~\cite{Burbano:2025pef}.

In this work, we carry out the first RESOs calculation on quantum hardware, taking a major step towards capitalizing on the power of quantum computers to study a broader class of reactions than are presently accessible with classical computing. 
We implement this procedure on a one-dimensional version of ``pionless EFT", a Effective Field Theory (EFT) for describing low-energy nuclear interactions.  Further details are provided in Ref.~\cite{long_paper}.

\paragraph*{Continuum Theory.} The  Lagrangian density of pionless EFT is given by~\cite{Kaplan:1996xu, Kaplan:1998tg, Kaplan:1998we, vanKolck:1998bw}
\begin{equation}
    \mathcal{L}
    = \psi^\dagger_\sigma \left(i\partial_t+\frac{\nabla^2}{2m}\right)\psi_\sigma
    -g\,\psi^\dagger_\uparrow\psi^\dagger_\downarrow
        \psi_\downarrow\psi_\uparrow\, ,
    \label{eq:lagrangian}
\end{equation}
where $\psi_\sigma$ is an anticommuting spin-1/2 fermionic field with a spin component $\sigma$, $m$ is its mass and $g$ is a dimensionless constant that characterizes the dynamics of the system.

We work in one spatial dimension, using ``nucleon'' and ``deuteron'' as shorthand for the fermion and its two-body bound state. In 1+1D, one can derive an exact analytic result for the continuum phase shift $\delta$, given by~\cite{long_paper}
\begin{equation}
    \cot\delta(E^\star)/p^\star=-\frac{2}{mg},
    \label{eq:exact_phase_shift}
\end{equation}
where $ p^\star=\sqrt{mE^\star}$ is the non-relativistic relative momentum in the center of mass frame (CMF) and $E^\star$ is the CMF energy.
This energy-independent ratio provides a direct benchmark for the hardware calculation.

For $g < 0$, the interaction is attractive, and the scattering amplitude has a pole at 
\begin{equation}
    E_B^\star=-\frac{mg^2}{4}\, \, ,
    \label{eq:continuum_binding_energy}
\end{equation}
where $E_B^\star$ corresponds to the negative of the binding energy of the deuteron.
To generate a system with a bound state, we will use $g=-0.2$ and $m=940~\mathrm{MeV}$, leading to $E^\star_B=-9.4$~MeV.

\paragraph*{Lattice Theory.} To make the theory amenable to quantum hardware, we discretize space as
$x_n=na$, with $a$ being the spatial lattice spacing, $n=0,\ldots,N_x-1$, and we relate the continuum and lattice
fields according to~\footnote{For useful references describing the study of this theory in a lattice, we point the reader to Refs.~\cite{Lee:2008fa,Endres:2011er,Lahde:2019npb, Rothman:2025uza, Chandrasekharan:2024iao, Singh:2018mnm, Lee:2004qd, Korber:2019cuq}.}
\begin{equation}
    \psi_\sigma(t,x_n)=\frac{1}{\sqrt{a}}\,c_{n,\sigma}(t),
    \label{eq:field_disc}
\end{equation}
resulting in the lattice field satisfying
$\{c_{n,\sigma},c^\dagger_{n',\sigma'}\}=\delta_{nn'}\delta_{\sigma\sigma'}$. We impose periodic boundary conditions on the fields, which constrains momenta $P$ to integer multiples of $2\pi/L$, where $L=a N_x$ is the spatial extent of the lattice.

The resulting lattice Hamiltonian is~\footnote{Note that this Hamiltonian is the one-dimensional Fermi-Hubbard model which describes the behavior of strongly correlated electrons in 1D materials~\cite{Hubbard:1963,Linke:2017xlv,Arute:2020ypn,Madhusudhana:2021qyp,Stanisic:2021irm,Chen:2023ukf,Paul:2024ldn,Srinivasan:2024fvq,Vilchez-Estevez:2025zjm,Chowdhury:2025tue,Hartnett:2026abf,Kalam:2026yxv}.}
\begin{align}
    a_t H
    &= \sum_{x,\sigma}\left(2 c^\dag_{x,\sigma} c_{x,\sigma}-c^\dagger_{x+1,\sigma}c_{x,\sigma}-c^\dagger_{x,\sigma}c_{x+1,\sigma}\right)
    \nonumber\\
    &\hspace{2cm}
    +a_tV\sum_x c^\dag_{x,\uparrow}c_{x,\uparrow}c^\dag_{x,\downarrow}c_{x,\downarrow},
    \label{eq:ham}
\end{align}
where $a_t = 2ma^2$   is an effective temporal lattice spacing that makes the Hamiltonian dimensionless, and $V$ is defined such that in the continuum limit  $aV$ approaches $g$. 
For the quantum-hardware calculation, the interaction
is fixed to $a_t V=-5$. 
To connect the lattice model to the continuum theory, we choose to tune $V$ to reproduce $E_B^\star$ in the infinite volume; this sets the scale of the problem to be $a \approx 2.3 \, {\rm fm}$.

\begin{figure}[t]
    \centering
    \includegraphics[width=0.5\textwidth]{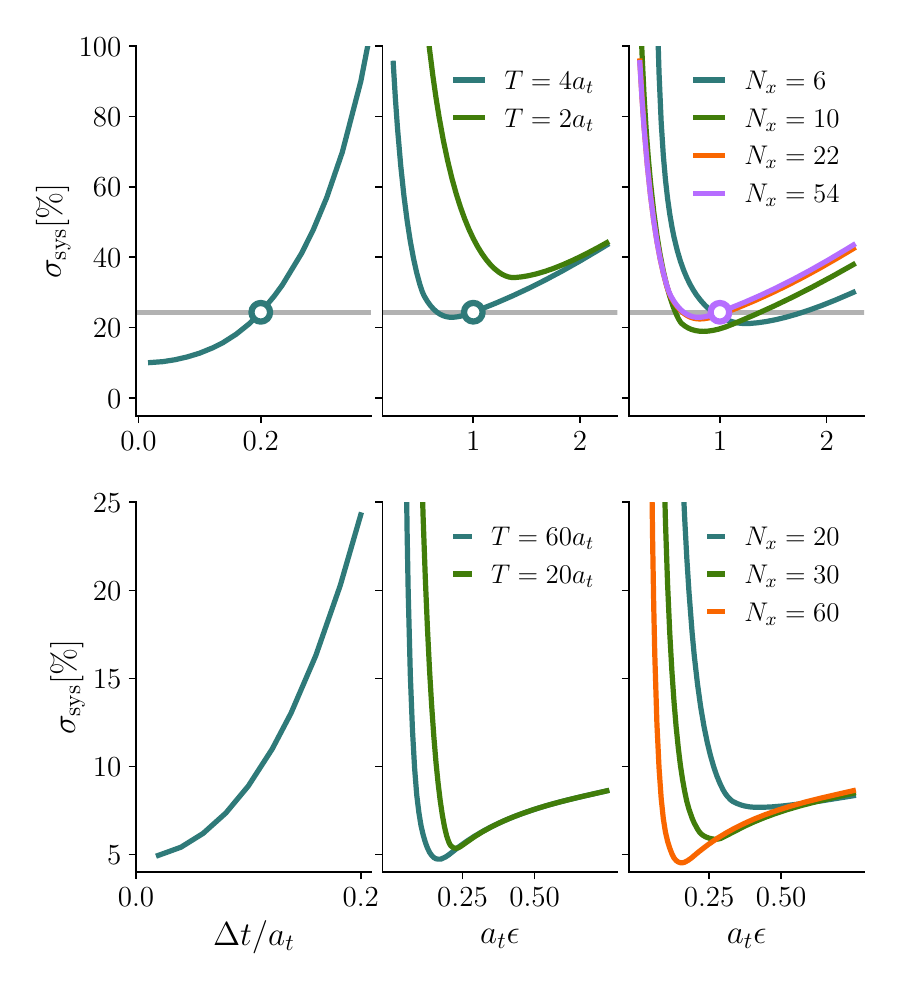}
    \caption{Average over the energy range $a_tE^\star = [0,2]$ of the error of the estimator in Eq.~\eqref{eq:delta_eff} relative to the analytical result in Eq.~\eqref{eq:lattice_ps}. 
    In the first row, unless stated in the plot, all parameters coincide with those used in the quantum hardware simulations with $N_x = 54$, which correspond to the markers and the horizontal lines. 
    The second row instead has nominal parameters $T=60a_t$, $N_x = 54$, $a_t\epsilon = 0.2$ and $\Delta t = 0.02a_t$.
    All results in this figure were obtained through tensor networks.
    }
    \label{fig:tensor_net}
\end{figure}

In a finite space, one loses Galilean invariance. As a result, one needs to know how to relate observables across different frames. At a finite lattice spacing, the CMF energy can be related to the other kinematic variables through the following two identities~\cite{long_paper}, 
\begin{align}
    a_t E^\star
    &= a_t E - 4 \left[ 1 - \cos\left( \frac{a P}{2} \right) \right] \, .
    \label{eq:lattice_disp}
    \\
    &=
    4 \cos \left( \frac{a P}{2} \right) \left[ 1 - \cos(a p^\star) \right] \, .
    \end{align}
Finally, these ingredients allow us to express the discretized version of Eq.~\eqref{eq:exact_phase_shift} as~\cite{long_paper} 
\begin{equation}
    \frac{\cot\delta(E^\star)}{\sin(ap^\star)}
    = - \frac{4}{a_t V} \cos \left( \frac{a P}{2} \right)  \, .
    \label{eq:lattice_ps}
\end{equation}

\begin{figure*}[t]
    \centering
    \includegraphics[width=\textwidth]{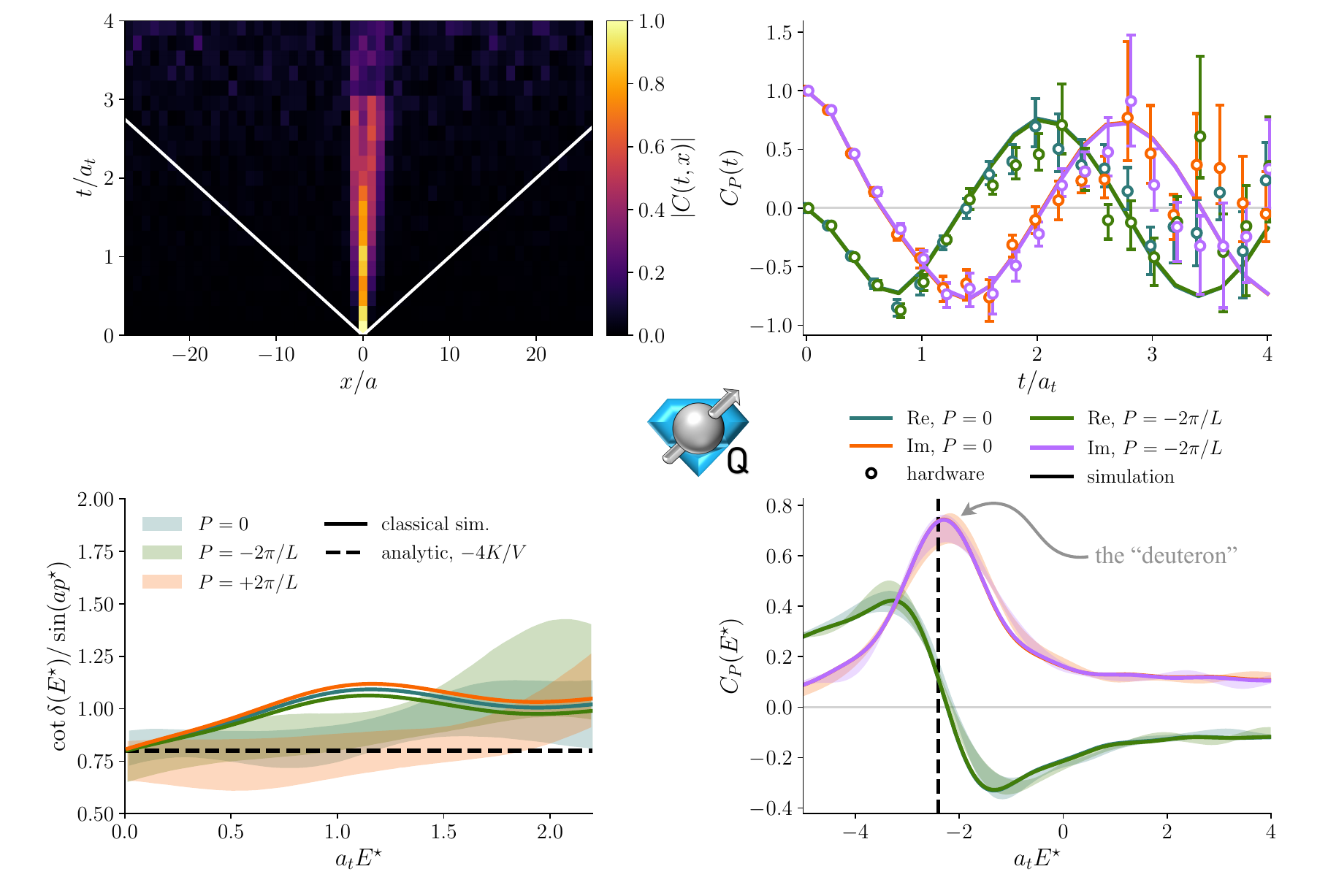}
    \caption{Shown are the three main stages of the correlation functions obtained from the quantum hardware. The top left panel shows the correlation functions in space and time for $N_x = 54$ spatial lattices. The white line shows the light cone of the quantum circuit used to simulate time evolution.
    The top right panel shows the result after Fourier transforming to definite spatial momentum for two momenta. The bottom right panel shows the correlator after Fourier transforming to energy for a fixed total $a_t \epsilon = 1$. The dashed line labels the location of the negative binding energy of the ``deuteron" bound state. The bottom left panel shows the scattering phase shifts computed from these correlators.  
    }
    \label{fig:raw_corr}
\end{figure*}

\paragraph*{Real-time estimator.} Having defined our target observable, we now proceed to describe the correlation functions used to compute it. We use a local two-nucleon interpolating field $\mathcal D(t,x)=c_{x,\downarrow}(t)c_{x,\uparrow}(t)$, which will create any spinless two-nucleon state, in addition to the deuteron. 

The basic quantity measured on the quantum hardware is the finite-volume real-time correlator
\begin{equation}
    C(t_j,x_n)
    =
    i\mel{0}{\mathcal D(t_j,x_n)
    \mathcal D^\dagger(0,0)}{0}_L,
    \label{eq:spacetime_correlator}
\end{equation}
where $t_j=j\Delta t$ and $T=N_t\Delta t$ is the maximum evolution time. In our study, we fix $N_t=20$ and $\Delta t / a_t = 0.2$ for the Trotter step.

We first project to total momentum,
\begin{equation}
 C_{P}(t_j)=\sum_{n=0}^{N_x-1}e^{-iPx_n}C(t_j,x_n),
 \label{eq:mom_proj}
\end{equation}
and then we project to energy using a standard Fourier transform, 
\begin{equation}
    C_P(a_t(E+i\epsilon))= \sum_{j=0}^{N_t} 
      w_j e^{i(E+i\epsilon) t_j}
    C_{P}(t_j),
    \label{eq:corr}
\end{equation}
where the integration weights $w_j$ correspond to Simpson’s rule, and we introduced an $\epsilon$ in the Fourier transform. This $\epsilon$ plays a critical role in the RESOs prescription~\cite{Briceno:2020rar, Carrillo:2024chu, Burbano:2025pef} by ensuring the Fourier transform converges while tempering finite-volume artifacts. The errors associated with both infrared regulators, namely the spatial extent $L$ and the maximal time extent considered $T$, scale as $\mathcal{O}(e^{-\epsilon L})$ and $\mathcal{O}(e^{-\epsilon T})$, respectively. At the same time, $\epsilon$ evaluates the correlation functions for a complex energy. Consequently, to compare with experimentally accessible observables, which are accessed using real energies, while taming these systematic errors, one wants to take the limit where $\epsilon$ is as small as possible, while keeping $\epsilon \gg L^{-1}, T^{-1}$. In the subsequent analysis, we fix $\epsilon T = 4$ to suppress finite-$T$ effects, which are the dominant source of systematic error, as can be seen in Fig.~\ref{fig:tensor_net}.

Given the correlation function defined in Eq.~\eqref{eq:corr}, one can show that in the infinite-volume limit the ratio of its real and imaginary parts yields the scattering phase shift~\cite{long_paper}. Away from this limit, we rely on the RESOs prescription, which ensures that the finite-volume observables converge quickly to their infinite-volume counterparts when averaging over different spatial momenta, $P$. With these two points in mind, we reconstruct our target observable using ~\cite{long_paper}
\begin{equation}
    \frac{\cot\delta(E^\star)}{\sin(ap^\star)}
    \equiv
        -\left\langle
    \frac{1}{\sin(ap^\star)}\frac{{\rm Im}\,C_P(a_tE) }
         {{\rm Re}\,C_P(a_tE) }
          \right\rangle_{P},
    \label{eq:delta_eff}
\end{equation}
where the average over different spatial momenta includes $LP/2\pi =0,\pm 1$.  

\begin{figure*}[t]
    \centering
    \includegraphics[width=0.92\textwidth]{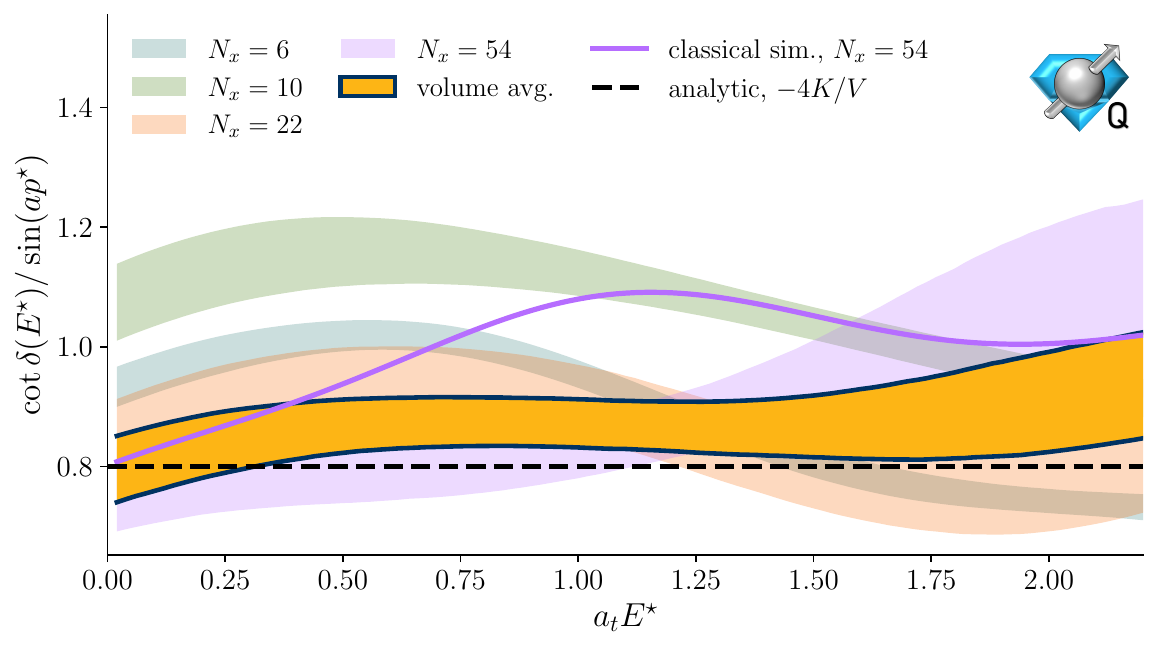}
    \caption{Shown are the resulting estimators for $\cot\delta/\sin(ap^\star)$. Each color corresponds to a different physical volume. The shaded regions show a $68\%$ confidence interval for the results from the quantum hardware and the solid lines show noiseless classical simulation of the Trotterized time evolution. The result is averaged over $P=0,\pm\frac{2\pi}{N_x}$ boosts. The solid gold region shows the $N_x=22$ and $N_x=54$ hardware results averaged together. The dashed line shows the analytic infinite-$N_x$ and $N_t$ prediction given in Eq.~\eqref{eq:lattice_ps}. 
    }
    \label{fig:main_results}
\end{figure*} 

\paragraph*{Benchmarks.} 
Before running on quantum hardware, a series of benchmarks, reported in Ref.~\cite{long_paper}, were performed. This included analytically solving for target observables in the discretized theory, numerically diagonalizing the Hamiltonian, and performing numerical Trotterized tensor-network simulations. Classical simulations on small volumes allowed us to assess Trotterization errors and test the convergence of the prescriptions. 
Figure~\ref{fig:tensor_net} shows the convergence of the estimators defined in Eq.~\eqref{eq:delta_eff}, both for parameters resembling those of the quantum hardware results in this paper, as well as a more forward-looking scenario where we can perform time evolution faithfully for a larger $T$.  
Further convergence tests and algorithmic details are given in Ref.~\cite{long_paper}

\paragraph*{Quantum implementation \& results.} Having passed this series of benchmarks, we proceeded to evaluate the correlation functions in IBM's quantum processor called~ibm\_boston, from the Heron r3 family of superconducting quantum processors. For this, we mapped the two fermion species to $2N_x$ qubits with a Jordan-Wigner transformation and used one ancilla to extract the real and imaginary parts of Eq.~\eqref{eq:corr} using the Hadamard test~\cite{10.1098/rspa.1998.0164}. For our largest volume, we used $N_x=54$, which required 109 qubits.

The Jordan-Wigner mapping with periodic boundary conditions leads to a hopping term that contains a string of $\hat{Z}$ operators that has support over the entire lattice. Because Fermion particle number ($n_f$) is conserved, we replace this operator by a sign of $(-1)^{n_f}$, effectively imposing (anti-)periodic qubit boundary conditions for (even) odd particle number. As interactions remain local in the qubit representation, very short-time dynamics can be computed exactly irrespective of system size. As a consequence, we can leverage it to estimate a hardware-induced $\epsilon$ shift. This shift can, in turn, be used for error mitigation for longer-time dynamics. For more details see Ref.~\cite{long_paper}.

Figure~\ref{fig:raw_corr} summarizes the path from circuit measurements to the energy-dependent correlation functions. 
The top-left panel shows the correlation function in space and time, corresponding to Eq.~\eqref{eq:spacetime_correlator}, obtained from the hardware.
The top right panel shows the Fourier transform in space to definite spatial momentum, corresponding to the correlation functions in Eq.~\eqref{eq:mom_proj}.

In the bottom right panel of Fig.~\ref{fig:raw_corr}, we see the results of the correlators after Fourier transforming to momentum and energy, corresponding to those defined in Eq.~\eqref{eq:corr}. Because energy is a continuous parameter introduced in the post-processing analysis of the quantum hardware results, we are able to evaluate these correlation functions for any energy value. The first feature that stands out is the presence of a peak in the correlation function below threshold. This is direct evidence of a dynamically generated bound state, namely the deuteron-like state in this theory, which is located at the dashed line.~\footnote{The fact that this is a finite-sized peak, as opposed to a pole, is due to the non-zero $\epsilon$.} Also, despite Galilean invariance being absent in a finite volume, because our volumes are relatively large, the correlation functions are only mildly sensitive to the total spatial momentum. Finally, in the bottom left panel, we show the determination of the estimator of the scattering observable for the three total momenta considered in the largest volume as a function of energy.

 Figure~\ref{fig:main_results} shows the boost-averaged estimator for the scattering parameters, according to Eq.~\eqref{eq:delta_eff}. For all volumes considered ($N_x = 6, 10, 22, 54$), we average over the spatial momenta, $LP/2\pi =0,\pm 1$. The hardware results and their corresponding statistical error are shown as continuous bands. These are compared with the classical simulations for $N_x=54$, shown as a solid line. From these, one sees no significant finite volume error for the time extents explored. Given our results for the two larger volumes are statistically consistent, we also show the result obtained after averaging over them. These are shown as a solid gold band. As can be seen, these are less than two standard deviations from the analytic result, shown as dashed lines.

\paragraph*{Outlook \& conclusion.}  This work presents the first quantum computation of a scattering observable using the RESOs protocol~\cite{Briceno:2020rar,Briceno:2023xcm,Carrillo:2024chu,Burbano:2025pef}. Although this work has focused on the simplest non-trivial two-particle scattering amplitude, the RESOs prescription in principle gives access to $n\to n'$ scattering amplitudes for any desired $n$ and $n'$, as well as any physical processes where external probes are being inserted. A useful future example is virtual Compton scattering~\cite{Briceno:2020rar,Carrillo:2024chu}, which gives access to parton distribution functions~\cite{Mueller:2019qqj,Lamm:2019uyc,Kreshchuk:2020dla,Echevarria:2020wct,Li:2021kcs,Qian:2021jxp,Grieninger:2024cdl,Chen:2025zeh,Zou:2026cfk}, as well as generalized parton distributions~\cite{Ji:1996nm}, both of which are a major part of the ongoing Jefferson Lab program and the future Electron Ion Collider.

Realizing the universal potential of the RESOs protocol requires two developments beyond the calculation presented here. The first is a formulation of lattice QCD that can be time-evolved on quantum hardware, including a qubit encoding of the SU($3$) gauge field with controlled errors~\cite{Byrnes:2005qx,Alexandru:2019nsa,Ciavarella:2021nmj,Ji:2022qvr,Ciavarella:2023mfc,Ciavarella:2024fzw,Gustafson:2024kym,Assi:2024pdn,Kadam:2024ifg,Balaji:2025afl,Illa:2025dou,Balaji:2025yua,Kadam:2025trs,Chen:2026hnh,Modi:2026syn,Siew:2026fax,Hidalgo:2026zsz,Yao:2026rya}, and circuits for preparing the vacuum~\cite{Atas:2021ext,Ciavarella:2021lel,Farrell:2023fgd,Farrell:2024fit,Ciavarella:2024lsp,Balaji:2025yua}. The second is an efficient method for evaluating the $(n+n')$-point real-time correlators that enter the estimator for $n\to n'$ scattering. Existing approaches to compute energy and momentum resolved correlators on quantum computers scale exponentially with the number of operator insertions, and this needs to be mitigated to apply RESOs at scale~\cite{Ortiz:2000gc,Somma:2001kjh,Pedernales:2014izf,Roggero:2018hrn,Rall:2020rsu,Kokcu:2023vwg,Wang:2025ojn}. Neither development is specific to RESOs, and progress on both is being made for other applications of quantum simulation to nuclear and high-energy physics. With the two developments above, the same procedure applies without modification to the amplitudes that lie beyond the reach of Euclidean lattice methods.

\section*{Acknowledgments}

\begin{acknowledgments}

This work was partly supported by the Laboratory Directed Research and Development Program of Lawrence Berkeley National Laboratory under U.S. Department of Energy Contract No. DE-AC02-05CH11231. It was also partly supported by the US Department of Energy, Office of Science, National Quantum Information Science Research Centers, Quantum Systems Accelerator (Award No. DE-SCL0000121) and partially through Quantum Information Science Enabled Discovery (QuantISED) for High Energy Physics (KA2401032).
This research used resources of the National Energy Research
Scientific Computing Center, a DOE Office of Science User Facility
supported by the Office of Science of the U.S. Department of Energy
under Contract No. DE-AC02-05CH11231 using NERSC award
NERSC DDR-ERCAP0038362. This work was also partly supported by the University of California, Berkeley Research Opportunity Program.
The work at the University of Maryland was supported by the U.S. Department of Energy (DOE), Office of Science, Office of Nuclear Physics (award no. DE-SC0026067) and Maryland Center for Fundamental Physics, Department of Physics, and College of Computer, Mathematical, and Natural Sciences at the University of Maryland. The authors acknowledge the use of Claude Opus 5 in the formatting of the plots in this paper. Simulation codes were cross-checked with the assistance of GPT-6 Astra.

\end{acknowledgments}

\bibliography{bibi}

\begin{thebibliography}{179}%
\makeatletter
\providecommand \@ifxundefined [1]{%
 \@ifx{#1\undefined}
}%
\providecommand \@ifnum [1]{%
 \ifnum #1\expandafter \@firstoftwo
 \else \expandafter \@secondoftwo
 \fi
}%
\providecommand \@ifx [1]{%
 \ifx #1\expandafter \@firstoftwo
 \else \expandafter \@secondoftwo
 \fi
}%
\providecommand \natexlab [1]{#1}%
\providecommand \enquote  [1]{``#1''}%
\providecommand \bibnamefont  [1]{#1}%
\providecommand \bibfnamefont [1]{#1}%
\providecommand \citenamefont [1]{#1}%
\providecommand \href@noop [0]{\@secondoftwo}%
\providecommand \href [0]{\begingroup \@sanitize@url \@href}%
\providecommand \@href[1]{\@@startlink{#1}\@@href}%
\providecommand \@@href[1]{\endgroup#1\@@endlink}%
\providecommand \@sanitize@url [0]{\catcode `\\12\catcode `\$12\catcode
  `\&12\catcode `\#12\catcode `\^12\catcode `\_12\catcode `\%12\relax}%
\providecommand \@@startlink[1]{}%
\providecommand \@@endlink[0]{}%
\providecommand \url  [0]{\begingroup\@sanitize@url \@url }%
\providecommand \@url [1]{\endgroup\@href {#1}{\urlprefix }}%
\providecommand \urlprefix  [0]{URL }%
\providecommand \Eprint [0]{\href }%
\providecommand \doibase [0]{http://dx.doi.org/}%
\providecommand \selectlanguage [0]{\@gobble}%
\providecommand \bibinfo  [0]{\@secondoftwo}%
\providecommand \bibfield  [0]{\@secondoftwo}%
\providecommand \translation [1]{[#1]}%
\providecommand \BibitemOpen [0]{}%
\providecommand \bibitemStop [0]{}%
\providecommand \bibitemNoStop [0]{.\EOS\space}%
\providecommand \EOS [0]{\spacefactor3000\relax}%
\providecommand \BibitemShut  [1]{\csname bibitem#1\endcsname}%
\let\auto@bib@innerbib\@empty
\bibitem [{\citenamefont {Briceno}\ \emph {et~al.}(2018)\citenamefont
  {Briceno}, \citenamefont {Dudek},\ and\ \citenamefont
  {Young}}]{Briceno:2017max}%
  \BibitemOpen
  \bibfield  {author} {\bibinfo {author} {\bibfnamefont {R.~A.}\ \bibnamefont
  {Briceno}}, \bibinfo {author} {\bibfnamefont {J.~J.}\ \bibnamefont {Dudek}},
  \ and\ \bibinfo {author} {\bibfnamefont {R.~D.}\ \bibnamefont {Young}},\
  }\href {\doibase 10.1103/RevModPhys.90.025001} {\bibfield  {journal}
  {\bibinfo  {journal} {Rev. Mod. Phys.}\ }\textbf {\bibinfo {volume} {90}},\
  \bibinfo {pages} {025001} (\bibinfo {year} {2018})},\ \Eprint
  {http://arxiv.org/abs/1706.06223} {arXiv:1706.06223} \BibitemShut {NoStop}%
\bibitem [{\citenamefont {Lebed}\ \emph {et~al.}(2017)\citenamefont {Lebed},
  \citenamefont {Mitchell},\ and\ \citenamefont {Swanson}}]{Lebed:2016hpi}%
  \BibitemOpen
  \bibfield  {author} {\bibinfo {author} {\bibfnamefont {R.~F.}\ \bibnamefont
  {Lebed}}, \bibinfo {author} {\bibfnamefont {R.~E.}\ \bibnamefont {Mitchell}},
  \ and\ \bibinfo {author} {\bibfnamefont {E.~S.}\ \bibnamefont {Swanson}},\
  }\href {\doibase 10.1016/j.ppnp.2016.11.003} {\bibfield  {journal} {\bibinfo
  {journal} {Prog. Part. Nucl. Phys.}\ }\textbf {\bibinfo {volume} {93}},\
  \bibinfo {pages} {143} (\bibinfo {year} {2017})},\ \Eprint
  {http://arxiv.org/abs/1610.04528} {arXiv:1610.04528 [hep-ph]} \BibitemShut
  {NoStop}%
\bibitem [{\citenamefont {Accardi}\ \emph {et~al.}(2016)\citenamefont {Accardi}
  \emph {et~al.}}]{Accardi:2012qut}%
  \BibitemOpen
  \bibfield  {author} {\bibinfo {author} {\bibfnamefont {A.}~\bibnamefont
  {Accardi}} \emph {et~al.},\ }\href {\doibase 10.1140/epja/i2016-16268-9}
  {\bibfield  {journal} {\bibinfo  {journal} {Eur. Phys. J. A}\ }\textbf
  {\bibinfo {volume} {52}},\ \bibinfo {pages} {268} (\bibinfo {year} {2016})},\
  \Eprint {http://arxiv.org/abs/1212.1701} {arXiv:1212.1701 [nucl-ex]}
  \BibitemShut {NoStop}%
\bibitem [{\citenamefont {Abdul~Khalek}\ \emph {et~al.}(2022)\citenamefont
  {Abdul~Khalek} \emph {et~al.}}]{AbdulKhalek:2022hcn}%
  \BibitemOpen
  \bibfield  {author} {\bibinfo {author} {\bibfnamefont {R.}~\bibnamefont
  {Abdul~Khalek}} \emph {et~al.},\ }\href@noop {} {\enquote {\bibinfo {title}
  {{Snowmass 2021 White Paper: Electron Ion Collider for High Energy
  Physics}},}\ } (\bibinfo {year} {2022}),\ \Eprint
  {http://arxiv.org/abs/2203.13199} {arXiv:2203.13199 [hep-ph]} \BibitemShut
  {NoStop}%
\bibitem [{\citenamefont {Acciarri}\ \emph {et~al.}(2016)\citenamefont
  {Acciarri} \emph {et~al.}}]{DUNE:2016hlj}%
  \BibitemOpen
  \bibfield  {author} {\bibinfo {author} {\bibfnamefont {R.}~\bibnamefont
  {Acciarri}} \emph {et~al.} (\bibinfo {collaboration} {DUNE}),\ }\href@noop {}
  {\enquote {\bibinfo {title} {{Long-Baseline Neutrino Facility (LBNF) and Deep
  Underground Neutrino Experiment (DUNE)}: {Conceptual Design Report, Volume 1:
  The LBNF and DUNE Projects}},}\ } (\bibinfo {year} {2016}),\ \Eprint
  {http://arxiv.org/abs/1601.05471} {arXiv:1601.05471 [physics.ins-det]}
  \BibitemShut {NoStop}%
\bibitem [{\citenamefont {Abi}\ \emph {et~al.}(2021)\citenamefont {Abi} \emph
  {et~al.}}]{DUNE:2020fgq}%
  \BibitemOpen
  \bibfield  {author} {\bibinfo {author} {\bibfnamefont {B.}~\bibnamefont
  {Abi}} \emph {et~al.} (\bibinfo {collaboration} {DUNE}),\ }\href {\doibase
  10.1140/epjc/s10052-021-09007-w} {\bibfield  {journal} {\bibinfo  {journal}
  {Eur. Phys. J. C}\ }\textbf {\bibinfo {volume} {81}},\ \bibinfo {pages} {322}
  (\bibinfo {year} {2021})},\ \Eprint {http://arxiv.org/abs/2008.12769}
  {arXiv:2008.12769 [hep-ex]} \BibitemShut {NoStop}%
\bibitem [{\citenamefont {Abgrall}\ \emph {et~al.}(2021)\citenamefont {Abgrall}
  \emph {et~al.}}]{LEGEND:2021bnm}%
  \BibitemOpen
  \bibfield  {author} {\bibinfo {author} {\bibfnamefont {N.}~\bibnamefont
  {Abgrall}} \emph {et~al.} (\bibinfo {collaboration} {LEGEND}),\ }\href@noop
  {} {\  (\bibinfo {year} {2021})},\ \Eprint {http://arxiv.org/abs/2107.11462}
  {arXiv:2107.11462 [physics.ins-det]} \BibitemShut {NoStop}%
\bibitem [{\citenamefont {Armstrong}\ \emph {et~al.}(2019)\citenamefont
  {Armstrong} \emph {et~al.}}]{CUPID:2019imh}%
  \BibitemOpen
  \bibfield  {author} {\bibinfo {author} {\bibfnamefont {W.~R.}\ \bibnamefont
  {Armstrong}} \emph {et~al.} (\bibinfo {collaboration} {CUPID}),\ }\href@noop
  {} {\  (\bibinfo {year} {2019})},\ \Eprint {http://arxiv.org/abs/1907.09376}
  {arXiv:1907.09376 [physics.ins-det]} \BibitemShut {NoStop}%
\bibitem [{\citenamefont {Abe}\ \emph {et~al.}(2025)\citenamefont {Abe} \emph
  {et~al.}}]{KamLAND-Zen:2024eml}%
  \BibitemOpen
  \bibfield  {author} {\bibinfo {author} {\bibfnamefont {S.}~\bibnamefont
  {Abe}} \emph {et~al.} (\bibinfo {collaboration} {KamLAND-Zen}),\ }\href
  {\doibase 10.1103/jkf6-48j8} {\bibfield  {journal} {\bibinfo  {journal}
  {Phys. Rev. Lett.}\ }\textbf {\bibinfo {volume} {135}},\ \bibinfo {pages}
  {262501} (\bibinfo {year} {2025})},\ \Eprint
  {http://arxiv.org/abs/2406.11438} {arXiv:2406.11438 [hep-ex]} \BibitemShut
  {NoStop}%
\bibitem [{\citenamefont {Agostini}\ \emph {et~al.}(2020)\citenamefont
  {Agostini} \emph {et~al.}}]{GERDA:2020xhi}%
  \BibitemOpen
  \bibfield  {author} {\bibinfo {author} {\bibfnamefont {M.}~\bibnamefont
  {Agostini}} \emph {et~al.} (\bibinfo {collaboration} {GERDA}),\ }\href
  {\doibase 10.1103/PhysRevLett.125.252502} {\bibfield  {journal} {\bibinfo
  {journal} {Phys. Rev. Lett.}\ }\textbf {\bibinfo {volume} {125}},\ \bibinfo
  {pages} {252502} (\bibinfo {year} {2020})},\ \Eprint
  {http://arxiv.org/abs/2009.06079} {arXiv:2009.06079 [nucl-ex]} \BibitemShut
  {NoStop}%
\bibitem [{\citenamefont {Wilson}\ \emph {et~al.}(2015)\citenamefont {Wilson},
  \citenamefont {Dudek}, \citenamefont {Edwards},\ and\ \citenamefont
  {Thomas}}]{Wilson:2014cna}%
  \BibitemOpen
  \bibfield  {author} {\bibinfo {author} {\bibfnamefont {D.~J.}\ \bibnamefont
  {Wilson}}, \bibinfo {author} {\bibfnamefont {J.~J.}\ \bibnamefont {Dudek}},
  \bibinfo {author} {\bibfnamefont {R.~G.}\ \bibnamefont {Edwards}}, \ and\
  \bibinfo {author} {\bibfnamefont {C.~E.}\ \bibnamefont {Thomas}},\ }\href
  {\doibase 10.1103/PhysRevD.91.054008} {\bibfield  {journal} {\bibinfo
  {journal} {Phys. Rev. D}\ }\textbf {\bibinfo {volume} {91}},\ \bibinfo
  {pages} {054008} (\bibinfo {year} {2015})},\ \Eprint
  {http://arxiv.org/abs/1411.2004} {arXiv:1411.2004 [hep-ph]} \BibitemShut
  {NoStop}%
\bibitem [{\citenamefont {Dudek}\ \emph {et~al.}(2016)\citenamefont {Dudek},
  \citenamefont {Edwards},\ and\ \citenamefont {Wilson}}]{Dudek:2016cru}%
  \BibitemOpen
  \bibfield  {author} {\bibinfo {author} {\bibfnamefont {J.~J.}\ \bibnamefont
  {Dudek}}, \bibinfo {author} {\bibfnamefont {R.~G.}\ \bibnamefont {Edwards}},
  \ and\ \bibinfo {author} {\bibfnamefont {D.~J.}\ \bibnamefont {Wilson}}
  (\bibinfo {collaboration} {Hadron Spectrum}),\ }\href {\doibase
  10.1103/PhysRevD.93.094506} {\bibfield  {journal} {\bibinfo  {journal} {Phys.
  Rev. D}\ }\textbf {\bibinfo {volume} {93}},\ \bibinfo {pages} {094506}
  (\bibinfo {year} {2016})},\ \Eprint {http://arxiv.org/abs/1602.05122}
  {arXiv:1602.05122 [hep-ph]} \BibitemShut {NoStop}%
\bibitem [{\citenamefont {Woss}\ \emph {et~al.}(2021)\citenamefont {Woss},
  \citenamefont {Dudek}, \citenamefont {Edwards}, \citenamefont {Thomas},\ and\
  \citenamefont {Wilson}}]{Woss:2020ayi}%
  \BibitemOpen
  \bibfield  {author} {\bibinfo {author} {\bibfnamefont {A.~J.}\ \bibnamefont
  {Woss}}, \bibinfo {author} {\bibfnamefont {J.~J.}\ \bibnamefont {Dudek}},
  \bibinfo {author} {\bibfnamefont {R.~G.}\ \bibnamefont {Edwards}}, \bibinfo
  {author} {\bibfnamefont {C.~E.}\ \bibnamefont {Thomas}}, \ and\ \bibinfo
  {author} {\bibfnamefont {D.~J.}\ \bibnamefont {Wilson}} (\bibinfo
  {collaboration} {Hadron Spectrum}),\ }\href {\doibase
  10.1103/PhysRevD.103.054502} {\bibfield  {journal} {\bibinfo  {journal}
  {Phys. Rev. D}\ }\textbf {\bibinfo {volume} {103}},\ \bibinfo {pages}
  {054502} (\bibinfo {year} {2021})},\ \Eprint
  {http://arxiv.org/abs/2009.10034} {arXiv:2009.10034 [hep-lat]} \BibitemShut
  {NoStop}%
\bibitem [{\citenamefont {Boyle}\ \emph {et~al.}(2025)\citenamefont {Boyle},
  \citenamefont {Erben}, \citenamefont {G{\"u}lpers}, \citenamefont {Hansen},
  \citenamefont {Joswig}, \citenamefont {Marshall}, \citenamefont {Lachini},\
  and\ \citenamefont {Portelli}}]{Boyle:2024hvv}%
  \BibitemOpen
  \bibfield  {author} {\bibinfo {author} {\bibfnamefont {P.}~\bibnamefont
  {Boyle}}, \bibinfo {author} {\bibfnamefont {F.}~\bibnamefont {Erben}},
  \bibinfo {author} {\bibfnamefont {V.}~\bibnamefont {G{\"u}lpers}}, \bibinfo
  {author} {\bibfnamefont {M.~T.}\ \bibnamefont {Hansen}}, \bibinfo {author}
  {\bibfnamefont {F.}~\bibnamefont {Joswig}}, \bibinfo {author} {\bibfnamefont
  {M.}~\bibnamefont {Marshall}}, \bibinfo {author} {\bibfnamefont {N.~P.}\
  \bibnamefont {Lachini}}, \ and\ \bibinfo {author} {\bibfnamefont
  {A.}~\bibnamefont {Portelli}},\ }\href {\doibase
  10.1103/PhysRevLett.134.111901} {\bibfield  {journal} {\bibinfo  {journal}
  {Phys. Rev. Lett.}\ }\textbf {\bibinfo {volume} {134}},\ \bibinfo {pages}
  {111901} (\bibinfo {year} {2025})},\ \Eprint
  {http://arxiv.org/abs/2406.19194} {arXiv:2406.19194 [hep-lat]} \BibitemShut
  {NoStop}%
\bibitem [{\citenamefont {Pitanga~Lachini}\ and\ \citenamefont
  {Brice{\~n}o}(2026)}]{PitangaLachini:2026lyd}%
  \BibitemOpen
  \bibfield  {author} {\bibinfo {author} {\bibfnamefont {N.}~\bibnamefont
  {Pitanga~Lachini}}\ and\ \bibinfo {author} {\bibfnamefont {R.~A.}\
  \bibnamefont {Brice{\~n}o}},\ }\href@noop {} {\enquote {\bibinfo {title}
  {{Resolving the $T_{cc}^+$ with $\pi$ Exchange from Lattice QCD}},}\ }
  (\bibinfo {year} {2026}),\ \Eprint {http://arxiv.org/abs/2607.19009}
  {arXiv:2607.19009 [hep-lat]} \BibitemShut {NoStop}%
\bibitem [{\citenamefont {Mai}\ and\ \citenamefont
  {Doring}(2019)}]{Mai:2018djl}%
  \BibitemOpen
  \bibfield  {author} {\bibinfo {author} {\bibfnamefont {M.}~\bibnamefont
  {Mai}}\ and\ \bibinfo {author} {\bibfnamefont {M.}~\bibnamefont {Doring}},\
  }\href {\doibase 10.1103/PhysRevLett.122.062503} {\bibfield  {journal}
  {\bibinfo  {journal} {Phys. Rev. Lett.}\ }\textbf {\bibinfo {volume} {122}},\
  \bibinfo {pages} {062503} (\bibinfo {year} {2019})},\ \Eprint
  {http://arxiv.org/abs/1807.04746} {arXiv:1807.04746 [hep-lat]} \BibitemShut
  {NoStop}%
\bibitem [{\citenamefont {Hansen}\ \emph {et~al.}(2021)\citenamefont {Hansen},
  \citenamefont {Brice{\~n}o}, \citenamefont {Edwards}, \citenamefont
  {Thomas},\ and\ \citenamefont {Wilson}}]{Hansen:2020otl}%
  \BibitemOpen
  \bibfield  {author} {\bibinfo {author} {\bibfnamefont {M.~T.}\ \bibnamefont
  {Hansen}}, \bibinfo {author} {\bibfnamefont {R.~A.}\ \bibnamefont
  {Brice{\~n}o}}, \bibinfo {author} {\bibfnamefont {R.~G.}\ \bibnamefont
  {Edwards}}, \bibinfo {author} {\bibfnamefont {C.~E.}\ \bibnamefont {Thomas}},
  \ and\ \bibinfo {author} {\bibfnamefont {D.~J.}\ \bibnamefont {Wilson}}
  (\bibinfo {collaboration} {Hadron Spectrum}),\ }\href {\doibase
  10.1103/PhysRevLett.126.012001} {\bibfield  {journal} {\bibinfo  {journal}
  {Phys. Rev. Lett.}\ }\textbf {\bibinfo {volume} {126}},\ \bibinfo {pages}
  {012001} (\bibinfo {year} {2021})},\ \Eprint
  {http://arxiv.org/abs/2009.04931} {arXiv:2009.04931 [hep-lat]} \BibitemShut
  {NoStop}%
\bibitem [{\citenamefont {Dawid}\ \emph {et~al.}(2025)\citenamefont {Dawid},
  \citenamefont {Draper}, \citenamefont {Hanlon}, \citenamefont {H{\"o}rz},
  \citenamefont {Morningstar}, \citenamefont {Romero-L{\'o}pez}, \citenamefont
  {Sharpe},\ and\ \citenamefont {Skinner}}]{Dawid:2025zxc}%
  \BibitemOpen
  \bibfield  {author} {\bibinfo {author} {\bibfnamefont {S.~M.}\ \bibnamefont
  {Dawid}}, \bibinfo {author} {\bibfnamefont {Z.~T.}\ \bibnamefont {Draper}},
  \bibinfo {author} {\bibfnamefont {A.~D.}\ \bibnamefont {Hanlon}}, \bibinfo
  {author} {\bibfnamefont {B.}~\bibnamefont {H{\"o}rz}}, \bibinfo {author}
  {\bibfnamefont {C.}~\bibnamefont {Morningstar}}, \bibinfo {author}
  {\bibfnamefont {F.}~\bibnamefont {Romero-L{\'o}pez}}, \bibinfo {author}
  {\bibfnamefont {S.~R.}\ \bibnamefont {Sharpe}}, \ and\ \bibinfo {author}
  {\bibfnamefont {S.}~\bibnamefont {Skinner}},\ }\href {\doibase
  10.1103/6nql-yrhw} {\bibfield  {journal} {\bibinfo  {journal} {Phys. Rev.
  Lett.}\ }\textbf {\bibinfo {volume} {135}},\ \bibinfo {pages} {021903}
  (\bibinfo {year} {2025})},\ \Eprint {http://arxiv.org/abs/2502.14348}
  {arXiv:2502.14348 [hep-lat]} \BibitemShut {NoStop}%
\bibitem [{\citenamefont {Yan}\ \emph {et~al.}(2026)\citenamefont {Yan},
  \citenamefont {Mai}, \citenamefont {Garofalo}, \citenamefont {Feng},
  \citenamefont {D{\"o}ring}, \citenamefont {Liu}, \citenamefont {Liu},
  \citenamefont {Mei{\ss}ner},\ and\ \citenamefont {Urbach}}]{Yan:2025mdm}%
  \BibitemOpen
  \bibfield  {author} {\bibinfo {author} {\bibfnamefont {H.}~\bibnamefont
  {Yan}}, \bibinfo {author} {\bibfnamefont {M.}~\bibnamefont {Mai}}, \bibinfo
  {author} {\bibfnamefont {M.}~\bibnamefont {Garofalo}}, \bibinfo {author}
  {\bibfnamefont {Y.}~\bibnamefont {Feng}}, \bibinfo {author} {\bibfnamefont
  {M.}~\bibnamefont {D{\"o}ring}}, \bibinfo {author} {\bibfnamefont
  {C.}~\bibnamefont {Liu}}, \bibinfo {author} {\bibfnamefont {L.}~\bibnamefont
  {Liu}}, \bibinfo {author} {\bibfnamefont {U.-G.}\ \bibnamefont
  {Mei{\ss}ner}}, \ and\ \bibinfo {author} {\bibfnamefont {C.}~\bibnamefont
  {Urbach}},\ }\href {\doibase 10.1103/vfr3-5lsb} {\bibfield  {journal}
  {\bibinfo  {journal} {Phys. Rev. Lett.}\ }\textbf {\bibinfo {volume} {136}},\
  \bibinfo {pages} {141901} (\bibinfo {year} {2026})},\ \Eprint
  {http://arxiv.org/abs/2510.09476} {arXiv:2510.09476 [hep-lat]} \BibitemShut
  {NoStop}%
\bibitem [{\citenamefont {Yan}\ \emph {et~al.}(2024)\citenamefont {Yan},
  \citenamefont {Mai}, \citenamefont {Garofalo}, \citenamefont {Mei{\ss}ner},
  \citenamefont {Liu}, \citenamefont {Liu},\ and\ \citenamefont
  {Urbach}}]{Yan:2024gwp}%
  \BibitemOpen
  \bibfield  {author} {\bibinfo {author} {\bibfnamefont {H.}~\bibnamefont
  {Yan}}, \bibinfo {author} {\bibfnamefont {M.}~\bibnamefont {Mai}}, \bibinfo
  {author} {\bibfnamefont {M.}~\bibnamefont {Garofalo}}, \bibinfo {author}
  {\bibfnamefont {U.-G.}\ \bibnamefont {Mei{\ss}ner}}, \bibinfo {author}
  {\bibfnamefont {C.}~\bibnamefont {Liu}}, \bibinfo {author} {\bibfnamefont
  {L.}~\bibnamefont {Liu}}, \ and\ \bibinfo {author} {\bibfnamefont
  {C.}~\bibnamefont {Urbach}},\ }\href {\doibase
  10.1103/PhysRevLett.133.211906} {\bibfield  {journal} {\bibinfo  {journal}
  {Phys. Rev. Lett.}\ }\textbf {\bibinfo {volume} {133}},\ \bibinfo {pages}
  {211906} (\bibinfo {year} {2024})},\ \Eprint
  {http://arxiv.org/abs/2407.16659} {arXiv:2407.16659 [hep-lat]} \BibitemShut
  {NoStop}%
\bibitem [{\citenamefont {Luscher}(1986{\natexlab{a}})}]{Luscher:1985dn}%
  \BibitemOpen
  \bibfield  {author} {\bibinfo {author} {\bibfnamefont {M.}~\bibnamefont
  {Luscher}},\ }\href {\doibase 10.1007/BF01211589} {\bibfield  {journal}
  {\bibinfo  {journal} {Commun. Math. Phys.}\ }\textbf {\bibinfo {volume}
  {104}},\ \bibinfo {pages} {177} (\bibinfo {year}
  {1986}{\natexlab{a}})}\BibitemShut {NoStop}%
\bibitem [{\citenamefont {Luscher}(1986{\natexlab{b}})}]{Luscher:1986pf}%
  \BibitemOpen
  \bibfield  {author} {\bibinfo {author} {\bibfnamefont {M.}~\bibnamefont
  {Luscher}},\ }\href {\doibase 10.1007/BF01211097} {\bibfield  {journal}
  {\bibinfo  {journal} {Commun. Math. Phys.}\ }\textbf {\bibinfo {volume}
  {105}},\ \bibinfo {pages} {153} (\bibinfo {year}
  {1986}{\natexlab{b}})}\BibitemShut {NoStop}%
\bibitem [{\citenamefont {Luscher}(1991)}]{Luscher:1990ux}%
  \BibitemOpen
  \bibfield  {author} {\bibinfo {author} {\bibfnamefont {M.}~\bibnamefont
  {Luscher}},\ }\href {\doibase 10.1016/0550-3213(91)90366-6} {\bibfield
  {journal} {\bibinfo  {journal} {Nucl. Phys. B}\ }\textbf {\bibinfo {volume}
  {354}},\ \bibinfo {pages} {531} (\bibinfo {year} {1991})}\BibitemShut
  {NoStop}%
\bibitem [{\citenamefont {Kim}\ \emph {et~al.}(2005)\citenamefont {Kim},
  \citenamefont {Sachrajda},\ and\ \citenamefont {Sharpe}}]{Kim:2005gf}%
  \BibitemOpen
  \bibfield  {author} {\bibinfo {author} {\bibfnamefont {C.~h.}\ \bibnamefont
  {Kim}}, \bibinfo {author} {\bibfnamefont {C.~T.}\ \bibnamefont {Sachrajda}},
  \ and\ \bibinfo {author} {\bibfnamefont {S.~R.}\ \bibnamefont {Sharpe}},\
  }\href {\doibase 10.1016/j.nuclphysb.2005.08.029} {\bibfield  {journal}
  {\bibinfo  {journal} {Nucl. Phys. B}\ }\textbf {\bibinfo {volume} {727}},\
  \bibinfo {pages} {218} (\bibinfo {year} {2005})},\ \Eprint
  {http://arxiv.org/abs/hep-lat/0507006} {arXiv:hep-lat/0507006} \BibitemShut
  {NoStop}%
\bibitem [{\citenamefont {Briceno}(2014)}]{Briceno:2014oea}%
  \BibitemOpen
  \bibfield  {author} {\bibinfo {author} {\bibfnamefont {R.~A.}\ \bibnamefont
  {Briceno}},\ }\href {\doibase 10.1103/PhysRevD.89.074507} {\bibfield
  {journal} {\bibinfo  {journal} {Phys. Rev. D}\ }\textbf {\bibinfo {volume}
  {89}},\ \bibinfo {pages} {074507} (\bibinfo {year} {2014})},\ \Eprint
  {http://arxiv.org/abs/1401.3312} {arXiv:1401.3312 [hep-lat]} \BibitemShut
  {NoStop}%
\bibitem [{\citenamefont {Hansen}\ and\ \citenamefont
  {Sharpe}(2014)}]{Hansen:2014eka}%
  \BibitemOpen
  \bibfield  {author} {\bibinfo {author} {\bibfnamefont {M.~T.}\ \bibnamefont
  {Hansen}}\ and\ \bibinfo {author} {\bibfnamefont {S.~R.}\ \bibnamefont
  {Sharpe}},\ }\href {\doibase 10.1103/PhysRevD.90.116003} {\bibfield
  {journal} {\bibinfo  {journal} {Phys. Rev. D}\ }\textbf {\bibinfo {volume}
  {90}},\ \bibinfo {pages} {116003} (\bibinfo {year} {2014})},\ \Eprint
  {http://arxiv.org/abs/1408.5933} {arXiv:1408.5933 [hep-lat]} \BibitemShut
  {NoStop}%
\bibitem [{\citenamefont {Hansen}\ and\ \citenamefont
  {Sharpe}(2015)}]{Hansen:2015zga}%
  \BibitemOpen
  \bibfield  {author} {\bibinfo {author} {\bibfnamefont {M.~T.}\ \bibnamefont
  {Hansen}}\ and\ \bibinfo {author} {\bibfnamefont {S.~R.}\ \bibnamefont
  {Sharpe}},\ }\href {\doibase 10.1103/PhysRevD.92.114509} {\bibfield
  {journal} {\bibinfo  {journal} {Phys. Rev. D}\ }\textbf {\bibinfo {volume}
  {92}},\ \bibinfo {pages} {114509} (\bibinfo {year} {2015})},\ \Eprint
  {http://arxiv.org/abs/1504.04248} {arXiv:1504.04248 [hep-lat]} \BibitemShut
  {NoStop}%
\bibitem [{\citenamefont {Mai}\ and\ \citenamefont
  {D\"oring}(2017)}]{Mai:2017bge}%
  \BibitemOpen
  \bibfield  {author} {\bibinfo {author} {\bibfnamefont {M.}~\bibnamefont
  {Mai}}\ and\ \bibinfo {author} {\bibfnamefont {M.}~\bibnamefont {D\"oring}},\
  }\href {\doibase 10.1140/epja/i2017-12440-1} {\bibfield  {journal} {\bibinfo
  {journal} {Eur. Phys. J. A}\ }\textbf {\bibinfo {volume} {53}},\ \bibinfo
  {pages} {240} (\bibinfo {year} {2017})},\ \Eprint
  {http://arxiv.org/abs/1709.08222} {arXiv:1709.08222 [hep-lat]} \BibitemShut
  {NoStop}%
\bibitem [{\citenamefont {D{\"o}ring}\ \emph {et~al.}(2018)\citenamefont
  {D{\"o}ring}, \citenamefont {Hammer}, \citenamefont {Mai}, \citenamefont
  {Pang}, \citenamefont {Rusetsky},\ and\ \citenamefont {Wu}}]{Doring:2018xxx}%
  \BibitemOpen
  \bibfield  {author} {\bibinfo {author} {\bibfnamefont {M.}~\bibnamefont
  {D{\"o}ring}}, \bibinfo {author} {\bibfnamefont {H.~W.}\ \bibnamefont
  {Hammer}}, \bibinfo {author} {\bibfnamefont {M.}~\bibnamefont {Mai}},
  \bibinfo {author} {\bibfnamefont {J.~Y.}\ \bibnamefont {Pang}}, \bibinfo
  {author} {\bibfnamefont {{\textsection}.~A.}\ \bibnamefont {Rusetsky}}, \
  and\ \bibinfo {author} {\bibfnamefont {J.}~\bibnamefont {Wu}},\ }\href
  {\doibase 10.1103/PhysRevD.97.114508} {\bibfield  {journal} {\bibinfo
  {journal} {Phys. Rev. D}\ }\textbf {\bibinfo {volume} {97}},\ \bibinfo
  {pages} {114508} (\bibinfo {year} {2018})},\ \Eprint
  {http://arxiv.org/abs/1802.03362} {arXiv:1802.03362 [hep-lat]} \BibitemShut
  {NoStop}%
\bibitem [{\citenamefont {Jackura}(2023)}]{Jackura:2022gib}%
  \BibitemOpen
  \bibfield  {author} {\bibinfo {author} {\bibfnamefont {A.~W.}\ \bibnamefont
  {Jackura}},\ }\href {\doibase 10.1103/PhysRevD.108.034505} {\bibfield
  {journal} {\bibinfo  {journal} {Phys. Rev. D}\ }\textbf {\bibinfo {volume}
  {108}},\ \bibinfo {pages} {034505} (\bibinfo {year} {2023})},\ \Eprint
  {http://arxiv.org/abs/2208.10587} {arXiv:2208.10587 [hep-lat]} \BibitemShut
  {NoStop}%
\bibitem [{\citenamefont {Raposo}\ \emph {et~al.}(2025)\citenamefont {Raposo},
  \citenamefont {Brice{\~n}o}, \citenamefont {Hansen},\ and\ \citenamefont
  {Jackura}}]{Raposo:2025dkb}%
  \BibitemOpen
  \bibfield  {author} {\bibinfo {author} {\bibfnamefont {A.~B.}\ \bibnamefont
  {Raposo}}, \bibinfo {author} {\bibfnamefont {R.~A.}\ \bibnamefont
  {Brice{\~n}o}}, \bibinfo {author} {\bibfnamefont {M.~T.}\ \bibnamefont
  {Hansen}}, \ and\ \bibinfo {author} {\bibfnamefont {A.~W.}\ \bibnamefont
  {Jackura}},\ }\href {\doibase 10.1007/JHEP06(2025)186} {\bibfield  {journal}
  {\bibinfo  {journal} {JHEP}\ }\textbf {\bibinfo {volume} {06}},\ \bibinfo
  {pages} {186} (\bibinfo {year} {2025})},\ \Eprint
  {http://arxiv.org/abs/2502.19375} {arXiv:2502.19375 [hep-lat]} \BibitemShut
  {NoStop}%
\bibitem [{\citenamefont {Hansen}\ and\ \citenamefont
  {Sharpe}(2019)}]{Hansen:2019nir}%
  \BibitemOpen
  \bibfield  {author} {\bibinfo {author} {\bibfnamefont {M.~T.}\ \bibnamefont
  {Hansen}}\ and\ \bibinfo {author} {\bibfnamefont {S.~R.}\ \bibnamefont
  {Sharpe}},\ }\href {\doibase 10.1146/annurev-nucl-101918-023723} {\bibfield
  {journal} {\bibinfo  {journal} {Ann. Rev. Nucl. Part. Sci.}\ }\textbf
  {\bibinfo {volume} {69}},\ \bibinfo {pages} {65} (\bibinfo {year} {2019})},\
  \Eprint {http://arxiv.org/abs/1901.00483} {arXiv:1901.00483 [hep-lat]}
  \BibitemShut {NoStop}%
\bibitem [{\citenamefont {Mai}\ \emph {et~al.}(2021)\citenamefont {Mai},
  \citenamefont {D\"oring},\ and\ \citenamefont {Rusetsky}}]{Mai:2021lwb}%
  \BibitemOpen
  \bibfield  {author} {\bibinfo {author} {\bibfnamefont {M.}~\bibnamefont
  {Mai}}, \bibinfo {author} {\bibfnamefont {M.}~\bibnamefont {D\"oring}}, \
  and\ \bibinfo {author} {\bibfnamefont {A.}~\bibnamefont {Rusetsky}},\ }\href
  {\doibase 10.1140/epjs/s11734-021-00146-5} {\bibfield  {journal} {\bibinfo
  {journal} {Eur. Phys. J. ST}\ }\textbf {\bibinfo {volume} {230}},\ \bibinfo
  {pages} {1623} (\bibinfo {year} {2021})},\ \Eprint
  {http://arxiv.org/abs/2103.00577} {arXiv:2103.00577 [hep-lat]} \BibitemShut
  {NoStop}%
\bibitem [{\citenamefont {Bauer}\ \emph
  {et~al.}(2023{\natexlab{a}})\citenamefont {Bauer} \emph
  {et~al.}}]{PRXQuantum.4.027001}%
  \BibitemOpen
  \bibfield  {author} {\bibinfo {author} {\bibfnamefont {C.~W.}\ \bibnamefont
  {Bauer}} \emph {et~al.},\ }\href {\doibase 10.1103/PRXQuantum.4.027001}
  {\bibfield  {journal} {\bibinfo  {journal} {PRX Quantum}\ }\textbf {\bibinfo
  {volume} {4}},\ \bibinfo {pages} {027001} (\bibinfo {year}
  {2023}{\natexlab{a}})},\ \Eprint {http://arxiv.org/abs/2204.03381}
  {arXiv:2204.03381 [quant-ph]} \BibitemShut {NoStop}%
\bibitem [{\citenamefont {Camps}\ \emph {et~al.}(2025)\citenamefont {Camps},
  \citenamefont {Rrapaj}, \citenamefont {Klymko}, \citenamefont {Kim},
  \citenamefont {Gott}, \citenamefont {Darbha}, \citenamefont {Balewski},
  \citenamefont {Austin},\ and\ \citenamefont {Wright}}]{osti_2588210}%
  \BibitemOpen
  \bibfield  {author} {\bibinfo {author} {\bibfnamefont {D.}~\bibnamefont
  {Camps}}, \bibinfo {author} {\bibfnamefont {E.}~\bibnamefont {Rrapaj}},
  \bibinfo {author} {\bibfnamefont {K.}~\bibnamefont {Klymko}}, \bibinfo
  {author} {\bibfnamefont {H.}~\bibnamefont {Kim}}, \bibinfo {author}
  {\bibfnamefont {K.}~\bibnamefont {Gott}}, \bibinfo {author} {\bibfnamefont
  {S.}~\bibnamefont {Darbha}}, \bibinfo {author} {\bibfnamefont
  {J.}~\bibnamefont {Balewski}}, \bibinfo {author} {\bibfnamefont
  {B.}~\bibnamefont {Austin}}, \ and\ \bibinfo {author} {\bibfnamefont {N.~J.}\
  \bibnamefont {Wright}},\ }\href {\doibase 10.2172/2588210} {\emph {\bibinfo
  {title} {Quantum Computing Technology Roadmaps and Capability Assessment for
  Scientific Computing - An analysis of use cases from the NERSC workload}}},\
  \bibinfo {type} {Tech. Rep.}\ (\bibinfo  {institution} {Lawrence Berkeley
  National Laboratory (LBNL), Berkeley, CA (United States)},\ \bibinfo {year}
  {2025})\BibitemShut {NoStop}%
\bibitem [{\citenamefont {Awschalom}\ \emph {et~al.}(2025)\citenamefont
  {Awschalom}, \citenamefont {Bernien}, \citenamefont {Hanson}, \citenamefont
  {Oliver},\ and\ \citenamefont {Vučković}}]{doi:10.1126/science.adz8659}%
  \BibitemOpen
  \bibfield  {author} {\bibinfo {author} {\bibfnamefont {D.~D.}\ \bibnamefont
  {Awschalom}}, \bibinfo {author} {\bibfnamefont {H.}~\bibnamefont {Bernien}},
  \bibinfo {author} {\bibfnamefont {R.}~\bibnamefont {Hanson}}, \bibinfo
  {author} {\bibfnamefont {W.~D.}\ \bibnamefont {Oliver}}, \ and\ \bibinfo
  {author} {\bibfnamefont {J.}~\bibnamefont {Vučković}},\ }\href {\doibase
  10.1126/science.adz8659} {\bibfield  {journal} {\bibinfo  {journal}
  {Science}\ }\textbf {\bibinfo {volume} {390}},\ \bibinfo {pages} {1004}
  (\bibinfo {year} {2025})},\ \Eprint
  {http://arxiv.org/abs/https://www.science.org/doi/pdf/10.1126/science.adz8659}
  {https://www.science.org/doi/pdf/10.1126/science.adz8659} \BibitemShut
  {NoStop}%
\bibitem [{\citenamefont {Jordan}\ \emph {et~al.}(2012)\citenamefont {Jordan},
  \citenamefont {Lee},\ and\ \citenamefont {Preskill}}]{Jordan:2012xnu}%
  \BibitemOpen
  \bibfield  {author} {\bibinfo {author} {\bibfnamefont {S.~P.}\ \bibnamefont
  {Jordan}}, \bibinfo {author} {\bibfnamefont {K.~S.~M.}\ \bibnamefont {Lee}},
  \ and\ \bibinfo {author} {\bibfnamefont {J.}~\bibnamefont {Preskill}},\
  }\href {\doibase 10.1126/science.1217069} {\bibfield  {journal} {\bibinfo
  {journal} {Science}\ }\textbf {\bibinfo {volume} {336}},\ \bibinfo {pages}
  {1130} (\bibinfo {year} {2012})},\ \Eprint {http://arxiv.org/abs/1111.3633}
  {arXiv:1111.3633 [quant-ph]} \BibitemShut {NoStop}%
\bibitem [{\citenamefont {Jordan}\ \emph
  {et~al.}(2014{\natexlab{a}})\citenamefont {Jordan}, \citenamefont {Lee},\
  and\ \citenamefont {Preskill}}]{Jordan:2011ci}%
  \BibitemOpen
  \bibfield  {author} {\bibinfo {author} {\bibfnamefont {S.~P.}\ \bibnamefont
  {Jordan}}, \bibinfo {author} {\bibfnamefont {K.~S.~M.}\ \bibnamefont {Lee}},
  \ and\ \bibinfo {author} {\bibfnamefont {J.}~\bibnamefont {Preskill}},\
  }\href@noop {} {\bibfield  {journal} {\bibinfo  {journal} {Quant. Inf.
  Comput.}\ }\textbf {\bibinfo {volume} {14}},\ \bibinfo {pages} {1014}
  (\bibinfo {year} {2014}{\natexlab{a}})},\ \Eprint
  {http://arxiv.org/abs/1112.4833} {arXiv:1112.4833 [hep-th]} \BibitemShut
  {NoStop}%
\bibitem [{\citenamefont {Jordan}\ \emph
  {et~al.}(2014{\natexlab{b}})\citenamefont {Jordan}, \citenamefont {Lee},\
  and\ \citenamefont {Preskill}}]{Jordan:2014tma}%
  \BibitemOpen
  \bibfield  {author} {\bibinfo {author} {\bibfnamefont {S.~P.}\ \bibnamefont
  {Jordan}}, \bibinfo {author} {\bibfnamefont {K.~S.~M.}\ \bibnamefont {Lee}},
  \ and\ \bibinfo {author} {\bibfnamefont {J.}~\bibnamefont {Preskill}},\
  }\href@noop {} {\enquote {\bibinfo {title} {{Quantum Algorithms for Fermionic
  Quantum Field Theories}},}\ } (\bibinfo {year} {2014}{\natexlab{b}}),\
  \Eprint {http://arxiv.org/abs/1404.7115} {arXiv:1404.7115 [hep-th]}
  \BibitemShut {NoStop}%
\bibitem [{\citenamefont {Bennewitz}\ \emph {et~al.}(2025)\citenamefont
  {Bennewitz} \emph {et~al.}}]{Bennewitz:2025nhz}%
  \BibitemOpen
  \bibfield  {author} {\bibinfo {author} {\bibfnamefont {E.~R.}\ \bibnamefont
  {Bennewitz}} \emph {et~al.},\ }\href {\doibase 10.22331/q-2025-06-17-1773}
  {\bibfield  {journal} {\bibinfo  {journal} {Quantum}\ }\textbf {\bibinfo
  {volume} {9}},\ \bibinfo {pages} {1773} (\bibinfo {year} {2025})},\ \Eprint
  {http://arxiv.org/abs/2403.07061} {arXiv:2403.07061 [quant-ph]} \BibitemShut
  {NoStop}%
\bibitem [{\citenamefont {Davoudi}\ \emph {et~al.}(2025)\citenamefont
  {Davoudi}, \citenamefont {Hsieh},\ and\ \citenamefont
  {Kadam}}]{Davoudi:2025rdv}%
  \BibitemOpen
  \bibfield  {author} {\bibinfo {author} {\bibfnamefont {Z.}~\bibnamefont
  {Davoudi}}, \bibinfo {author} {\bibfnamefont {C.-C.}\ \bibnamefont {Hsieh}},
  \ and\ \bibinfo {author} {\bibfnamefont {S.~V.}\ \bibnamefont {Kadam}},\
  }\href@noop {} {\enquote {\bibinfo {title} {{Quantum computation of hadron
  scattering in a lattice gauge theory}},}\ } (\bibinfo {year} {2025}),\
  \Eprint {http://arxiv.org/abs/2505.20408} {arXiv:2505.20408 [quant-ph]}
  \BibitemShut {NoStop}%
\bibitem [{\citenamefont {Davoudi}\ \emph {et~al.}(2024)\citenamefont
  {Davoudi}, \citenamefont {Hsieh},\ and\ \citenamefont
  {Kadam}}]{Davoudi:2024wyv}%
  \BibitemOpen
  \bibfield  {author} {\bibinfo {author} {\bibfnamefont {Z.}~\bibnamefont
  {Davoudi}}, \bibinfo {author} {\bibfnamefont {C.-C.}\ \bibnamefont {Hsieh}},
  \ and\ \bibinfo {author} {\bibfnamefont {S.~V.}\ \bibnamefont {Kadam}},\
  }\href {\doibase 10.22331/q-2024-11-11-1520} {\bibfield  {journal} {\bibinfo
  {journal} {Quantum}\ }\textbf {\bibinfo {volume} {8}},\ \bibinfo {pages}
  {1520} (\bibinfo {year} {2024})},\ \Eprint {http://arxiv.org/abs/2402.00840}
  {arXiv:2402.00840 [quant-ph]} \BibitemShut {NoStop}%
\bibitem [{\citenamefont {Guo}\ \emph {et~al.}(2026{\natexlab{a}})\citenamefont
  {Guo}, \citenamefont {LeVan}, \citenamefont {Lee},\ and\ \citenamefont
  {Zhao}}]{Guo:2026nuc}%
  \BibitemOpen
  \bibfield  {author} {\bibinfo {author} {\bibfnamefont {P.}~\bibnamefont
  {Guo}}, \bibinfo {author} {\bibfnamefont {P.}~\bibnamefont {LeVan}}, \bibinfo
  {author} {\bibfnamefont {F.~X.}\ \bibnamefont {Lee}}, \ and\ \bibinfo
  {author} {\bibfnamefont {Y.}~\bibnamefont {Zhao}},\ }\href@noop {} {\enquote
  {\bibinfo {title} {{Scattering phase shift in quantum mechanics on quantum
  computers: non-Hermitian systems and imaginary-time simulations}},}\ }
  (\bibinfo {year} {2026}{\natexlab{a}}),\ \Eprint
  {http://arxiv.org/abs/2604.00127} {arXiv:2604.00127 [quant-ph]} \BibitemShut
  {NoStop}%
\bibitem [{\citenamefont {Guo}\ \emph {et~al.}(2026{\natexlab{b}})\citenamefont
  {Guo}, \citenamefont {LeVan}, \citenamefont {Lee},\ and\ \citenamefont
  {Zhao}}]{Guo:2026qkx}%
  \BibitemOpen
  \bibfield  {author} {\bibinfo {author} {\bibfnamefont {P.}~\bibnamefont
  {Guo}}, \bibinfo {author} {\bibfnamefont {P.}~\bibnamefont {LeVan}}, \bibinfo
  {author} {\bibfnamefont {F.~X.}\ \bibnamefont {Lee}}, \ and\ \bibinfo
  {author} {\bibfnamefont {Y.}~\bibnamefont {Zhao}},\ }\href {\doibase
  10.1103/ds7x-7bkp} {\bibfield  {journal} {\bibinfo  {journal} {Phys. Rev. D}\
  }\textbf {\bibinfo {volume} {113}},\ \bibinfo {pages} {054512} (\bibinfo
  {year} {2026}{\natexlab{b}})},\ \Eprint {http://arxiv.org/abs/2601.04092}
  {arXiv:2601.04092 [quant-ph]} \BibitemShut {NoStop}%
\bibitem [{\citenamefont {Guo}(2025)}]{Guo:2025vgk}%
  \BibitemOpen
  \bibfield  {author} {\bibinfo {author} {\bibfnamefont {P.}~\bibnamefont
  {Guo}},\ }\href {\doibase 10.1103/f9mm-vz8w} {\bibfield  {journal} {\bibinfo
  {journal} {Phys. Rev. Res.}\ }\textbf {\bibinfo {volume} {7}},\ \bibinfo
  {pages} {043164} (\bibinfo {year} {2025})},\ \Eprint
  {http://arxiv.org/abs/2504.14474} {arXiv:2504.14474 [quant-ph]} \BibitemShut
  {NoStop}%
\bibitem [{\citenamefont {Gustafson}\ \emph {et~al.}(2019)\citenamefont
  {Gustafson}, \citenamefont {Meurice},\ and\ \citenamefont
  {Unmuth-Yockey}}]{Gustafson:2019mpk}%
  \BibitemOpen
  \bibfield  {author} {\bibinfo {author} {\bibfnamefont {E.}~\bibnamefont
  {Gustafson}}, \bibinfo {author} {\bibfnamefont {Y.}~\bibnamefont {Meurice}},
  \ and\ \bibinfo {author} {\bibfnamefont {J.}~\bibnamefont {Unmuth-Yockey}},\
  }\href {\doibase 10.1103/PhysRevD.99.094503} {\bibfield  {journal} {\bibinfo
  {journal} {Phys. Rev. D}\ }\textbf {\bibinfo {volume} {99}},\ \bibinfo
  {pages} {094503} (\bibinfo {year} {2019})},\ \Eprint
  {http://arxiv.org/abs/1901.05944} {arXiv:1901.05944 [hep-lat]} \BibitemShut
  {NoStop}%
\bibitem [{\citenamefont {Davoudi}\ \emph {et~al.}(2021)\citenamefont
  {Davoudi}, \citenamefont {Linke},\ and\ \citenamefont
  {Pagano}}]{Davoudi:2021ney}%
  \BibitemOpen
  \bibfield  {author} {\bibinfo {author} {\bibfnamefont {Z.}~\bibnamefont
  {Davoudi}}, \bibinfo {author} {\bibfnamefont {N.~M.}\ \bibnamefont {Linke}},
  \ and\ \bibinfo {author} {\bibfnamefont {G.}~\bibnamefont {Pagano}},\ }\href
  {\doibase 10.1103/PhysRevResearch.3.043072} {\bibfield  {journal} {\bibinfo
  {journal} {Phys. Rev. Res.}\ }\textbf {\bibinfo {volume} {3}},\ \bibinfo
  {pages} {043072} (\bibinfo {year} {2021})},\ \Eprint
  {http://arxiv.org/abs/2104.09346} {arXiv:2104.09346 [quant-ph]} \BibitemShut
  {NoStop}%
\bibitem [{\citenamefont {Bauer}\ and\ \citenamefont
  {Grabowska}(2023)}]{Bauer:2021gek}%
  \BibitemOpen
  \bibfield  {author} {\bibinfo {author} {\bibfnamefont {C.~W.}\ \bibnamefont
  {Bauer}}\ and\ \bibinfo {author} {\bibfnamefont {D.~M.}\ \bibnamefont
  {Grabowska}},\ }\href {\doibase 10.1103/PhysRevD.107.L031503} {\bibfield
  {journal} {\bibinfo  {journal} {Phys. Rev. D}\ }\textbf {\bibinfo {volume}
  {107}},\ \bibinfo {pages} {L031503} (\bibinfo {year} {2023})},\ \Eprint
  {http://arxiv.org/abs/2111.08015} {arXiv:2111.08015 [hep-ph]} \BibitemShut
  {NoStop}%
\bibitem [{\citenamefont {Bauer}\ \emph
  {et~al.}(2023{\natexlab{b}})\citenamefont {Bauer} \emph
  {et~al.}}]{Bauer:2022hpo}%
  \BibitemOpen
  \bibfield  {author} {\bibinfo {author} {\bibfnamefont {C.~W.}\ \bibnamefont
  {Bauer}} \emph {et~al.},\ }\href {\doibase 10.1103/PRXQuantum.4.027001}
  {\bibfield  {journal} {\bibinfo  {journal} {PRX Quantum}\ }\textbf {\bibinfo
  {volume} {4}},\ \bibinfo {pages} {027001} (\bibinfo {year}
  {2023}{\natexlab{b}})},\ \Eprint {http://arxiv.org/abs/2204.03381}
  {arXiv:2204.03381 [quant-ph]} \BibitemShut {NoStop}%
\bibitem [{\citenamefont {Davoudi}\ \emph {et~al.}(2022)\citenamefont {Davoudi}
  \emph {et~al.}}]{Davoudi:2022bnl}%
  \BibitemOpen
  \bibfield  {author} {\bibinfo {author} {\bibfnamefont {Z.}~\bibnamefont
  {Davoudi}} \emph {et~al.},\ }in\ \href@noop {} {\emph {\bibinfo {booktitle}
  {{Snowmass 2021}}}}\ (\bibinfo {year} {2022})\ \Eprint
  {http://arxiv.org/abs/2209.10758} {arXiv:2209.10758 [hep-lat]} \BibitemShut
  {NoStop}%
\bibitem [{\citenamefont {Catterall}\ \emph {et~al.}(2022)\citenamefont
  {Catterall} \emph {et~al.}}]{Catterall:2022wjq}%
  \BibitemOpen
  \bibfield  {author} {\bibinfo {author} {\bibfnamefont {S.}~\bibnamefont
  {Catterall}} \emph {et~al.},\ }in\ \href {\doibase 10.2172/1892238} {\emph
  {\bibinfo {booktitle} {{Snowmass 2021}}}}\ (\bibinfo {year} {2022})\ \Eprint
  {http://arxiv.org/abs/2209.14839} {arXiv:2209.14839 [quant-ph]} \BibitemShut
  {NoStop}%
\bibitem [{\citenamefont {Di~Meglio}\ \emph {et~al.}(2024)\citenamefont
  {Di~Meglio} \emph {et~al.}}]{DiMeglio:2023nsa}%
  \BibitemOpen
  \bibfield  {author} {\bibinfo {author} {\bibfnamefont {A.}~\bibnamefont
  {Di~Meglio}} \emph {et~al.},\ }\href {\doibase 10.1103/PRXQuantum.5.037001}
  {\bibfield  {journal} {\bibinfo  {journal} {PRX Quantum}\ }\textbf {\bibinfo
  {volume} {5}},\ \bibinfo {pages} {037001} (\bibinfo {year} {2024})},\ \Eprint
  {http://arxiv.org/abs/2307.03236} {arXiv:2307.03236 [quant-ph]} \BibitemShut
  {NoStop}%
\bibitem [{\citenamefont {Halimeh}\ \emph {et~al.}(2025)\citenamefont
  {Halimeh}, \citenamefont {Mueller}, \citenamefont {Knolle}, \citenamefont
  {Papi{\'c}},\ and\ \citenamefont {Davoudi}}]{Halimeh:2025vvp}%
  \BibitemOpen
  \bibfield  {author} {\bibinfo {author} {\bibfnamefont {J.~C.}\ \bibnamefont
  {Halimeh}}, \bibinfo {author} {\bibfnamefont {N.}~\bibnamefont {Mueller}},
  \bibinfo {author} {\bibfnamefont {J.}~\bibnamefont {Knolle}}, \bibinfo
  {author} {\bibfnamefont {Z.}~\bibnamefont {Papi{\'c}}}, \ and\ \bibinfo
  {author} {\bibfnamefont {Z.}~\bibnamefont {Davoudi}},\ }\href@noop {}
  {\enquote {\bibinfo {title} {{Quantum simulation of out-of-equilibrium
  dynamics in gauge theories}},}\ } (\bibinfo {year} {2025}),\ \Eprint
  {http://arxiv.org/abs/2509.03586} {arXiv:2509.03586 [quant-ph]} \BibitemShut
  {NoStop}%
\bibitem [{\citenamefont {Davoudi}(2025)}]{Davoudi:2025kxb}%
  \BibitemOpen
  \bibfield  {author} {\bibinfo {author} {\bibfnamefont {Z.}~\bibnamefont
  {Davoudi}},\ }\href@noop {} {\enquote {\bibinfo {title} {{TASI/CERN/KITP
  Lecture Notes on ''Toward Quantum Computing Gauge Theories of Nature''}},}\ }
  (\bibinfo {year} {2025}),\ \Eprint {http://arxiv.org/abs/2507.15840}
  {arXiv:2507.15840 [hep-lat]} \BibitemShut {NoStop}%
\bibitem [{\citenamefont {Bauer}(2025)}]{Bauer:2025nzf}%
  \BibitemOpen
  \bibfield  {author} {\bibinfo {author} {\bibfnamefont {C.~W.}\ \bibnamefont
  {Bauer}},\ }\href {\doibase 10.1007/JHEP11(2025)108} {\bibfield  {journal}
  {\bibinfo  {journal} {JHEP}\ }\textbf {\bibinfo {volume} {11}},\ \bibinfo
  {pages} {108} (\bibinfo {year} {2025})},\ \Eprint
  {http://arxiv.org/abs/2503.16602} {arXiv:2503.16602 [hep-ph]} \BibitemShut
  {NoStop}%
\bibitem [{\citenamefont {Raychowdhury}\ and\ \citenamefont
  {Stryker}(2020{\natexlab{a}})}]{Raychowdhury:2019iki}%
  \BibitemOpen
  \bibfield  {author} {\bibinfo {author} {\bibfnamefont {I.}~\bibnamefont
  {Raychowdhury}}\ and\ \bibinfo {author} {\bibfnamefont {J.~R.}\ \bibnamefont
  {Stryker}},\ }\href {\doibase 10.1103/PhysRevD.101.114502} {\bibfield
  {journal} {\bibinfo  {journal} {Phys. Rev. D}\ }\textbf {\bibinfo {volume}
  {101}},\ \bibinfo {pages} {114502} (\bibinfo {year} {2020}{\natexlab{a}})},\
  \Eprint {http://arxiv.org/abs/1912.06133} {arXiv:1912.06133 [hep-lat]}
  \BibitemShut {NoStop}%
\bibitem [{\citenamefont {Klco}\ and\ \citenamefont
  {Savage}(2019)}]{Klco:2018zqz}%
  \BibitemOpen
  \bibfield  {author} {\bibinfo {author} {\bibfnamefont {N.}~\bibnamefont
  {Klco}}\ and\ \bibinfo {author} {\bibfnamefont {M.~J.}\ \bibnamefont
  {Savage}},\ }\href {\doibase 10.1103/PhysRevA.99.052335} {\bibfield
  {journal} {\bibinfo  {journal} {Phys. Rev. A}\ }\textbf {\bibinfo {volume}
  {99}},\ \bibinfo {pages} {052335} (\bibinfo {year} {2019})},\ \Eprint
  {http://arxiv.org/abs/1808.10378} {arXiv:1808.10378 [quant-ph]} \BibitemShut
  {NoStop}%
\bibitem [{\citenamefont {Ciavarella}\ \emph {et~al.}(2021)\citenamefont
  {Ciavarella}, \citenamefont {Klco},\ and\ \citenamefont
  {Savage}}]{Ciavarella:2021nmj}%
  \BibitemOpen
  \bibfield  {author} {\bibinfo {author} {\bibfnamefont {A.}~\bibnamefont
  {Ciavarella}}, \bibinfo {author} {\bibfnamefont {N.}~\bibnamefont {Klco}}, \
  and\ \bibinfo {author} {\bibfnamefont {M.~J.}\ \bibnamefont {Savage}},\
  }\href {\doibase 10.1103/PhysRevD.103.094501} {\bibfield  {journal} {\bibinfo
   {journal} {Phys. Rev. D}\ }\textbf {\bibinfo {volume} {103}},\ \bibinfo
  {pages} {094501} (\bibinfo {year} {2021})},\ \Eprint
  {http://arxiv.org/abs/2101.10227} {arXiv:2101.10227 [quant-ph]} \BibitemShut
  {NoStop}%
\bibitem [{\citenamefont {Farrell}\ \emph
  {et~al.}(2024{\natexlab{a}})\citenamefont {Farrell}, \citenamefont {Illa},
  \citenamefont {Ciavarella},\ and\ \citenamefont {Savage}}]{Farrell:2023fgd}%
  \BibitemOpen
  \bibfield  {author} {\bibinfo {author} {\bibfnamefont {R.~C.}\ \bibnamefont
  {Farrell}}, \bibinfo {author} {\bibfnamefont {M.}~\bibnamefont {Illa}},
  \bibinfo {author} {\bibfnamefont {A.~N.}\ \bibnamefont {Ciavarella}}, \ and\
  \bibinfo {author} {\bibfnamefont {M.~J.}\ \bibnamefont {Savage}},\ }\href
  {\doibase 10.1103/PRXQuantum.5.020315} {\bibfield  {journal} {\bibinfo
  {journal} {PRX Quantum}\ }\textbf {\bibinfo {volume} {5}},\ \bibinfo {pages}
  {020315} (\bibinfo {year} {2024}{\natexlab{a}})},\ \Eprint
  {http://arxiv.org/abs/2308.04481} {arXiv:2308.04481 [quant-ph]} \BibitemShut
  {NoStop}%
\bibitem [{\citenamefont {Farrell}\ \emph
  {et~al.}(2024{\natexlab{b}})\citenamefont {Farrell}, \citenamefont {Illa},
  \citenamefont {Ciavarella},\ and\ \citenamefont {Savage}}]{Farrell:2024fit}%
  \BibitemOpen
  \bibfield  {author} {\bibinfo {author} {\bibfnamefont {R.~C.}\ \bibnamefont
  {Farrell}}, \bibinfo {author} {\bibfnamefont {M.}~\bibnamefont {Illa}},
  \bibinfo {author} {\bibfnamefont {A.~N.}\ \bibnamefont {Ciavarella}}, \ and\
  \bibinfo {author} {\bibfnamefont {M.~J.}\ \bibnamefont {Savage}},\ }\href
  {\doibase 10.1103/PhysRevD.109.114510} {\bibfield  {journal} {\bibinfo
  {journal} {Phys.Rev.D}\ }\textbf {\bibinfo {volume} {109}},\ \bibinfo {pages}
  {114510} (\bibinfo {year} {2024}{\natexlab{b}})},\ \Eprint
  {http://arxiv.org/abs/2401.08044} {arXiv:2401.08044 [quant-ph]} \BibitemShut
  {NoStop}%
\bibitem [{\citenamefont {Farrell}\ \emph {et~al.}(2025)\citenamefont
  {Farrell}, \citenamefont {Zemlevskiy}, \citenamefont {Illa},\ and\
  \citenamefont {Preskill}}]{Farrell:2025nkx}%
  \BibitemOpen
  \bibfield  {author} {\bibinfo {author} {\bibfnamefont {R.~C.}\ \bibnamefont
  {Farrell}}, \bibinfo {author} {\bibfnamefont {N.~A.}\ \bibnamefont
  {Zemlevskiy}}, \bibinfo {author} {\bibfnamefont {M.}~\bibnamefont {Illa}}, \
  and\ \bibinfo {author} {\bibfnamefont {J.}~\bibnamefont {Preskill}},\
  }\href@noop {} {\enquote {\bibinfo {title} {{Digital quantum simulations of
  scattering in quantum field theories using W states}},}\ } (\bibinfo {year}
  {2025}),\ \Eprint {http://arxiv.org/abs/2505.03111} {arXiv:2505.03111
  [quant-ph]} \BibitemShut {NoStop}%
\bibitem [{\citenamefont {Schuhmacher}\ \emph {et~al.}(2025)\citenamefont
  {Schuhmacher}, \citenamefont {Su}, \citenamefont {Osborne}, \citenamefont
  {Gandon}, \citenamefont {Halimeh},\ and\ \citenamefont
  {Tavernelli}}]{Schuhmacher:2025ehh}%
  \BibitemOpen
  \bibfield  {author} {\bibinfo {author} {\bibfnamefont {J.}~\bibnamefont
  {Schuhmacher}}, \bibinfo {author} {\bibfnamefont {G.-X.}\ \bibnamefont {Su}},
  \bibinfo {author} {\bibfnamefont {J.~J.}\ \bibnamefont {Osborne}}, \bibinfo
  {author} {\bibfnamefont {A.}~\bibnamefont {Gandon}}, \bibinfo {author}
  {\bibfnamefont {J.~C.}\ \bibnamefont {Halimeh}}, \ and\ \bibinfo {author}
  {\bibfnamefont {I.}~\bibnamefont {Tavernelli}},\ }\href@noop {} {\enquote
  {\bibinfo {title} {{Observation of hadron scattering in a lattice gauge
  theory on a quantum computer}},}\ } (\bibinfo {year} {2025}),\ \Eprint
  {http://arxiv.org/abs/2505.20387} {arXiv:2505.20387 [quant-ph]} \BibitemShut
  {NoStop}%
\bibitem [{\citenamefont {Bauer}\ \emph {et~al.}(2025)\citenamefont {Bauer},
  \citenamefont {Ale}, \citenamefont {Laurell}, \citenamefont {Huang},
  \citenamefont {Watabe}, \citenamefont {Tennant},\ and\ \citenamefont
  {Siopsis}}]{Bauer:2025oqz}%
  \BibitemOpen
  \bibfield  {author} {\bibinfo {author} {\bibfnamefont {N.}~\bibnamefont
  {Bauer}}, \bibinfo {author} {\bibfnamefont {V.}~\bibnamefont {Ale}}, \bibinfo
  {author} {\bibfnamefont {P.}~\bibnamefont {Laurell}}, \bibinfo {author}
  {\bibfnamefont {S.}~\bibnamefont {Huang}}, \bibinfo {author} {\bibfnamefont
  {S.}~\bibnamefont {Watabe}}, \bibinfo {author} {\bibfnamefont {D.~A.}\
  \bibnamefont {Tennant}}, \ and\ \bibinfo {author} {\bibfnamefont
  {G.}~\bibnamefont {Siopsis}},\ }\href {\doibase 10.1103/PhysRevA.111.022442}
  {\bibfield  {journal} {\bibinfo  {journal} {Phys. Rev. A}\ }\textbf {\bibinfo
  {volume} {111}},\ \bibinfo {pages} {022442} (\bibinfo {year}
  {2025})}\BibitemShut {NoStop}%
\bibitem [{\citenamefont {Jha}\ \emph {et~al.}(2024)\citenamefont {Jha},
  \citenamefont {Ringer}, \citenamefont {Siopsis},\ and\ \citenamefont
  {Thompson}}]{Jha:2023ecu}%
  \BibitemOpen
  \bibfield  {author} {\bibinfo {author} {\bibfnamefont {R.~G.}\ \bibnamefont
  {Jha}}, \bibinfo {author} {\bibfnamefont {F.}~\bibnamefont {Ringer}},
  \bibinfo {author} {\bibfnamefont {G.}~\bibnamefont {Siopsis}}, \ and\
  \bibinfo {author} {\bibfnamefont {S.}~\bibnamefont {Thompson}},\ }\href
  {\doibase 10.1103/PhysRevA.109.052412} {\bibfield  {journal} {\bibinfo
  {journal} {Phys. Rev. A}\ }\textbf {\bibinfo {volume} {109}},\ \bibinfo
  {pages} {052412} (\bibinfo {year} {2024})},\ \Eprint
  {http://arxiv.org/abs/2310.12512} {arXiv:2310.12512 [quant-ph]} \BibitemShut
  {NoStop}%
\bibitem [{\citenamefont {Thompson}\ and\ \citenamefont
  {Siopsis}(2022)}]{Thompson:2021eze}%
  \BibitemOpen
  \bibfield  {author} {\bibinfo {author} {\bibfnamefont {S.}~\bibnamefont
  {Thompson}}\ and\ \bibinfo {author} {\bibfnamefont {G.}~\bibnamefont
  {Siopsis}},\ }\href {\doibase 10.1088/2058-9565/ac5f5a} {\bibfield  {journal}
  {\bibinfo  {journal} {Quantum Sci. Technol.}\ }\textbf {\bibinfo {volume}
  {7}},\ \bibinfo {pages} {035001} (\bibinfo {year} {2022})},\ \Eprint
  {http://arxiv.org/abs/2110.13046} {arXiv:2110.13046 [quant-ph]} \BibitemShut
  {NoStop}%
\bibitem [{\citenamefont {Marshall}\ \emph {et~al.}(2015)\citenamefont
  {Marshall}, \citenamefont {Pooser}, \citenamefont {Siopsis},\ and\
  \citenamefont {Weedbrook}}]{Marshall:2015mna}%
  \BibitemOpen
  \bibfield  {author} {\bibinfo {author} {\bibfnamefont {K.}~\bibnamefont
  {Marshall}}, \bibinfo {author} {\bibfnamefont {R.}~\bibnamefont {Pooser}},
  \bibinfo {author} {\bibfnamefont {G.}~\bibnamefont {Siopsis}}, \ and\
  \bibinfo {author} {\bibfnamefont {C.}~\bibnamefont {Weedbrook}},\ }\href
  {\doibase 10.1103/PhysRevA.92.063825} {\bibfield  {journal} {\bibinfo
  {journal} {Phys. Rev. A}\ }\textbf {\bibinfo {volume} {92}},\ \bibinfo
  {pages} {063825} (\bibinfo {year} {2015})},\ \Eprint
  {http://arxiv.org/abs/1503.08121} {arXiv:1503.08121 [quant-ph]} \BibitemShut
  {NoStop}%
\bibitem [{\citenamefont {Ale}\ \emph {et~al.}(2026)\citenamefont {Ale},
  \citenamefont {Rainaldi}, \citenamefont {Rico}, \citenamefont {Ringer},\ and\
  \citenamefont {Siopsis}}]{Ale:2025sxz}%
  \BibitemOpen
  \bibfield  {author} {\bibinfo {author} {\bibfnamefont {V.}~\bibnamefont
  {Ale}}, \bibinfo {author} {\bibfnamefont {T.}~\bibnamefont {Rainaldi}},
  \bibinfo {author} {\bibfnamefont {E.}~\bibnamefont {Rico}}, \bibinfo {author}
  {\bibfnamefont {F.}~\bibnamefont {Ringer}}, \ and\ \bibinfo {author}
  {\bibfnamefont {G.}~\bibnamefont {Siopsis}},\ }\href {\doibase
  10.1007/JHEP04(2026)122} {\bibfield  {journal} {\bibinfo  {journal} {JHEP}\
  }\textbf {\bibinfo {volume} {04}},\ \bibinfo {pages} {122} (\bibinfo {year}
  {2026})},\ \Eprint {http://arxiv.org/abs/2511.14506} {arXiv:2511.14506
  [quant-ph]} \BibitemShut {NoStop}%
\bibitem [{\citenamefont {de~Jong}\ \emph {et~al.}(2022)\citenamefont
  {de~Jong}, \citenamefont {Lee}, \citenamefont {Mulligan}, \citenamefont
  {P{\l}osko{\'n}}, \citenamefont {Ringer},\ and\ \citenamefont
  {Yao}}]{deJong:2021wsd}%
  \BibitemOpen
  \bibfield  {author} {\bibinfo {author} {\bibfnamefont {W.~A.}\ \bibnamefont
  {de~Jong}}, \bibinfo {author} {\bibfnamefont {K.}~\bibnamefont {Lee}},
  \bibinfo {author} {\bibfnamefont {J.}~\bibnamefont {Mulligan}}, \bibinfo
  {author} {\bibfnamefont {M.}~\bibnamefont {P{\l}osko{\'n}}}, \bibinfo
  {author} {\bibfnamefont {F.}~\bibnamefont {Ringer}}, \ and\ \bibinfo {author}
  {\bibfnamefont {X.}~\bibnamefont {Yao}},\ }\href {\doibase
  10.1103/PhysRevD.106.054508} {\bibfield  {journal} {\bibinfo  {journal}
  {Phys. Rev. D}\ }\textbf {\bibinfo {volume} {106}},\ \bibinfo {pages}
  {054508} (\bibinfo {year} {2022})},\ \Eprint
  {http://arxiv.org/abs/2106.08394} {arXiv:2106.08394 [quant-ph]} \BibitemShut
  {NoStop}%
\bibitem [{\citenamefont {Choi}\ \emph {et~al.}(2021)\citenamefont {Choi},
  \citenamefont {Lee}, \citenamefont {Bonitati}, \citenamefont {Qian},\ and\
  \citenamefont {Watkins}}]{Choi:2020pdg}%
  \BibitemOpen
  \bibfield  {author} {\bibinfo {author} {\bibfnamefont {K.}~\bibnamefont
  {Choi}}, \bibinfo {author} {\bibfnamefont {D.}~\bibnamefont {Lee}}, \bibinfo
  {author} {\bibfnamefont {J.}~\bibnamefont {Bonitati}}, \bibinfo {author}
  {\bibfnamefont {Z.}~\bibnamefont {Qian}}, \ and\ \bibinfo {author}
  {\bibfnamefont {J.}~\bibnamefont {Watkins}},\ }\href {\doibase
  10.1103/PhysRevLett.127.040505} {\bibfield  {journal} {\bibinfo  {journal}
  {Phys. Rev. Lett.}\ }\textbf {\bibinfo {volume} {127}},\ \bibinfo {pages}
  {040505} (\bibinfo {year} {2021})},\ \Eprint
  {http://arxiv.org/abs/2009.04092} {arXiv:2009.04092 [quant-ph]} \BibitemShut
  {NoStop}%
\bibitem [{\citenamefont {Ciavarella}\ \emph {et~al.}(2022)\citenamefont
  {Ciavarella}, \citenamefont {Klco},\ and\ \citenamefont
  {Savage}}]{Ciavarella:2022zhe}%
  \BibitemOpen
  \bibfield  {author} {\bibinfo {author} {\bibfnamefont {A.}~\bibnamefont
  {Ciavarella}}, \bibinfo {author} {\bibfnamefont {N.}~\bibnamefont {Klco}}, \
  and\ \bibinfo {author} {\bibfnamefont {M.~J.}\ \bibnamefont {Savage}}\
  }(\bibinfo {year} {2022})\ \Eprint {http://arxiv.org/abs/2203.11988}
  {arXiv:2203.11988 [quant-ph]} \BibitemShut {NoStop}%
\bibitem [{\citenamefont {Ciavarella}\ \emph {et~al.}(2023)\citenamefont
  {Ciavarella}, \citenamefont {Caspar}, \citenamefont {Singh},\ and\
  \citenamefont {Savage}}]{Ciavarella:2022qdx}%
  \BibitemOpen
  \bibfield  {author} {\bibinfo {author} {\bibfnamefont {A.~N.}\ \bibnamefont
  {Ciavarella}}, \bibinfo {author} {\bibfnamefont {S.}~\bibnamefont {Caspar}},
  \bibinfo {author} {\bibfnamefont {H.}~\bibnamefont {Singh}}, \ and\ \bibinfo
  {author} {\bibfnamefont {M.~J.}\ \bibnamefont {Savage}},\ }\href {\doibase
  10.1103/PhysRevA.107.042404} {\bibfield  {journal} {\bibinfo  {journal}
  {Phys.Rev.A}\ }\textbf {\bibinfo {volume} {107}},\ \bibinfo {pages} {042404}
  (\bibinfo {year} {2023})},\ \Eprint {http://arxiv.org/abs/2211.07684}
  {arXiv:2211.07684 [quant-ph]} \BibitemShut {NoStop}%
\bibitem [{\citenamefont {Ciavarella}(2023)}]{Ciavarella:2023mfc}%
  \BibitemOpen
  \bibfield  {author} {\bibinfo {author} {\bibfnamefont {A.~N.}\ \bibnamefont
  {Ciavarella}},\ }\href {\doibase 10.1103/PhysRevD.108.094513} {\bibfield
  {journal} {\bibinfo  {journal} {Phys.Rev.D}\ }\textbf {\bibinfo {volume}
  {108}},\ \bibinfo {pages} {094513} (\bibinfo {year} {2023})},\ \Eprint
  {http://arxiv.org/abs/2307.05593} {arXiv:2307.05593 [hep-lat]} \BibitemShut
  {NoStop}%
\bibitem [{\citenamefont {Ciavarella}\ and\ \citenamefont
  {Bauer}(2024)}]{Ciavarella:2024fzw}%
  \BibitemOpen
  \bibfield  {author} {\bibinfo {author} {\bibfnamefont {A.~N.}\ \bibnamefont
  {Ciavarella}}\ and\ \bibinfo {author} {\bibfnamefont {C.~W.}\ \bibnamefont
  {Bauer}},\ }\href {\doibase 10.1103/PhysRevLett.133.111901} {\bibfield
  {journal} {\bibinfo  {journal} {Phys. Rev. Lett.}\ }\textbf {\bibinfo
  {volume} {133}},\ \bibinfo {pages} {111901} (\bibinfo {year} {2024})},\
  \Eprint {http://arxiv.org/abs/2402.10265} {arXiv:2402.10265 [hep-ph]}
  \BibitemShut {NoStop}%
\bibitem [{\citenamefont {Ciavarella}\ \emph
  {et~al.}(2025{\natexlab{a}})\citenamefont {Ciavarella}, \citenamefont
  {Burbano},\ and\ \citenamefont {Bauer}}]{Ciavarella:2025bsg}%
  \BibitemOpen
  \bibfield  {author} {\bibinfo {author} {\bibfnamefont {A.~N.}\ \bibnamefont
  {Ciavarella}}, \bibinfo {author} {\bibfnamefont {I.~M.}\ \bibnamefont
  {Burbano}}, \ and\ \bibinfo {author} {\bibfnamefont {C.~W.}\ \bibnamefont
  {Bauer}},\ }\href {\doibase 10.1103/ylqb-phv5} {\bibfield  {journal}
  {\bibinfo  {journal} {Phys. Rev. D}\ }\textbf {\bibinfo {volume} {112}},\
  \bibinfo {pages} {054514} (\bibinfo {year} {2025}{\natexlab{a}})},\ \Eprint
  {http://arxiv.org/abs/2503.11888} {arXiv:2503.11888 [hep-lat]} \BibitemShut
  {NoStop}%
\bibitem [{\citenamefont {Ciavarella}\ \emph
  {et~al.}(2025{\natexlab{b}})\citenamefont {Ciavarella}, \citenamefont
  {Hariprakash}, \citenamefont {Halimeh},\ and\ \citenamefont
  {Bauer}}]{Ciavarella:2025tdl}%
  \BibitemOpen
  \bibfield  {author} {\bibinfo {author} {\bibfnamefont {A.~N.}\ \bibnamefont
  {Ciavarella}}, \bibinfo {author} {\bibfnamefont {S.}~\bibnamefont
  {Hariprakash}}, \bibinfo {author} {\bibfnamefont {J.~C.}\ \bibnamefont
  {Halimeh}}, \ and\ \bibinfo {author} {\bibfnamefont {C.~W.}\ \bibnamefont
  {Bauer}},\ }\href@noop {} {\enquote {\bibinfo {title} {{Truncation
  uncertainties for accurate quantum simulations of lattice gauge theories}},}\
  } (\bibinfo {year} {2025}{\natexlab{b}}),\ \Eprint
  {http://arxiv.org/abs/2508.00061} {arXiv:2508.00061 [quant-ph]} \BibitemShut
  {NoStop}%
\bibitem [{\citenamefont {Modi}\ \emph {et~al.}(2026)\citenamefont {Modi},
  \citenamefont {Ciavarella}, \citenamefont {Halimeh},\ and\ \citenamefont
  {Bauer}}]{Modi:2026syn}%
  \BibitemOpen
  \bibfield  {author} {\bibinfo {author} {\bibfnamefont {N.~S.}\ \bibnamefont
  {Modi}}, \bibinfo {author} {\bibfnamefont {A.~N.}\ \bibnamefont
  {Ciavarella}}, \bibinfo {author} {\bibfnamefont {J.~C.}\ \bibnamefont
  {Halimeh}}, \ and\ \bibinfo {author} {\bibfnamefont {C.~W.}\ \bibnamefont
  {Bauer}},\ }\href@noop {} {\enquote {\bibinfo {title} {{Large Nc Truncations
  for SU(Nc) Lattice Yang-Mills Theory with Fermions}},}\ } (\bibinfo {year}
  {2026}),\ \Eprint {http://arxiv.org/abs/2602.02344} {arXiv:2602.02344
  [hep-lat]} \BibitemShut {NoStop}%
\bibitem [{\citenamefont {Balaji}\ \emph {et~al.}(2025)\citenamefont {Balaji},
  \citenamefont {Conefrey-Shinozaki}, \citenamefont {Draper}, \citenamefont
  {Elhaderi}, \citenamefont {Gupta}, \citenamefont {Hidalgo}, \citenamefont
  {Lytle},\ and\ \citenamefont {Rinaldi}}]{Balaji:2025afl}%
  \BibitemOpen
  \bibfield  {author} {\bibinfo {author} {\bibfnamefont {P.}~\bibnamefont
  {Balaji}}, \bibinfo {author} {\bibfnamefont {C.}~\bibnamefont
  {Conefrey-Shinozaki}}, \bibinfo {author} {\bibfnamefont {P.}~\bibnamefont
  {Draper}}, \bibinfo {author} {\bibfnamefont {J.~K.}\ \bibnamefont
  {Elhaderi}}, \bibinfo {author} {\bibfnamefont {D.}~\bibnamefont {Gupta}},
  \bibinfo {author} {\bibfnamefont {L.}~\bibnamefont {Hidalgo}}, \bibinfo
  {author} {\bibfnamefont {A.}~\bibnamefont {Lytle}}, \ and\ \bibinfo {author}
  {\bibfnamefont {E.}~\bibnamefont {Rinaldi}},\ }\href {\doibase
  10.1103/k8f6-yft8} {\bibfield  {journal} {\bibinfo  {journal} {Phys.Rev.D}\
  }\textbf {\bibinfo {volume} {112}},\ \bibinfo {pages} {054511} (\bibinfo
  {year} {2025})},\ \Eprint {http://arxiv.org/abs/2503.08866} {arXiv:2503.08866
  [hep-lat]} \BibitemShut {NoStop}%
\bibitem [{\citenamefont {Balaji}\ \emph {et~al.}(2026)\citenamefont {Balaji},
  \citenamefont {Conefrey-Shinozaki}, \citenamefont {Draper}, \citenamefont
  {Elhaderi}, \citenamefont {Gupta}, \citenamefont {Hidalgo},\ and\
  \citenamefont {Lytle}}]{Balaji:2025yua}%
  \BibitemOpen
  \bibfield  {author} {\bibinfo {author} {\bibfnamefont {P.}~\bibnamefont
  {Balaji}}, \bibinfo {author} {\bibfnamefont {C.}~\bibnamefont
  {Conefrey-Shinozaki}}, \bibinfo {author} {\bibfnamefont {P.}~\bibnamefont
  {Draper}}, \bibinfo {author} {\bibfnamefont {J.~K.}\ \bibnamefont
  {Elhaderi}}, \bibinfo {author} {\bibfnamefont {D.}~\bibnamefont {Gupta}},
  \bibinfo {author} {\bibfnamefont {L.}~\bibnamefont {Hidalgo}}, \ and\
  \bibinfo {author} {\bibfnamefont {A.}~\bibnamefont {Lytle}},\ }\href
  {\doibase 10.1103/m719-7tdf} {\bibfield  {journal} {\bibinfo  {journal}
  {Phys.Rev.D}\ }\textbf {\bibinfo {volume} {113}},\ \bibinfo {pages} {094505}
  (\bibinfo {year} {2026})},\ \Eprint {http://arxiv.org/abs/2509.25865}
  {arXiv:2509.25865 [hep-lat]} \BibitemShut {NoStop}%
\bibitem [{\citenamefont {Draper}(2026)}]{Draper:2026bcj}%
  \BibitemOpen
  \bibfield  {author} {\bibinfo {author} {\bibfnamefont {P.}~\bibnamefont
  {Draper}},\ }\href@noop {} {\enquote {\bibinfo {title} {{Block Encoding
  Non-Abelian Lattice Gauge Theory}},}\ } (\bibinfo {year} {2026}),\ \Eprint
  {http://arxiv.org/abs/2608.17115} {arXiv:2608.17115 [quant-ph]} \BibitemShut
  {NoStop}%
\bibitem [{\citenamefont {Hidalgo}\ and\ \citenamefont
  {Draper}(2026)}]{Hidalgo:2026zsz}%
  \BibitemOpen
  \bibfield  {author} {\bibinfo {author} {\bibfnamefont {L.}~\bibnamefont
  {Hidalgo}}\ and\ \bibinfo {author} {\bibfnamefont {P.}~\bibnamefont
  {Draper}},\ }\href@noop {} {\  (\bibinfo {year} {2026})},\ \Eprint
  {http://arxiv.org/abs/2607.26445} {arXiv:2607.26445 [quant-ph]} \BibitemShut
  {NoStop}%
\bibitem [{\citenamefont {Raychowdhury}(2019)}]{Raychowdhury:2018tfj}%
  \BibitemOpen
  \bibfield  {author} {\bibinfo {author} {\bibfnamefont {I.}~\bibnamefont
  {Raychowdhury}},\ }\href {\doibase 10.1140/epjc/s10052-019-6753-0} {\bibfield
   {journal} {\bibinfo  {journal} {Eur.Phys.J.C}\ }\textbf {\bibinfo {volume}
  {79}},\ \bibinfo {pages} {235} (\bibinfo {year} {2019})},\ \Eprint
  {http://arxiv.org/abs/1804.01304} {arXiv:1804.01304 [hep-lat]} \BibitemShut
  {NoStop}%
\bibitem [{\citenamefont {Raychowdhury}\ and\ \citenamefont
  {Stryker}(2020{\natexlab{b}})}]{Raychowdhury:2018osk}%
  \BibitemOpen
  \bibfield  {author} {\bibinfo {author} {\bibfnamefont {I.}~\bibnamefont
  {Raychowdhury}}\ and\ \bibinfo {author} {\bibfnamefont {J.~R.}\ \bibnamefont
  {Stryker}},\ }\href {\doibase 10.1103/PhysRevResearch.2.033039} {\bibfield
  {journal} {\bibinfo  {journal} {Phys.Rev.Res.}\ }\textbf {\bibinfo {volume}
  {2}},\ \bibinfo {pages} {033039} (\bibinfo {year} {2020}{\natexlab{b}})},\
  \Eprint {http://arxiv.org/abs/1812.07554} {arXiv:1812.07554 [hep-lat]}
  \BibitemShut {NoStop}%
\bibitem [{\citenamefont {Kadam}\ \emph {et~al.}(2023)\citenamefont {Kadam},
  \citenamefont {Raychowdhury},\ and\ \citenamefont {Stryker}}]{Kadam:2022ipf}%
  \BibitemOpen
  \bibfield  {author} {\bibinfo {author} {\bibfnamefont {S.~V.}\ \bibnamefont
  {Kadam}}, \bibinfo {author} {\bibfnamefont {I.}~\bibnamefont {Raychowdhury}},
  \ and\ \bibinfo {author} {\bibfnamefont {J.~R.}\ \bibnamefont {Stryker}},\
  }\href {\doibase 10.1103/PhysRevD.107.094513} {\bibfield  {journal} {\bibinfo
   {journal} {Phys. Rev. D}\ }\textbf {\bibinfo {volume} {107}},\ \bibinfo
  {pages} {094513} (\bibinfo {year} {2023})},\ \Eprint
  {http://arxiv.org/abs/2212.04490} {arXiv:2212.04490 [hep-lat]} \BibitemShut
  {NoStop}%
\bibitem [{\citenamefont {Davoudi}\ \emph {et~al.}(2023)\citenamefont
  {Davoudi}, \citenamefont {Shaw},\ and\ \citenamefont
  {Stryker}}]{Davoudi:2022xmb}%
  \BibitemOpen
  \bibfield  {author} {\bibinfo {author} {\bibfnamefont {Z.}~\bibnamefont
  {Davoudi}}, \bibinfo {author} {\bibfnamefont {A.~F.}\ \bibnamefont {Shaw}}, \
  and\ \bibinfo {author} {\bibfnamefont {J.~R.}\ \bibnamefont {Stryker}},\
  }\href {\doibase 10.22331/q-2023-12-20-1213} {\bibfield  {journal} {\bibinfo
  {journal} {Quantum}\ }\textbf {\bibinfo {volume} {7}},\ \bibinfo {pages}
  {1213} (\bibinfo {year} {2023})},\ \Eprint {http://arxiv.org/abs/2212.14030}
  {arXiv:2212.14030 [hep-lat]} \BibitemShut {NoStop}%
\bibitem [{\citenamefont {Kadam}\ \emph
  {et~al.}(2025{\natexlab{a}})\citenamefont {Kadam}, \citenamefont {Naskar},
  \citenamefont {Raychowdhury},\ and\ \citenamefont {Stryker}}]{Kadam:2024ifg}%
  \BibitemOpen
  \bibfield  {author} {\bibinfo {author} {\bibfnamefont {S.~V.}\ \bibnamefont
  {Kadam}}, \bibinfo {author} {\bibfnamefont {A.}~\bibnamefont {Naskar}},
  \bibinfo {author} {\bibfnamefont {I.}~\bibnamefont {Raychowdhury}}, \ and\
  \bibinfo {author} {\bibfnamefont {J.~R.}\ \bibnamefont {Stryker}},\ }\href
  {\doibase 10.1103/PhysRevD.111.074516} {\bibfield  {journal} {\bibinfo
  {journal} {Phys. Rev. D}\ }\textbf {\bibinfo {volume} {111}},\ \bibinfo
  {pages} {074516} (\bibinfo {year} {2025}{\natexlab{a}})},\ \Eprint
  {http://arxiv.org/abs/2407.19181} {arXiv:2407.19181 [hep-lat]} \BibitemShut
  {NoStop}%
\bibitem [{\citenamefont {Burbano}\ and\ \citenamefont
  {Bauer}(2025)}]{Burbano:2024uvn}%
  \BibitemOpen
  \bibfield  {author} {\bibinfo {author} {\bibfnamefont {I.~M.}\ \bibnamefont
  {Burbano}}\ and\ \bibinfo {author} {\bibfnamefont {C.~W.}\ \bibnamefont
  {Bauer}},\ }\href {\doibase 10.1007/JHEP12(2025)060} {\bibfield  {journal}
  {\bibinfo  {journal} {JHEP}\ }\textbf {\bibinfo {volume} {12}},\ \bibinfo
  {pages} {060} (\bibinfo {year} {2025})},\ \Eprint
  {http://arxiv.org/abs/2409.13812} {arXiv:2409.13812 [hep-lat]} \BibitemShut
  {NoStop}%
\bibitem [{\citenamefont {Kadam}\ \emph
  {et~al.}(2025{\natexlab{b}})\citenamefont {Kadam}, \citenamefont {Naskar},
  \citenamefont {Raychowdhury},\ and\ \citenamefont {Stryker}}]{Kadam:2025trs}%
  \BibitemOpen
  \bibfield  {author} {\bibinfo {author} {\bibfnamefont {S.~V.}\ \bibnamefont
  {Kadam}}, \bibinfo {author} {\bibfnamefont {A.}~\bibnamefont {Naskar}},
  \bibinfo {author} {\bibfnamefont {I.}~\bibnamefont {Raychowdhury}}, \ and\
  \bibinfo {author} {\bibfnamefont {J.~R.}\ \bibnamefont {Stryker}},\
  }\href@noop {} {\enquote {\bibinfo {title} {{Loop-string-hadron approach to
  SU(3) lattice Yang-Mills theory, II: Operator representation for the
  trivalent vertex}},}\ } (\bibinfo {year} {2025}{\natexlab{b}}),\ \Eprint
  {http://arxiv.org/abs/2512.11796} {arXiv:2512.11796 [hep-lat]} \BibitemShut
  {NoStop}%
\bibitem [{\citenamefont {Ilcic}\ and\ \citenamefont
  {Raychowdhury}(2025)}]{Ilcic:2025gel}%
  \BibitemOpen
  \bibfield  {author} {\bibinfo {author} {\bibfnamefont {F.}~\bibnamefont
  {Ilcic}}\ and\ \bibinfo {author} {\bibfnamefont {I.}~\bibnamefont
  {Raychowdhury}},\ }\href@noop {} {\  (\bibinfo {year} {2025})},\ \Eprint
  {http://arxiv.org/abs/2512.13035} {arXiv:2512.13035 [hep-lat]} \BibitemShut
  {NoStop}%
\bibitem [{\citenamefont {Das}\ \emph {et~al.}(2026)\citenamefont {Das},
  \citenamefont {Ebner}, \citenamefont {Kadam}, \citenamefont {Raychowdhury},
  \citenamefont {Schäfer},\ and\ \citenamefont {Yao}}]{Das:2025utp}%
  \BibitemOpen
  \bibfield  {author} {\bibinfo {author} {\bibfnamefont {D.}~\bibnamefont
  {Das}}, \bibinfo {author} {\bibfnamefont {L.}~\bibnamefont {Ebner}}, \bibinfo
  {author} {\bibfnamefont {S.~V.}\ \bibnamefont {Kadam}}, \bibinfo {author}
  {\bibfnamefont {I.}~\bibnamefont {Raychowdhury}}, \bibinfo {author}
  {\bibfnamefont {A.}~\bibnamefont {Schäfer}}, \ and\ \bibinfo {author}
  {\bibfnamefont {X.}~\bibnamefont {Yao}},\ }\href {\doibase 10.1103/19c2-k9x9}
  {\bibfield  {journal} {\bibinfo  {journal} {Phys.Rev.D}\ }\textbf {\bibinfo
  {volume} {113}},\ \bibinfo {pages} {074514} (\bibinfo {year} {2026})},\
  \Eprint {http://arxiv.org/abs/2509.18269} {arXiv:2509.18269 [hep-th]}
  \BibitemShut {NoStop}%
\bibitem [{\citenamefont {Gupta}\ \emph {et~al.}(2026)\citenamefont {Gupta},
  \citenamefont {Mathew}, \citenamefont {Kadam}, \citenamefont {Stryker},
  \citenamefont {Bapat}, \citenamefont {Mueller}, \citenamefont {Davoudi},\
  and\ \citenamefont {Raychowdhury}}]{Gupta:2026tcg}%
  \BibitemOpen
  \bibfield  {author} {\bibinfo {author} {\bibfnamefont {N.}~\bibnamefont
  {Gupta}}, \bibinfo {author} {\bibfnamefont {E.}~\bibnamefont {Mathew}},
  \bibinfo {author} {\bibfnamefont {S.~V.}\ \bibnamefont {Kadam}}, \bibinfo
  {author} {\bibfnamefont {J.~R.}\ \bibnamefont {Stryker}}, \bibinfo {author}
  {\bibfnamefont {A.}~\bibnamefont {Bapat}}, \bibinfo {author} {\bibfnamefont
  {N.}~\bibnamefont {Mueller}}, \bibinfo {author} {\bibfnamefont
  {Z.}~\bibnamefont {Davoudi}}, \ and\ \bibinfo {author} {\bibfnamefont
  {I.}~\bibnamefont {Raychowdhury}},\ }\href@noop {} {\  (\bibinfo {year}
  {2026})},\ \Eprint {http://arxiv.org/abs/2603.24698} {arXiv:2603.24698
  [hep-lat]} \BibitemShut {NoStop}%
\bibitem [{\citenamefont {Chandrasekharan}\ and\ \citenamefont
  {Wiese}(1997)}]{Chandrasekharan:1996ih}%
  \BibitemOpen
  \bibfield  {author} {\bibinfo {author} {\bibfnamefont {S.}~\bibnamefont
  {Chandrasekharan}}\ and\ \bibinfo {author} {\bibfnamefont {U.}~\bibnamefont
  {Wiese}},\ }\href {\doibase 10.1016/S0550-3213(97)00006-0} {\bibfield
  {journal} {\bibinfo  {journal} {Nucl.Phys.B}\ }\textbf {\bibinfo {volume}
  {492}},\ \bibinfo {pages} {455} (\bibinfo {year} {1997})},\ \Eprint
  {http://arxiv.org/abs/hep-lat/9609042} {arXiv:hep-lat/9609042 [hep-lat]}
  \BibitemShut {NoStop}%
\bibitem [{\citenamefont {Brower}\ \emph {et~al.}(1999)\citenamefont {Brower},
  \citenamefont {Chandrasekharan},\ and\ \citenamefont
  {Wiese}}]{Brower:1997ha}%
  \BibitemOpen
  \bibfield  {author} {\bibinfo {author} {\bibfnamefont {R.}~\bibnamefont
  {Brower}}, \bibinfo {author} {\bibfnamefont {S.}~\bibnamefont
  {Chandrasekharan}}, \ and\ \bibinfo {author} {\bibfnamefont {U.}~\bibnamefont
  {Wiese}},\ }\href {\doibase 10.1103/PhysRevD.60.094502} {\bibfield  {journal}
  {\bibinfo  {journal} {Phys.Rev.D}\ }\textbf {\bibinfo {volume} {60}},\
  \bibinfo {pages} {094502} (\bibinfo {year} {1999})},\ \Eprint
  {http://arxiv.org/abs/hep-th/9704106} {arXiv:hep-th/9704106 [hep-th]}
  \BibitemShut {NoStop}%
\bibitem [{\citenamefont {Brower}\ \emph {et~al.}(2004)\citenamefont {Brower},
  \citenamefont {Chandrasekharan}, \citenamefont {Riederer},\ and\
  \citenamefont {Wiese}}]{Brower:2003vy}%
  \BibitemOpen
  \bibfield  {author} {\bibinfo {author} {\bibfnamefont {R.}~\bibnamefont
  {Brower}}, \bibinfo {author} {\bibfnamefont {S.}~\bibnamefont
  {Chandrasekharan}}, \bibinfo {author} {\bibfnamefont {S.}~\bibnamefont
  {Riederer}}, \ and\ \bibinfo {author} {\bibfnamefont {U.}~\bibnamefont
  {Wiese}},\ }\href {\doibase 10.1016/j.nuclphysb.2004.06.007} {\bibfield
  {journal} {\bibinfo  {journal} {Nucl.Phys.B}\ }\textbf {\bibinfo {volume}
  {693}},\ \bibinfo {pages} {149} (\bibinfo {year} {2004})},\ \Eprint
  {http://arxiv.org/abs/hep-lat/0309182} {arXiv:hep-lat/0309182 [hep-lat]}
  \BibitemShut {NoStop}%
\bibitem [{\citenamefont {Zache}\ \emph {et~al.}(2022)\citenamefont {Zache},
  \citenamefont {Van~Damme}, \citenamefont {Halimeh}, \citenamefont {Hauke},\
  and\ \citenamefont {Banerjee}}]{Zache:2021ggw}%
  \BibitemOpen
  \bibfield  {author} {\bibinfo {author} {\bibfnamefont {T.~V.}\ \bibnamefont
  {Zache}}, \bibinfo {author} {\bibfnamefont {M.}~\bibnamefont {Van~Damme}},
  \bibinfo {author} {\bibfnamefont {J.~C.}\ \bibnamefont {Halimeh}}, \bibinfo
  {author} {\bibfnamefont {P.}~\bibnamefont {Hauke}}, \ and\ \bibinfo {author}
  {\bibfnamefont {D.}~\bibnamefont {Banerjee}},\ }\href {\doibase
  10.1103/PhysRevD.106.L091502} {\bibfield  {journal} {\bibinfo  {journal}
  {Phys.Rev.D}\ }\textbf {\bibinfo {volume} {106}},\ \bibinfo {pages} {L091502}
  (\bibinfo {year} {2022})},\ \Eprint {http://arxiv.org/abs/2104.00025}
  {arXiv:2104.00025 [hep-lat]} \BibitemShut {NoStop}%
\bibitem [{\citenamefont {Halimeh}\ \emph {et~al.}(2022)\citenamefont
  {Halimeh}, \citenamefont {Van~Damme}, \citenamefont {Zache}, \citenamefont
  {Banerjee},\ and\ \citenamefont {Hauke}}]{Halimeh:2021ufh}%
  \BibitemOpen
  \bibfield  {author} {\bibinfo {author} {\bibfnamefont {J.~C.}\ \bibnamefont
  {Halimeh}}, \bibinfo {author} {\bibfnamefont {M.}~\bibnamefont {Van~Damme}},
  \bibinfo {author} {\bibfnamefont {T.~V.}\ \bibnamefont {Zache}}, \bibinfo
  {author} {\bibfnamefont {D.}~\bibnamefont {Banerjee}}, \ and\ \bibinfo
  {author} {\bibfnamefont {P.}~\bibnamefont {Hauke}},\ }\href {\doibase
  10.22331/q-2022-12-19-878} {\bibfield  {journal} {\bibinfo  {journal}
  {Quantum}\ }\textbf {\bibinfo {volume} {6}},\ \bibinfo {pages} {878}
  (\bibinfo {year} {2022})},\ \Eprint {http://arxiv.org/abs/2112.04501}
  {arXiv:2112.04501 [cond-mat.quant-gas]} \BibitemShut {NoStop}%
\bibitem [{\citenamefont {Osborne}\ \emph {et~al.}(2026)\citenamefont
  {Osborne}, \citenamefont {Yang}, \citenamefont {McCulloch}, \citenamefont
  {Hauke},\ and\ \citenamefont {Halimeh}}]{Osborne:2023rzx}%
  \BibitemOpen
  \bibfield  {author} {\bibinfo {author} {\bibfnamefont {J.~J.}\ \bibnamefont
  {Osborne}}, \bibinfo {author} {\bibfnamefont {B.}~\bibnamefont {Yang}},
  \bibinfo {author} {\bibfnamefont {I.~P.}\ \bibnamefont {McCulloch}}, \bibinfo
  {author} {\bibfnamefont {P.}~\bibnamefont {Hauke}}, \ and\ \bibinfo {author}
  {\bibfnamefont {J.~C.}\ \bibnamefont {Halimeh}},\ }\href {\doibase
  10.1038/s42005-026-02805-2} {\bibfield  {journal} {\bibinfo  {journal}
  {Commun.Phys.}\ }\textbf {\bibinfo {volume} {9}},\ \bibinfo {pages} {285}
  (\bibinfo {year} {2026})},\ \Eprint {http://arxiv.org/abs/2305.06368}
  {arXiv:2305.06368 [cond-mat.quant-gas]} \BibitemShut {NoStop}%
\bibitem [{\citenamefont {Joshi}\ \emph
  {et~al.}(2025{\natexlab{a}})\citenamefont {Joshi}, \citenamefont {Meth},
  \citenamefont {Louw}, \citenamefont {Osborne}, \citenamefont {Mato},
  \citenamefont {Ringbauer},\ and\ \citenamefont {Halimeh}}]{Joshi:2025pgv}%
  \BibitemOpen
  \bibfield  {author} {\bibinfo {author} {\bibfnamefont {R.}~\bibnamefont
  {Joshi}}, \bibinfo {author} {\bibfnamefont {M.}~\bibnamefont {Meth}},
  \bibinfo {author} {\bibfnamefont {J.~C.}\ \bibnamefont {Louw}}, \bibinfo
  {author} {\bibfnamefont {J.~J.}\ \bibnamefont {Osborne}}, \bibinfo {author}
  {\bibfnamefont {K.}~\bibnamefont {Mato}}, \bibinfo {author} {\bibfnamefont
  {M.}~\bibnamefont {Ringbauer}}, \ and\ \bibinfo {author} {\bibfnamefont
  {J.~C.}\ \bibnamefont {Halimeh}},\ }\href@noop {} {\  (\bibinfo {year}
  {2025}{\natexlab{a}})},\ \Eprint {http://arxiv.org/abs/2507.12589}
  {arXiv:2507.12589 [quant-ph]} \BibitemShut {NoStop}%
\bibitem [{\citenamefont {Cao}\ \emph {et~al.}(2026)\citenamefont {Cao},
  \citenamefont {Joshi}, \citenamefont {Tian}, \citenamefont {Srivatsa},\ and\
  \citenamefont {Halimeh}}]{Cao:2026qky}%
  \BibitemOpen
  \bibfield  {author} {\bibinfo {author} {\bibfnamefont {J.}~\bibnamefont
  {Cao}}, \bibinfo {author} {\bibfnamefont {R.}~\bibnamefont {Joshi}}, \bibinfo
  {author} {\bibfnamefont {Y.}~\bibnamefont {Tian}}, \bibinfo {author}
  {\bibfnamefont {N.}~\bibnamefont {Srivatsa}}, \ and\ \bibinfo {author}
  {\bibfnamefont {J.~C.}\ \bibnamefont {Halimeh}},\ }\href@noop {} {\
  (\bibinfo {year} {2026})},\ \Eprint {http://arxiv.org/abs/2601.16166}
  {arXiv:2601.16166 [hep-lat]} \BibitemShut {NoStop}%
\bibitem [{\citenamefont {Gandon}\ \emph {et~al.}(2026)\citenamefont {Gandon},
  \citenamefont {Mariani}, \citenamefont {Banerjee}, \citenamefont {Huffman},
  \citenamefont {Kanwar}, \citenamefont {Tacchino}, \citenamefont {Wiese},\
  and\ \citenamefont {Tavernelli}}]{Gandon:2026das}%
  \BibitemOpen
  \bibfield  {author} {\bibinfo {author} {\bibfnamefont {A.}~\bibnamefont
  {Gandon}}, \bibinfo {author} {\bibfnamefont {A.}~\bibnamefont {Mariani}},
  \bibinfo {author} {\bibfnamefont {D.}~\bibnamefont {Banerjee}}, \bibinfo
  {author} {\bibfnamefont {E.}~\bibnamefont {Huffman}}, \bibinfo {author}
  {\bibfnamefont {G.}~\bibnamefont {Kanwar}}, \bibinfo {author} {\bibfnamefont
  {F.}~\bibnamefont {Tacchino}}, \bibinfo {author} {\bibfnamefont {U.-J.}\
  \bibnamefont {Wiese}}, \ and\ \bibinfo {author} {\bibfnamefont
  {I.}~\bibnamefont {Tavernelli}},\ }\href@noop {} {\enquote {\bibinfo {title}
  {{String dynamics of a (2+1)D U(1) quantum link model on a digital quantum
  computer}},}\ } (\bibinfo {year} {2026}),\ \Eprint
  {http://arxiv.org/abs/2606.19601} {arXiv:2606.19601 [quant-ph]} \BibitemShut
  {NoStop}%
\bibitem [{\citenamefont {Joshi}\ \emph {et~al.}(2026)\citenamefont {Joshi},
  \citenamefont {Tian}, \citenamefont {Hemery}, \citenamefont {Srivatsa},
  \citenamefont {Osborne}, \citenamefont {Dreyer}, \citenamefont {Rinaldi},\
  and\ \citenamefont {Halimeh}}]{Joshi:2026hfe}%
  \BibitemOpen
  \bibfield  {author} {\bibinfo {author} {\bibfnamefont {R.}~\bibnamefont
  {Joshi}}, \bibinfo {author} {\bibfnamefont {Y.}~\bibnamefont {Tian}},
  \bibinfo {author} {\bibfnamefont {K.}~\bibnamefont {Hemery}}, \bibinfo
  {author} {\bibfnamefont {N.}~\bibnamefont {Srivatsa}}, \bibinfo {author}
  {\bibfnamefont {J.~J.}\ \bibnamefont {Osborne}}, \bibinfo {author}
  {\bibfnamefont {H.}~\bibnamefont {Dreyer}}, \bibinfo {author} {\bibfnamefont
  {E.}~\bibnamefont {Rinaldi}}, \ and\ \bibinfo {author} {\bibfnamefont
  {J.~C.}\ \bibnamefont {Halimeh}},\ }\href@noop {} {\enquote {\bibinfo {title}
  {{Observation of genuine $2+1$D string dynamics in a U$(1)$ lattice gauge
  theory with a tunable plaquette term on a trapped-ion quantum computer}},}\ }
  (\bibinfo {year} {2026}),\ \Eprint {http://arxiv.org/abs/2604.07436}
  {arXiv:2604.07436 [quant-ph]} \BibitemShut {NoStop}%
\bibitem [{\citenamefont {Rule}\ and\ \citenamefont
  {Stetcu}(2026)}]{Rule:2026brk}%
  \BibitemOpen
  \bibfield  {author} {\bibinfo {author} {\bibfnamefont {E.}~\bibnamefont
  {Rule}}\ and\ \bibinfo {author} {\bibfnamefont {I.}~\bibnamefont {Stetcu}},\
  }\href@noop {} {\  (\bibinfo {year} {2026})},\ \Eprint
  {http://arxiv.org/abs/2603.26881} {arXiv:2603.26881 [nucl-th]} \BibitemShut
  {NoStop}%
\bibitem [{\citenamefont {Gustafson}\ \emph {et~al.}(2021)\citenamefont
  {Gustafson}, \citenamefont {Zhu}, \citenamefont {Dreher}, \citenamefont
  {Linke},\ and\ \citenamefont {Meurice}}]{Gustafson:2021imb}%
  \BibitemOpen
  \bibfield  {author} {\bibinfo {author} {\bibfnamefont {E.}~\bibnamefont
  {Gustafson}}, \bibinfo {author} {\bibfnamefont {Y.}~\bibnamefont {Zhu}},
  \bibinfo {author} {\bibfnamefont {P.}~\bibnamefont {Dreher}}, \bibinfo
  {author} {\bibfnamefont {N.~M.}\ \bibnamefont {Linke}}, \ and\ \bibinfo
  {author} {\bibfnamefont {Y.}~\bibnamefont {Meurice}},\ }\href {\doibase
  10.1103/PhysRevD.104.054507} {\bibfield  {journal} {\bibinfo  {journal}
  {Phys.Rev.D}\ }\textbf {\bibinfo {volume} {104}},\ \bibinfo {pages} {054507}
  (\bibinfo {year} {2021})},\ \Eprint {http://arxiv.org/abs/2103.06848}
  {arXiv:2103.06848 [hep-lat]} \BibitemShut {NoStop}%
\bibitem [{\citenamefont {Parks}\ \emph {et~al.}(2024)\citenamefont {Parks},
  \citenamefont {Carignan-Dugas}, \citenamefont {Gustafson}, \citenamefont
  {Meurice},\ and\ \citenamefont {Dreher}}]{Parks:2022kdb}%
  \BibitemOpen
  \bibfield  {author} {\bibinfo {author} {\bibfnamefont {Z.}~\bibnamefont
  {Parks}}, \bibinfo {author} {\bibfnamefont {A.}~\bibnamefont
  {Carignan-Dugas}}, \bibinfo {author} {\bibfnamefont {E.}~\bibnamefont
  {Gustafson}}, \bibinfo {author} {\bibfnamefont {Y.}~\bibnamefont {Meurice}},
  \ and\ \bibinfo {author} {\bibfnamefont {P.}~\bibnamefont {Dreher}},\ }\href
  {\doibase 10.1103/PhysRevD.109.014505} {\bibfield  {journal} {\bibinfo
  {journal} {Phys.Rev.D}\ }\textbf {\bibinfo {volume} {109}},\ \bibinfo {pages}
  {014505} (\bibinfo {year} {2024})},\ \Eprint
  {http://arxiv.org/abs/2212.05333} {arXiv:2212.05333 [quant-ph]} \BibitemShut
  {NoStop}%
\bibitem [{\citenamefont {Belyansky}\ \emph {et~al.}(2024)\citenamefont
  {Belyansky}, \citenamefont {Whitsitt}, \citenamefont {Mueller}, \citenamefont
  {Fahimniya}, \citenamefont {Bennewitz}, \citenamefont {Davoudi},\ and\
  \citenamefont {Gorshkov}}]{Belyansky:2023rgh}%
  \BibitemOpen
  \bibfield  {author} {\bibinfo {author} {\bibfnamefont {R.}~\bibnamefont
  {Belyansky}}, \bibinfo {author} {\bibfnamefont {S.}~\bibnamefont {Whitsitt}},
  \bibinfo {author} {\bibfnamefont {N.}~\bibnamefont {Mueller}}, \bibinfo
  {author} {\bibfnamefont {A.}~\bibnamefont {Fahimniya}}, \bibinfo {author}
  {\bibfnamefont {E.~R.}\ \bibnamefont {Bennewitz}}, \bibinfo {author}
  {\bibfnamefont {Z.}~\bibnamefont {Davoudi}}, \ and\ \bibinfo {author}
  {\bibfnamefont {A.~V.}\ \bibnamefont {Gorshkov}},\ }\href {\doibase
  10.1103/PhysRevLett.132.091903} {\bibfield  {journal} {\bibinfo  {journal}
  {Phys.Rev.Lett.}\ }\textbf {\bibinfo {volume} {132}},\ \bibinfo {pages}
  {091903} (\bibinfo {year} {2024})},\ \Eprint
  {http://arxiv.org/abs/2307.02522} {arXiv:2307.02522 [quant-ph]} \BibitemShut
  {NoStop}%
\bibitem [{\citenamefont {Chai}\ \emph
  {et~al.}(2025{\natexlab{a}})\citenamefont {Chai}, \citenamefont {Crippa},
  \citenamefont {Jansen}, \citenamefont {Kühn}, \citenamefont {Pascuzzi},
  \citenamefont {Tacchino},\ and\ \citenamefont {Tavernelli}}]{Chai:2023qpq}%
  \BibitemOpen
  \bibfield  {author} {\bibinfo {author} {\bibfnamefont {Y.}~\bibnamefont
  {Chai}}, \bibinfo {author} {\bibfnamefont {A.}~\bibnamefont {Crippa}},
  \bibinfo {author} {\bibfnamefont {K.}~\bibnamefont {Jansen}}, \bibinfo
  {author} {\bibfnamefont {S.}~\bibnamefont {Kühn}}, \bibinfo {author}
  {\bibfnamefont {V.~R.}\ \bibnamefont {Pascuzzi}}, \bibinfo {author}
  {\bibfnamefont {F.}~\bibnamefont {Tacchino}}, \ and\ \bibinfo {author}
  {\bibfnamefont {I.}~\bibnamefont {Tavernelli}},\ }\href {\doibase
  10.22331/q-2025-02-19-1638} {\bibfield  {journal} {\bibinfo  {journal}
  {Quantum}\ }\textbf {\bibinfo {volume} {9}},\ \bibinfo {pages} {1638}
  (\bibinfo {year} {2025}{\natexlab{a}})},\ \Eprint
  {http://arxiv.org/abs/2312.02272} {arXiv:2312.02272 [quant-ph]} \BibitemShut
  {NoStop}%
\bibitem [{\citenamefont {Yusf}\ \emph {et~al.}(2025)\citenamefont {Yusf},
  \citenamefont {Gan}, \citenamefont {Moffat},\ and\ \citenamefont
  {Rupak}}]{Yusf:2024igb}%
  \BibitemOpen
  \bibfield  {author} {\bibinfo {author} {\bibfnamefont {M.}~\bibnamefont
  {Yusf}}, \bibinfo {author} {\bibfnamefont {L.}~\bibnamefont {Gan}}, \bibinfo
  {author} {\bibfnamefont {C.}~\bibnamefont {Moffat}}, \ and\ \bibinfo {author}
  {\bibfnamefont {G.}~\bibnamefont {Rupak}},\ }\href {\doibase
  10.1103/PhysRevC.111.034001} {\bibfield  {journal} {\bibinfo  {journal}
  {Phys.Rev.C}\ }\textbf {\bibinfo {volume} {111}},\ \bibinfo {pages} {034001}
  (\bibinfo {year} {2025})},\ \Eprint {http://arxiv.org/abs/2406.09231}
  {arXiv:2406.09231 [nucl-th]} \BibitemShut {NoStop}%
\bibitem [{\citenamefont {Jha}\ \emph {et~al.}(2025)\citenamefont {Jha},
  \citenamefont {Milsted}, \citenamefont {Neuenfeld}, \citenamefont
  {Preskill},\ and\ \citenamefont {Vieira}}]{Jha:2024jan}%
  \BibitemOpen
  \bibfield  {author} {\bibinfo {author} {\bibfnamefont {R.~G.}\ \bibnamefont
  {Jha}}, \bibinfo {author} {\bibfnamefont {A.}~\bibnamefont {Milsted}},
  \bibinfo {author} {\bibfnamefont {D.}~\bibnamefont {Neuenfeld}}, \bibinfo
  {author} {\bibfnamefont {J.}~\bibnamefont {Preskill}}, \ and\ \bibinfo
  {author} {\bibfnamefont {P.}~\bibnamefont {Vieira}},\ }\href {\doibase
  10.1103/9dxz-k5wb} {\bibfield  {journal} {\bibinfo  {journal}
  {Phys.Rev.Res.}\ }\textbf {\bibinfo {volume} {7}},\ \bibinfo {pages} {023266}
  (\bibinfo {year} {2025})},\ \Eprint {http://arxiv.org/abs/2411.13645}
  {arXiv:2411.13645 [hep-th]} \BibitemShut {NoStop}%
\bibitem [{\citenamefont {Zemlevskiy}(2025)}]{Zemlevskiy:2024vxt}%
  \BibitemOpen
  \bibfield  {author} {\bibinfo {author} {\bibfnamefont {N.~A.}\ \bibnamefont
  {Zemlevskiy}},\ }\href {\doibase 10.1103/qr72-51v1} {\bibfield  {journal}
  {\bibinfo  {journal} {Phys.Rev.D}\ }\textbf {\bibinfo {volume} {112}},\
  \bibinfo {pages} {034502} (\bibinfo {year} {2025})},\ \Eprint
  {http://arxiv.org/abs/2411.02486} {arXiv:2411.02486 [quant-ph]} \BibitemShut
  {NoStop}%
\bibitem [{\citenamefont {Abel}\ \emph {et~al.}(2025)\citenamefont {Abel},
  \citenamefont {Spannowsky},\ and\ \citenamefont {Williams}}]{Abel:2025zxb}%
  \BibitemOpen
  \bibfield  {author} {\bibinfo {author} {\bibfnamefont {S.}~\bibnamefont
  {Abel}}, \bibinfo {author} {\bibfnamefont {M.}~\bibnamefont {Spannowsky}}, \
  and\ \bibinfo {author} {\bibfnamefont {S.}~\bibnamefont {Williams}},\ }\href
  {\doibase 10.1103/q36d-w649} {\bibfield  {journal} {\bibinfo  {journal}
  {Phys.Rev.A}\ }\textbf {\bibinfo {volume} {112}},\ \bibinfo {pages} {012614}
  (\bibinfo {year} {2025})},\ \Eprint {http://arxiv.org/abs/2502.01767}
  {arXiv:2502.01767 [quant-ph]} \BibitemShut {NoStop}%
\bibitem [{\citenamefont {Chai}\ \emph
  {et~al.}(2025{\natexlab{b}})\citenamefont {Chai}, \citenamefont {Guo},\ and\
  \citenamefont {K\"uhn}}]{Chai:2025qhf}%
  \BibitemOpen
  \bibfield  {author} {\bibinfo {author} {\bibfnamefont {Y.}~\bibnamefont
  {Chai}}, \bibinfo {author} {\bibfnamefont {Y.}~\bibnamefont {Guo}}, \ and\
  \bibinfo {author} {\bibfnamefont {S.}~\bibnamefont {K\"uhn}},\ }\href@noop {}
  {\enquote {\bibinfo {title} {{Towards Quantum Simulation of Meson Scattering
  in a Z2 Lattice Gauge Theory}},}\ } (\bibinfo {year} {2025}{\natexlab{b}}),\
  \Eprint {http://arxiv.org/abs/2505.21240} {arXiv:2505.21240 [quant-ph]}
  \BibitemShut {NoStop}%
\bibitem [{\citenamefont {Joshi}\ \emph
  {et~al.}(2025{\natexlab{b}})\citenamefont {Joshi}, \citenamefont {Louw},
  \citenamefont {Meth}, \citenamefont {Osborne}, \citenamefont {Mato},
  \citenamefont {Su}, \citenamefont {Ringbauer},\ and\ \citenamefont
  {Halimeh}}]{Joshi:2025rha}%
  \BibitemOpen
  \bibfield  {author} {\bibinfo {author} {\bibfnamefont {R.}~\bibnamefont
  {Joshi}}, \bibinfo {author} {\bibfnamefont {J.~C.}\ \bibnamefont {Louw}},
  \bibinfo {author} {\bibfnamefont {M.}~\bibnamefont {Meth}}, \bibinfo {author}
  {\bibfnamefont {J.~J.}\ \bibnamefont {Osborne}}, \bibinfo {author}
  {\bibfnamefont {K.}~\bibnamefont {Mato}}, \bibinfo {author} {\bibfnamefont
  {G.-X.}\ \bibnamefont {Su}}, \bibinfo {author} {\bibfnamefont
  {M.}~\bibnamefont {Ringbauer}}, \ and\ \bibinfo {author} {\bibfnamefont
  {J.~C.}\ \bibnamefont {Halimeh}},\ }\href@noop {} {\enquote {\bibinfo {title}
  {{Probing Hadron Scattering in Lattice Gauge Theories on Qudit Quantum
  Computers}},}\ } (\bibinfo {year} {2025}{\natexlab{b}}),\ \Eprint
  {http://arxiv.org/abs/2507.12614} {arXiv:2507.12614 [quant-ph]} \BibitemShut
  {NoStop}%
\bibitem [{\citenamefont {Chai}\ \emph {et~al.}(2026)\citenamefont {Chai},
  \citenamefont {Gibbs}, \citenamefont {Pascuzzi}, \citenamefont {Holmes},
  \citenamefont {Kühn}, \citenamefont {Tacchino},\ and\ \citenamefont
  {Tavernelli}}]{Chai:2025kbi}%
  \BibitemOpen
  \bibfield  {author} {\bibinfo {author} {\bibfnamefont {Y.}~\bibnamefont
  {Chai}}, \bibinfo {author} {\bibfnamefont {J.}~\bibnamefont {Gibbs}},
  \bibinfo {author} {\bibfnamefont {V.~R.}\ \bibnamefont {Pascuzzi}}, \bibinfo
  {author} {\bibfnamefont {Z.}~\bibnamefont {Holmes}}, \bibinfo {author}
  {\bibfnamefont {S.}~\bibnamefont {Kühn}}, \bibinfo {author} {\bibfnamefont
  {F.}~\bibnamefont {Tacchino}}, \ and\ \bibinfo {author} {\bibfnamefont
  {I.}~\bibnamefont {Tavernelli}},\ }\href {\doibase
  10.1038/s41534-026-01311-1} {\bibfield  {journal} {\bibinfo  {journal} {npj
  Quantum Inf.}\ }\textbf {\bibinfo {volume} {12}},\ \bibinfo {pages} {100}
  (\bibinfo {year} {2026})},\ \Eprint {http://arxiv.org/abs/2507.17832}
  {arXiv:2507.17832 [quant-ph]} \BibitemShut {NoStop}%
\bibitem [{\citenamefont {Zemlevskiy}(2026)}]{Zemlevskiy:2026kpc}%
  \BibitemOpen
  \bibfield  {author} {\bibinfo {author} {\bibfnamefont {N.~A.}\ \bibnamefont
  {Zemlevskiy}},\ }\href@noop {} {\enquote {\bibinfo {title} {{Exclusive
  Scattering Channels from Entanglement Structure in Real-Time Simulations}},}\
  } (\bibinfo {year} {2026}),\ \Eprint {http://arxiv.org/abs/2603.15621}
  {arXiv:2603.15621 [quant-ph]} \BibitemShut {NoStop}%
\bibitem [{\citenamefont {Ingoldby}\ \emph {et~al.}(2025)\citenamefont
  {Ingoldby}, \citenamefont {Spannowsky}, \citenamefont {Sypchenko},
  \citenamefont {Williams},\ and\ \citenamefont {Wingate}}]{Ingoldby:2025bdb}%
  \BibitemOpen
  \bibfield  {author} {\bibinfo {author} {\bibfnamefont {J.}~\bibnamefont
  {Ingoldby}}, \bibinfo {author} {\bibfnamefont {M.}~\bibnamefont
  {Spannowsky}}, \bibinfo {author} {\bibfnamefont {T.}~\bibnamefont
  {Sypchenko}}, \bibinfo {author} {\bibfnamefont {S.}~\bibnamefont {Williams}},
  \ and\ \bibinfo {author} {\bibfnamefont {M.}~\bibnamefont {Wingate}},\
  }\href@noop {} {\enquote {\bibinfo {title} {{Real-Time Scattering on Quantum
  Computers via Hamiltonian Truncation}},}\ } (\bibinfo {year} {2025}),\
  \Eprint {http://arxiv.org/abs/2505.03878} {arXiv:2505.03878 [quant-ph]}
  \BibitemShut {NoStop}%
\bibitem [{\citenamefont {Wang}\ \emph
  {et~al.}(2025{\natexlab{a}})\citenamefont {Wang} \emph
  {et~al.}}]{Wang:2025ocn}%
  \BibitemOpen
  \bibfield  {author} {\bibinfo {author} {\bibfnamefont {Z.}~\bibnamefont
  {Wang}} \emph {et~al.},\ }\href@noop {} {\  (\bibinfo {year}
  {2025}{\natexlab{a}})},\ \Eprint {http://arxiv.org/abs/2508.20759}
  {arXiv:2508.20759 [quant-ph]} \BibitemShut {NoStop}%
\bibitem [{\citenamefont {Artiaco}\ \emph {et~al.}(2025)\citenamefont
  {Artiaco}, \citenamefont {Barata},\ and\ \citenamefont
  {Rico}}]{Artiaco:2025qqq}%
  \BibitemOpen
  \bibfield  {author} {\bibinfo {author} {\bibfnamefont {C.}~\bibnamefont
  {Artiaco}}, \bibinfo {author} {\bibfnamefont {J.}~\bibnamefont {Barata}}, \
  and\ \bibinfo {author} {\bibfnamefont {E.}~\bibnamefont {Rico}},\ }\href@noop
  {} {\enquote {\bibinfo {title} {{Out-of-Equilibrium Dynamics in a U(1)
  Lattice Gauge Theory via Local Information Flows: Scattering and String
  Breaking}},}\ } (\bibinfo {year} {2025}),\ \Eprint
  {http://arxiv.org/abs/2510.16101} {arXiv:2510.16101 [quant-ph]} \BibitemShut
  {NoStop}%
\bibitem [{\citenamefont {Surace}\ \emph {et~al.}(2026)\citenamefont {Surace},
  \citenamefont {Bseiso},\ and\ \citenamefont {Preskill}}]{Surace:2026rtg}%
  \BibitemOpen
  \bibfield  {author} {\bibinfo {author} {\bibfnamefont {F.~M.}\ \bibnamefont
  {Surace}}, \bibinfo {author} {\bibfnamefont {S.}~\bibnamefont {Bseiso}}, \
  and\ \bibinfo {author} {\bibfnamefont {J.}~\bibnamefont {Preskill}},\
  }\href@noop {} {\enquote {\bibinfo {title} {{State preparation and detection
  for quantum simulation of particle collisions}},}\ } (\bibinfo {year}
  {2026}),\ \Eprint {http://arxiv.org/abs/2607.26142} {arXiv:2607.26142
  [quant-ph]} \BibitemShut {NoStop}%
\bibitem [{\citenamefont {Farrell}\ \emph {et~al.}(2023)\citenamefont
  {Farrell}, \citenamefont {Chernyshev}, \citenamefont {Powell}, \citenamefont
  {Zemlevskiy}, \citenamefont {Illa},\ and\ \citenamefont
  {Savage}}]{Farrell:2022vyh}%
  \BibitemOpen
  \bibfield  {author} {\bibinfo {author} {\bibfnamefont {R.~C.}\ \bibnamefont
  {Farrell}}, \bibinfo {author} {\bibfnamefont {I.~A.}\ \bibnamefont
  {Chernyshev}}, \bibinfo {author} {\bibfnamefont {S.~J.~M.}\ \bibnamefont
  {Powell}}, \bibinfo {author} {\bibfnamefont {N.~A.}\ \bibnamefont
  {Zemlevskiy}}, \bibinfo {author} {\bibfnamefont {M.}~\bibnamefont {Illa}}, \
  and\ \bibinfo {author} {\bibfnamefont {M.~J.}\ \bibnamefont {Savage}},\
  }\href {\doibase 10.1103/PhysRevD.107.054513} {\bibfield  {journal} {\bibinfo
   {journal} {Phys. Rev. D}\ }\textbf {\bibinfo {volume} {107}},\ \bibinfo
  {pages} {054513} (\bibinfo {year} {2023})},\ \Eprint
  {http://arxiv.org/abs/2209.10781} {arXiv:2209.10781 [quant-ph]} \BibitemShut
  {NoStop}%
\bibitem [{\citenamefont {Chernyshev}\ \emph {et~al.}(2026)\citenamefont
  {Chernyshev}, \citenamefont {Farrell}, \citenamefont {Illa}, \citenamefont
  {Savage}, \citenamefont {Maksymov}, \citenamefont {Tripier}, \citenamefont
  {Lopez-Ruiz}, \citenamefont {Arrasmith}, \citenamefont {de~Sereville},
  \citenamefont {Brodutch}, \citenamefont {Girotto}, \citenamefont {Kaushik},\
  and\ \citenamefont {Roetteler}}]{Chernyshev:2025lil}%
  \BibitemOpen
  \bibfield  {author} {\bibinfo {author} {\bibfnamefont {I.~A.}\ \bibnamefont
  {Chernyshev}}, \bibinfo {author} {\bibfnamefont {R.~C.}\ \bibnamefont
  {Farrell}}, \bibinfo {author} {\bibfnamefont {M.}~\bibnamefont {Illa}},
  \bibinfo {author} {\bibfnamefont {M.~J.}\ \bibnamefont {Savage}}, \bibinfo
  {author} {\bibfnamefont {A.}~\bibnamefont {Maksymov}}, \bibinfo {author}
  {\bibfnamefont {F.}~\bibnamefont {Tripier}}, \bibinfo {author} {\bibfnamefont
  {M.~A.}\ \bibnamefont {Lopez-Ruiz}}, \bibinfo {author} {\bibfnamefont
  {A.}~\bibnamefont {Arrasmith}}, \bibinfo {author} {\bibfnamefont
  {Y.}~\bibnamefont {de~Sereville}}, \bibinfo {author} {\bibfnamefont
  {A.}~\bibnamefont {Brodutch}}, \bibinfo {author} {\bibfnamefont
  {C.}~\bibnamefont {Girotto}}, \bibinfo {author} {\bibfnamefont
  {A.}~\bibnamefont {Kaushik}}, \ and\ \bibinfo {author} {\bibfnamefont
  {M.}~\bibnamefont {Roetteler}},\ }\href {\doibase 10.1038/s41467-026-68536-8}
  {\bibfield  {journal} {\bibinfo  {journal} {Nature Commun.}\ }\textbf
  {\bibinfo {volume} {17}},\ \bibinfo {pages} {1826} (\bibinfo {year}
  {2026})},\ \Eprint {http://arxiv.org/abs/2506.05757} {arXiv:2506.05757
  [quant-ph]} \BibitemShut {NoStop}%
\bibitem [{\citenamefont {Briceño}\ \emph {et~al.}(2021)\citenamefont
  {Briceño}, \citenamefont {Guerrero}, \citenamefont {Hansen},\ and\
  \citenamefont {Sturzu}}]{Briceno:2020rar}%
  \BibitemOpen
  \bibfield  {author} {\bibinfo {author} {\bibfnamefont {R.~A.}\ \bibnamefont
  {Briceño}}, \bibinfo {author} {\bibfnamefont {J.~V.}\ \bibnamefont
  {Guerrero}}, \bibinfo {author} {\bibfnamefont {M.~T.}\ \bibnamefont
  {Hansen}}, \ and\ \bibinfo {author} {\bibfnamefont {A.~M.}\ \bibnamefont
  {Sturzu}},\ }\href {\doibase 10.1103/PhysRevD.103.014506} {\bibfield
  {journal} {\bibinfo  {journal} {Phys.Rev.D}\ }\textbf {\bibinfo {volume}
  {103}},\ \bibinfo {pages} {014506} (\bibinfo {year} {2021})},\ \Eprint
  {http://arxiv.org/abs/2007.01155} {arXiv:2007.01155 [hep-lat]} \BibitemShut
  {NoStop}%
\bibitem [{\citenamefont {Brice{\~n}o}\ \emph {et~al.}(2024)\citenamefont
  {Brice{\~n}o}, \citenamefont {Edwards}, \citenamefont {Eaton}, \citenamefont
  {Gonz{\'a}lez-Arciniegas}, \citenamefont {Pfister},\ and\ \citenamefont
  {Siopsis}}]{Briceno:2023xcm}%
  \BibitemOpen
  \bibfield  {author} {\bibinfo {author} {\bibfnamefont {R.~A.}\ \bibnamefont
  {Brice{\~n}o}}, \bibinfo {author} {\bibfnamefont {R.~G.}\ \bibnamefont
  {Edwards}}, \bibinfo {author} {\bibfnamefont {M.}~\bibnamefont {Eaton}},
  \bibinfo {author} {\bibfnamefont {C.}~\bibnamefont
  {Gonz{\'a}lez-Arciniegas}}, \bibinfo {author} {\bibfnamefont
  {O.}~\bibnamefont {Pfister}}, \ and\ \bibinfo {author} {\bibfnamefont
  {G.}~\bibnamefont {Siopsis}},\ }\href {\doibase
  10.1103/PhysRevResearch.6.043065} {\bibfield  {journal} {\bibinfo  {journal}
  {Phys. Rev. Res.}\ }\textbf {\bibinfo {volume} {6}},\ \bibinfo {pages}
  {043065} (\bibinfo {year} {2024})},\ \Eprint
  {http://arxiv.org/abs/2312.12613} {arXiv:2312.12613 [quant-ph]} \BibitemShut
  {NoStop}%
\bibitem [{\citenamefont {Carrillo}\ \emph {et~al.}(2024)\citenamefont
  {Carrillo}, \citenamefont {Briceño},\ and\ \citenamefont
  {Sturzu}}]{Carrillo:2024chu}%
  \BibitemOpen
  \bibfield  {author} {\bibinfo {author} {\bibfnamefont {M.~A.}\ \bibnamefont
  {Carrillo}}, \bibinfo {author} {\bibfnamefont {R.~A.}\ \bibnamefont
  {Briceño}}, \ and\ \bibinfo {author} {\bibfnamefont {A.~M.}\ \bibnamefont
  {Sturzu}},\ }\href {\doibase 10.1103/PhysRevD.110.054503} {\bibfield
  {journal} {\bibinfo  {journal} {Phys.Rev.D}\ }\textbf {\bibinfo {volume}
  {110}},\ \bibinfo {pages} {054503} (\bibinfo {year} {2024})},\ \Eprint
  {http://arxiv.org/abs/2406.06877} {arXiv:2406.06877 [hep-lat]} \BibitemShut
  {NoStop}%
\bibitem [{\citenamefont {Burbano}\ \emph {et~al.}(2026)\citenamefont
  {Burbano}, \citenamefont {Carrillo}, \citenamefont {Urek}, \citenamefont
  {Ciavarella},\ and\ \citenamefont {Briceño}}]{Burbano:2025pef}%
  \BibitemOpen
  \bibfield  {author} {\bibinfo {author} {\bibfnamefont {I.~M.}\ \bibnamefont
  {Burbano}}, \bibinfo {author} {\bibfnamefont {M.~A.}\ \bibnamefont
  {Carrillo}}, \bibinfo {author} {\bibfnamefont {R.}~\bibnamefont {Urek}},
  \bibinfo {author} {\bibfnamefont {A.~N.}\ \bibnamefont {Ciavarella}}, \ and\
  \bibinfo {author} {\bibfnamefont {R.~A.}\ \bibnamefont {Briceño}},\ }\href
  {\doibase 10.1103/dc17-4zjy} {\bibfield  {journal} {\bibinfo  {journal}
  {Phys.Rev.D}\ }\textbf {\bibinfo {volume} {113}},\ \bibinfo {pages} {L071502}
  (\bibinfo {year} {2026})},\ \Eprint {http://arxiv.org/abs/2506.06511}
  {arXiv:2506.06511 [hep-lat]} \BibitemShut {NoStop}%
\bibitem [{\citenamefont {Ciavarella}(2020)}]{Ciavarella:2020vqm}%
  \BibitemOpen
  \bibfield  {author} {\bibinfo {author} {\bibfnamefont {A.}~\bibnamefont
  {Ciavarella}},\ }\href {\doibase 10.1103/PhysRevD.102.094505} {\bibfield
  {journal} {\bibinfo  {journal} {Phys.Rev.D}\ }\textbf {\bibinfo {volume}
  {102}},\ \bibinfo {pages} {094505} (\bibinfo {year} {2020})},\ \Eprint
  {http://arxiv.org/abs/2007.04447} {arXiv:2007.04447 [hep-th]} \BibitemShut
  {NoStop}%
\bibitem [{\citenamefont {Brice{\~n}o}\ \emph {et~al.}(2026)\citenamefont
  {Brice{\~n}o}, \citenamefont {Burbano}, \citenamefont {Ciavarella},
  \citenamefont {Rrapaj}, \citenamefont {Richardson},\ and\ \citenamefont
  {Walker-Loud}}]{long_paper}%
  \BibitemOpen
  \bibfield  {author} {\bibinfo {author} {\bibfnamefont {R.~A.}\ \bibnamefont
  {Brice{\~n}o}}, \bibinfo {author} {\bibfnamefont {I.~M.}\ \bibnamefont
  {Burbano}}, \bibinfo {author} {\bibfnamefont {A.~N.}\ \bibnamefont
  {Ciavarella}}, \bibinfo {author} {\bibfnamefont {E.}~\bibnamefont {Rrapaj}},
  \bibinfo {author} {\bibfnamefont {T.~R.}\ \bibnamefont {Richardson}}, \ and\
  \bibinfo {author} {\bibfnamefont {A.}~\bibnamefont {Walker-Loud}},\
  }\href@noop {} {\enquote {\bibinfo {title} {Quantum computation of ${NN}$
  scattering in $1+1$ dimensions {\`a} la {RESOs}},}\ } (\bibinfo {year}
  {2026}),\ \bibinfo {note} {in preparation}\BibitemShut {NoStop}%
\bibitem [{\citenamefont {Kaplan}\ \emph {et~al.}(1996)\citenamefont {Kaplan},
  \citenamefont {Savage},\ and\ \citenamefont {Wise}}]{Kaplan:1996xu}%
  \BibitemOpen
  \bibfield  {author} {\bibinfo {author} {\bibfnamefont {D.~B.}\ \bibnamefont
  {Kaplan}}, \bibinfo {author} {\bibfnamefont {M.~J.}\ \bibnamefont {Savage}},
  \ and\ \bibinfo {author} {\bibfnamefont {M.~B.}\ \bibnamefont {Wise}},\
  }\href {\doibase 10.1016/0550-3213(96)00357-4} {\bibfield  {journal}
  {\bibinfo  {journal} {Nucl. Phys. B}\ }\textbf {\bibinfo {volume} {478}},\
  \bibinfo {pages} {629} (\bibinfo {year} {1996})},\ \Eprint
  {http://arxiv.org/abs/nucl-th/9605002} {arXiv:nucl-th/9605002} \BibitemShut
  {NoStop}%
\bibitem [{\citenamefont {Kaplan}\ \emph
  {et~al.}(1998{\natexlab{a}})\citenamefont {Kaplan}, \citenamefont {Savage},\
  and\ \citenamefont {Wise}}]{Kaplan:1998tg}%
  \BibitemOpen
  \bibfield  {author} {\bibinfo {author} {\bibfnamefont {D.~B.}\ \bibnamefont
  {Kaplan}}, \bibinfo {author} {\bibfnamefont {M.~J.}\ \bibnamefont {Savage}},
  \ and\ \bibinfo {author} {\bibfnamefont {M.~B.}\ \bibnamefont {Wise}},\
  }\href {\doibase 10.1016/S0370-2693(98)00210-X} {\bibfield  {journal}
  {\bibinfo  {journal} {Phys. Lett. B}\ }\textbf {\bibinfo {volume} {424}},\
  \bibinfo {pages} {390} (\bibinfo {year} {1998}{\natexlab{a}})},\ \Eprint
  {http://arxiv.org/abs/nucl-th/9801034} {arXiv:nucl-th/9801034} \BibitemShut
  {NoStop}%
\bibitem [{\citenamefont {Kaplan}\ \emph
  {et~al.}(1998{\natexlab{b}})\citenamefont {Kaplan}, \citenamefont {Savage},\
  and\ \citenamefont {Wise}}]{Kaplan:1998we}%
  \BibitemOpen
  \bibfield  {author} {\bibinfo {author} {\bibfnamefont {D.~B.}\ \bibnamefont
  {Kaplan}}, \bibinfo {author} {\bibfnamefont {M.~J.}\ \bibnamefont {Savage}},
  \ and\ \bibinfo {author} {\bibfnamefont {M.~B.}\ \bibnamefont {Wise}},\
  }\href {\doibase 10.1016/S0550-3213(98)00440-4} {\bibfield  {journal}
  {\bibinfo  {journal} {Nucl. Phys. B}\ }\textbf {\bibinfo {volume} {534}},\
  \bibinfo {pages} {329} (\bibinfo {year} {1998}{\natexlab{b}})},\ \Eprint
  {http://arxiv.org/abs/nucl-th/9802075} {arXiv:nucl-th/9802075} \BibitemShut
  {NoStop}%
\bibitem [{\citenamefont {van Kolck}(1999)}]{vanKolck:1998bw}%
  \BibitemOpen
  \bibfield  {author} {\bibinfo {author} {\bibfnamefont {U.}~\bibnamefont {van
  Kolck}},\ }\href {\doibase 10.1016/S0375-9474(98)00612-5} {\bibfield
  {journal} {\bibinfo  {journal} {Nucl. Phys. A}\ }\textbf {\bibinfo {volume}
  {645}},\ \bibinfo {pages} {273} (\bibinfo {year} {1999})},\ \Eprint
  {http://arxiv.org/abs/nucl-th/9808007} {arXiv:nucl-th/9808007} \BibitemShut
  {NoStop}%
\bibitem [{\citenamefont {Lee}(2009)}]{Lee:2008fa}%
  \BibitemOpen
  \bibfield  {author} {\bibinfo {author} {\bibfnamefont {D.}~\bibnamefont
  {Lee}},\ }\href {\doibase 10.1016/j.ppnp.2008.12.001} {\bibfield  {journal}
  {\bibinfo  {journal} {Prog. Part. Nucl. Phys.}\ }\textbf {\bibinfo {volume}
  {63}},\ \bibinfo {pages} {117} (\bibinfo {year} {2009})},\ \Eprint
  {http://arxiv.org/abs/0804.3501} {arXiv:0804.3501 [nucl-th]} \BibitemShut
  {NoStop}%
\bibitem [{\citenamefont {Endres}\ \emph {et~al.}(2011)\citenamefont {Endres},
  \citenamefont {Kaplan}, \citenamefont {Lee},\ and\ \citenamefont
  {Nicholson}}]{Endres:2011er}%
  \BibitemOpen
  \bibfield  {author} {\bibinfo {author} {\bibfnamefont {M.~G.}\ \bibnamefont
  {Endres}}, \bibinfo {author} {\bibfnamefont {D.~B.}\ \bibnamefont {Kaplan}},
  \bibinfo {author} {\bibfnamefont {J.-W.}\ \bibnamefont {Lee}}, \ and\
  \bibinfo {author} {\bibfnamefont {A.~N.}\ \bibnamefont {Nicholson}},\ }\href
  {\doibase 10.1103/PhysRevA.84.043644} {\bibfield  {journal} {\bibinfo
  {journal} {Phys. Rev. A}\ }\textbf {\bibinfo {volume} {84}},\ \bibinfo
  {pages} {043644} (\bibinfo {year} {2011})},\ \Eprint
  {http://arxiv.org/abs/1106.5725} {arXiv:1106.5725 [hep-lat]} \BibitemShut
  {NoStop}%
\bibitem [{\citenamefont {L{\"a}hde}\ and\ \citenamefont
  {Mei{\ss}ner}(2019)}]{Lahde:2019npb}%
  \BibitemOpen
  \bibfield  {author} {\bibinfo {author} {\bibfnamefont {T.~A.}\ \bibnamefont
  {L{\"a}hde}}\ and\ \bibinfo {author} {\bibfnamefont {U.-G.}\ \bibnamefont
  {Mei{\ss}ner}},\ }\href {\doibase 10.1007/978-3-030-14189-9} {\emph {\bibinfo
  {title} {{Nuclear Lattice Effective Field Theory}: {An introduction}}}},\
  Vol.\ \bibinfo {volume} {957}\ (\bibinfo  {publisher} {Springer},\ \bibinfo
  {year} {2019})\BibitemShut {NoStop}%
\bibitem [{\citenamefont {Rothman}\ \emph {et~al.}(2026)\citenamefont
  {Rothman}, \citenamefont {Johnson-Toth}, \citenamefont {Hagen}, \citenamefont
  {Heinz},\ and\ \citenamefont {Papenbrock}}]{Rothman:2025uza}%
  \BibitemOpen
  \bibfield  {author} {\bibinfo {author} {\bibfnamefont {M.}~\bibnamefont
  {Rothman}}, \bibinfo {author} {\bibfnamefont {B.}~\bibnamefont
  {Johnson-Toth}}, \bibinfo {author} {\bibfnamefont {G.}~\bibnamefont {Hagen}},
  \bibinfo {author} {\bibfnamefont {M.}~\bibnamefont {Heinz}}, \ and\ \bibinfo
  {author} {\bibfnamefont {T.}~\bibnamefont {Papenbrock}},\ }\href {\doibase
  10.1140/epja/s10050-025-01764-6} {\bibfield  {journal} {\bibinfo  {journal}
  {Eur. Phys. J. A}\ }\textbf {\bibinfo {volume} {62}},\ \bibinfo {pages} {28}
  (\bibinfo {year} {2026})},\ \Eprint {http://arxiv.org/abs/2509.08771}
  {arXiv:2509.08771 [nucl-th]} \BibitemShut {NoStop}%
\bibitem [{\citenamefont {Chandrasekharan}\ \emph {et~al.}(2024)\citenamefont
  {Chandrasekharan}, \citenamefont {Nguyen},\ and\ \citenamefont
  {Richardson}}]{Chandrasekharan:2024iao}%
  \BibitemOpen
  \bibfield  {author} {\bibinfo {author} {\bibfnamefont {S.}~\bibnamefont
  {Chandrasekharan}}, \bibinfo {author} {\bibfnamefont {S.~T.}\ \bibnamefont
  {Nguyen}}, \ and\ \bibinfo {author} {\bibfnamefont {T.~R.}\ \bibnamefont
  {Richardson}},\ }\href {\doibase 10.1103/PhysRevC.110.024002} {\bibfield
  {journal} {\bibinfo  {journal} {Phys. Rev. C}\ }\textbf {\bibinfo {volume}
  {110}},\ \bibinfo {pages} {024002} (\bibinfo {year} {2024})},\ \Eprint
  {http://arxiv.org/abs/2402.15377} {arXiv:2402.15377 [nucl-th]} \BibitemShut
  {NoStop}%
\bibitem [{\citenamefont {Singh}\ and\ \citenamefont
  {Chandrasekharan}(2019)}]{Singh:2018mnm}%
  \BibitemOpen
  \bibfield  {author} {\bibinfo {author} {\bibfnamefont {H.}~\bibnamefont
  {Singh}}\ and\ \bibinfo {author} {\bibfnamefont {S.}~\bibnamefont
  {Chandrasekharan}},\ }\href {\doibase 10.1103/PhysRevD.99.074511} {\bibfield
  {journal} {\bibinfo  {journal} {Phys. Rev. D}\ }\textbf {\bibinfo {volume}
  {99}},\ \bibinfo {pages} {074511} (\bibinfo {year} {2019})},\ \Eprint
  {http://arxiv.org/abs/1812.05080} {arXiv:1812.05080 [hep-lat]} \BibitemShut
  {NoStop}%
\bibitem [{\citenamefont {Lee}\ and\ \citenamefont
  {Sch{\"a}fer}(2005)}]{Lee:2004qd}%
  \BibitemOpen
  \bibfield  {author} {\bibinfo {author} {\bibfnamefont {D.}~\bibnamefont
  {Lee}}\ and\ \bibinfo {author} {\bibfnamefont {T.}~\bibnamefont
  {Sch{\"a}fer}},\ }\href {\doibase 10.1103/PhysRevC.72.024006} {\bibfield
  {journal} {\bibinfo  {journal} {Phys. Rev. C}\ }\textbf {\bibinfo {volume}
  {72}},\ \bibinfo {pages} {024006} (\bibinfo {year} {2005})},\ \Eprint
  {http://arxiv.org/abs/nucl-th/0412002} {arXiv:nucl-th/0412002} \BibitemShut
  {NoStop}%
\bibitem [{\citenamefont {K{\"o}rber}\ \emph {et~al.}(2019)\citenamefont
  {K{\"o}rber}, \citenamefont {Berkowitz},\ and\ \citenamefont
  {Luu}}]{Korber:2019cuq}%
  \BibitemOpen
  \bibfield  {author} {\bibinfo {author} {\bibfnamefont {C.}~\bibnamefont
  {K{\"o}rber}}, \bibinfo {author} {\bibfnamefont {E.}~\bibnamefont
  {Berkowitz}}, \ and\ \bibinfo {author} {\bibfnamefont {T.}~\bibnamefont
  {Luu}},\ }\href@noop {} {\enquote {\bibinfo {title} {{Renormalization of a
  Contact Interaction on a Lattice}},}\ } (\bibinfo {year} {2019}),\ \Eprint
  {http://arxiv.org/abs/1912.04425} {arXiv:1912.04425 [hep-lat]} \BibitemShut
  {NoStop}%
\bibitem [{\citenamefont {Hubbard}(1963)}]{Hubbard:1963}%
  \BibitemOpen
  \bibfield  {author} {\bibinfo {author} {\bibfnamefont {J.}~\bibnamefont
  {Hubbard}},\ }\href {\doibase 10.1098/rspa.1963.0204} {\bibfield  {journal}
  {\bibinfo  {journal} {Proc. Roy. Soc. Lond. A}\ }\textbf {\bibinfo {volume}
  {276}},\ \bibinfo {pages} {238} (\bibinfo {year} {1963})}\BibitemShut
  {NoStop}%
\bibitem [{\citenamefont {Linke}\ \emph {et~al.}(2018)\citenamefont {Linke},
  \citenamefont {Johri}, \citenamefont {Figgatt}, \citenamefont {Landsman},
  \citenamefont {Matsuura},\ and\ \citenamefont {Monroe}}]{Linke:2017xlv}%
  \BibitemOpen
  \bibfield  {author} {\bibinfo {author} {\bibfnamefont {N.~M.}\ \bibnamefont
  {Linke}}, \bibinfo {author} {\bibfnamefont {S.}~\bibnamefont {Johri}},
  \bibinfo {author} {\bibfnamefont {C.}~\bibnamefont {Figgatt}}, \bibinfo
  {author} {\bibfnamefont {K.~A.}\ \bibnamefont {Landsman}}, \bibinfo {author}
  {\bibfnamefont {A.~Y.}\ \bibnamefont {Matsuura}}, \ and\ \bibinfo {author}
  {\bibfnamefont {C.}~\bibnamefont {Monroe}},\ }\href {\doibase
  10.1103/PhysRevA.98.052334} {\bibfield  {journal} {\bibinfo  {journal} {Phys.
  Rev. A}\ }\textbf {\bibinfo {volume} {98}},\ \bibinfo {pages} {052334}
  (\bibinfo {year} {2018})},\ \Eprint {http://arxiv.org/abs/1712.08581}
  {arXiv:1712.08581 [quant-ph]} \BibitemShut {NoStop}%
\bibitem [{\citenamefont {Arute}\ \emph {et~al.}(2020)\citenamefont {Arute}
  \emph {et~al.}}]{Arute:2020ypn}%
  \BibitemOpen
  \bibfield  {author} {\bibinfo {author} {\bibfnamefont {F.}~\bibnamefont
  {Arute}} \emph {et~al.},\ }\href@noop {} {\  (\bibinfo {year} {2020})},\
  \Eprint {http://arxiv.org/abs/2010.07965} {arXiv:2010.07965 [quant-ph]}
  \BibitemShut {NoStop}%
\bibitem [{\citenamefont {Madhusudhana}\ \emph {et~al.}(2021)\citenamefont
  {Madhusudhana}, \citenamefont {Scherg}, \citenamefont {Kohlert},
  \citenamefont {Bloch},\ and\ \citenamefont
  {Aidelsburger}}]{Madhusudhana:2021qyp}%
  \BibitemOpen
  \bibfield  {author} {\bibinfo {author} {\bibfnamefont {B.~H.}\ \bibnamefont
  {Madhusudhana}}, \bibinfo {author} {\bibfnamefont {S.}~\bibnamefont
  {Scherg}}, \bibinfo {author} {\bibfnamefont {T.}~\bibnamefont {Kohlert}},
  \bibinfo {author} {\bibfnamefont {I.}~\bibnamefont {Bloch}}, \ and\ \bibinfo
  {author} {\bibfnamefont {M.}~\bibnamefont {Aidelsburger}},\ }\href {\doibase
  10.1103/PRXQuantum.2.040325} {\bibfield  {journal} {\bibinfo  {journal} {PRX
  Quantum}\ }\textbf {\bibinfo {volume} {2}},\ \bibinfo {pages} {040325}
  (\bibinfo {year} {2021})},\ \Eprint {http://arxiv.org/abs/2105.06372}
  {arXiv:2105.06372 [quant-ph]} \BibitemShut {NoStop}%
\bibitem [{\citenamefont {Stanisic}\ \emph {et~al.}(2022)\citenamefont
  {Stanisic}, \citenamefont {Bosse}, \citenamefont {Gambetta}, \citenamefont
  {Santos}, \citenamefont {Mruczkiewicz}, \citenamefont {O'Brien},
  \citenamefont {Ostby},\ and\ \citenamefont {Montanaro}}]{Stanisic:2021irm}%
  \BibitemOpen
  \bibfield  {author} {\bibinfo {author} {\bibfnamefont {S.}~\bibnamefont
  {Stanisic}}, \bibinfo {author} {\bibfnamefont {J.~L.}\ \bibnamefont {Bosse}},
  \bibinfo {author} {\bibfnamefont {F.~M.}\ \bibnamefont {Gambetta}}, \bibinfo
  {author} {\bibfnamefont {R.~A.}\ \bibnamefont {Santos}}, \bibinfo {author}
  {\bibfnamefont {W.}~\bibnamefont {Mruczkiewicz}}, \bibinfo {author}
  {\bibfnamefont {T.~E.}\ \bibnamefont {O'Brien}}, \bibinfo {author}
  {\bibfnamefont {E.}~\bibnamefont {Ostby}}, \ and\ \bibinfo {author}
  {\bibfnamefont {A.}~\bibnamefont {Montanaro}},\ }\href {\doibase
  10.1038/s41467-022-33335-4} {\bibfield  {journal} {\bibinfo  {journal}
  {Nature Commun.}\ }\textbf {\bibinfo {volume} {13}},\ \bibinfo {pages} {5743}
  (\bibinfo {year} {2022})},\ \Eprint {http://arxiv.org/abs/2112.02025}
  {arXiv:2112.02025 [quant-ph]} \BibitemShut {NoStop}%
\bibitem [{\citenamefont {Chen}\ \emph {et~al.}(2023)\citenamefont {Chen},
  \citenamefont {Zhang}, \citenamefont {Zhang}, \citenamefont {Su},
  \citenamefont {Lu}, \citenamefont {Zhang}, \citenamefont {Qiao},
  \citenamefont {Li}, \citenamefont {Zhang},\ and\ \citenamefont
  {Kim}}]{Chen:2023ukf}%
  \BibitemOpen
  \bibfield  {author} {\bibinfo {author} {\bibfnamefont {W.}~\bibnamefont
  {Chen}}, \bibinfo {author} {\bibfnamefont {S.}~\bibnamefont {Zhang}},
  \bibinfo {author} {\bibfnamefont {J.}~\bibnamefont {Zhang}}, \bibinfo
  {author} {\bibfnamefont {X.}~\bibnamefont {Su}}, \bibinfo {author}
  {\bibfnamefont {Y.}~\bibnamefont {Lu}}, \bibinfo {author} {\bibfnamefont
  {K.}~\bibnamefont {Zhang}}, \bibinfo {author} {\bibfnamefont
  {M.}~\bibnamefont {Qiao}}, \bibinfo {author} {\bibfnamefont {Y.}~\bibnamefont
  {Li}}, \bibinfo {author} {\bibfnamefont {J.-N.}\ \bibnamefont {Zhang}}, \
  and\ \bibinfo {author} {\bibfnamefont {K.}~\bibnamefont {Kim}},\ }\href
  {\doibase 10.1038/s41534-023-00784-8} {\bibfield  {journal} {\bibinfo
  {journal} {npj Quantum Inf.}\ }\textbf {\bibinfo {volume} {9}},\ \bibinfo
  {pages} {122} (\bibinfo {year} {2023})},\ \Eprint
  {http://arxiv.org/abs/2302.10436} {arXiv:2302.10436 [quant-ph]} \BibitemShut
  {NoStop}%
\bibitem [{\citenamefont {Paul}\ and\ \citenamefont
  {Mishra}(2024)}]{Paul:2024ldn}%
  \BibitemOpen
  \bibfield  {author} {\bibinfo {author} {\bibfnamefont {B.}~\bibnamefont
  {Paul}}\ and\ \bibinfo {author} {\bibfnamefont {T.}~\bibnamefont {Mishra}},\
  }\href {\doibase 10.1103/PhysRevB.110.L020302} {\bibfield  {journal}
  {\bibinfo  {journal} {Phys. Rev. B}\ }\textbf {\bibinfo {volume} {110}},\
  \bibinfo {pages} {L020302} (\bibinfo {year} {2024})},\ \Eprint
  {http://arxiv.org/abs/2403.02229} {arXiv:2403.02229 [quant-ph]} \BibitemShut
  {NoStop}%
\bibitem [{\citenamefont {Srinivasan}\ \emph {et~al.}(2024)\citenamefont
  {Srinivasan} \emph {et~al.}}]{Srinivasan:2024fvq}%
  \BibitemOpen
  \bibfield  {author} {\bibinfo {author} {\bibfnamefont {D.}~\bibnamefont
  {Srinivasan}} \emph {et~al.},\ }\href@noop {} {\  (\bibinfo {year} {2024})},\
  \Eprint {http://arxiv.org/abs/2411.07778} {arXiv:2411.07778 [quant-ph]}
  \BibitemShut {NoStop}%
\bibitem [{\citenamefont {Vilchez-Estevez}\ \emph {et~al.}(2025)\citenamefont
  {Vilchez-Estevez}, \citenamefont {Santos}, \citenamefont {Wang},\ and\
  \citenamefont {Gambetta}}]{Vilchez-Estevez:2025zjm}%
  \BibitemOpen
  \bibfield  {author} {\bibinfo {author} {\bibfnamefont {L.}~\bibnamefont
  {Vilchez-Estevez}}, \bibinfo {author} {\bibfnamefont {R.~A.}\ \bibnamefont
  {Santos}}, \bibinfo {author} {\bibfnamefont {S.~Y.}\ \bibnamefont {Wang}}, \
  and\ \bibinfo {author} {\bibfnamefont {F.~M.}\ \bibnamefont {Gambetta}},\
  }\href {\doibase 10.1103/ydfw-k83n} {\bibfield  {journal} {\bibinfo
  {journal} {Phys. Rev. B}\ }\textbf {\bibinfo {volume} {112}},\ \bibinfo
  {pages} {045143} (\bibinfo {year} {2025})},\ \bibinfo {note} {[Erratum:
  Phys.Rev.B 112, 239902 (2025)]},\ \Eprint {http://arxiv.org/abs/2501.04649}
  {arXiv:2501.04649 [quant-ph]} \BibitemShut {NoStop}%
\bibitem [{\citenamefont {Chowdhury}\ \emph {et~al.}(2026)\citenamefont
  {Chowdhury}, \citenamefont {Korepin}, \citenamefont {Pascuzzi},\ and\
  \citenamefont {Yu}}]{Chowdhury:2025tue}%
  \BibitemOpen
  \bibfield  {author} {\bibinfo {author} {\bibfnamefont {T.~A.}\ \bibnamefont
  {Chowdhury}}, \bibinfo {author} {\bibfnamefont {V.}~\bibnamefont {Korepin}},
  \bibinfo {author} {\bibfnamefont {V.~R.}\ \bibnamefont {Pascuzzi}}, \ and\
  \bibinfo {author} {\bibfnamefont {K.}~\bibnamefont {Yu}},\ }\href {\doibase
  10.1063/5.0306069} {\bibfield  {journal} {\bibinfo  {journal} {Appl. Phys.
  Rev.}\ }\textbf {\bibinfo {volume} {13}},\ \bibinfo {pages} {011434}
  (\bibinfo {year} {2026})},\ \Eprint {http://arxiv.org/abs/2509.14196}
  {arXiv:2509.14196 [quant-ph]} \BibitemShut {NoStop}%
\bibitem [{\citenamefont {Hartnett}\ \emph {et~al.}(2026)\citenamefont
  {Hartnett}, \citenamefont {Najafi}, \citenamefont {Khindanov}, \citenamefont
  {Liao}, \citenamefont {Schutzman}, \citenamefont {Hush}, \citenamefont
  {Biercuk},\ and\ \citenamefont {Baum}}]{Hartnett:2026abf}%
  \BibitemOpen
  \bibfield  {author} {\bibinfo {author} {\bibfnamefont {G.~S.}\ \bibnamefont
  {Hartnett}}, \bibinfo {author} {\bibfnamefont {K.~S.}\ \bibnamefont
  {Najafi}}, \bibinfo {author} {\bibfnamefont {A.}~\bibnamefont {Khindanov}},
  \bibinfo {author} {\bibfnamefont {H.}~\bibnamefont {Liao}}, \bibinfo {author}
  {\bibfnamefont {M.}~\bibnamefont {Schutzman}}, \bibinfo {author}
  {\bibfnamefont {M.~R.}\ \bibnamefont {Hush}}, \bibinfo {author}
  {\bibfnamefont {M.~J.}\ \bibnamefont {Biercuk}}, \ and\ \bibinfo {author}
  {\bibfnamefont {Y.}~\bibnamefont {Baum}},\ }\href@noop {} {\  (\bibinfo
  {year} {2026})},\ \Eprint {http://arxiv.org/abs/2605.04025} {arXiv:2605.04025
  [quant-ph]} \BibitemShut {NoStop}%
\bibitem [{\citenamefont {Kalam}\ \emph {et~al.}(2026)\citenamefont {Kalam},
  \citenamefont {Deb}, \citenamefont {Sakurai}, \citenamefont {Mishra},
  \citenamefont {Prasannaa},\ and\ \citenamefont {Das}}]{Kalam:2026yxv}%
  \BibitemOpen
  \bibfield  {author} {\bibinfo {author} {\bibfnamefont {A.}~\bibnamefont
  {Kalam}}, \bibinfo {author} {\bibfnamefont {P.}~\bibnamefont {Deb}}, \bibinfo
  {author} {\bibfnamefont {A.}~\bibnamefont {Sakurai}}, \bibinfo {author}
  {\bibfnamefont {T.}~\bibnamefont {Mishra}}, \bibinfo {author} {\bibfnamefont
  {V.~S.}\ \bibnamefont {Prasannaa}}, \ and\ \bibinfo {author} {\bibfnamefont
  {B.~P.}\ \bibnamefont {Das}},\ }\href@noop {} {\  (\bibinfo {year} {2026})},\
  \Eprint {http://arxiv.org/abs/2608.02245} {arXiv:2608.02245 [quant-ph]}
  \BibitemShut {NoStop}%
\bibitem [{\citenamefont {Cleve}\ \emph {et~al.}(1998)\citenamefont {Cleve},
  \citenamefont {Ekert}, \citenamefont {Macchiavello},\ and\ \citenamefont
  {Mosca}}]{10.1098/rspa.1998.0164}%
  \BibitemOpen
  \bibfield  {author} {\bibinfo {author} {\bibfnamefont {R.}~\bibnamefont
  {Cleve}}, \bibinfo {author} {\bibfnamefont {A.}~\bibnamefont {Ekert}},
  \bibinfo {author} {\bibfnamefont {C.}~\bibnamefont {Macchiavello}}, \ and\
  \bibinfo {author} {\bibfnamefont {M.}~\bibnamefont {Mosca}},\ }\href
  {\doibase 10.1098/rspa.1998.0164} {\bibfield  {journal} {\bibinfo  {journal}
  {Proceedings of the Royal Society A: Mathematical, Physical and Engineering
  Sciences}\ }\textbf {\bibinfo {volume} {454}},\ \bibinfo {pages} {339}
  (\bibinfo {year} {1998})},\ \Eprint
  {http://arxiv.org/abs/https://royalsocietypublishing.org/rspa/article-pdf/454/1969/339/633969/rspa.1998.0164.pdf}
  {https://royalsocietypublishing.org/rspa/article-pdf/454/1969/339/633969/rspa.1998.0164.pdf}
  \BibitemShut {NoStop}%
\bibitem [{\citenamefont {Mueller}\ \emph {et~al.}(2020)\citenamefont
  {Mueller}, \citenamefont {Tarasov},\ and\ \citenamefont
  {Venugopalan}}]{Mueller:2019qqj}%
  \BibitemOpen
  \bibfield  {author} {\bibinfo {author} {\bibfnamefont {N.}~\bibnamefont
  {Mueller}}, \bibinfo {author} {\bibfnamefont {A.}~\bibnamefont {Tarasov}}, \
  and\ \bibinfo {author} {\bibfnamefont {R.}~\bibnamefont {Venugopalan}},\
  }\href {\doibase 10.1103/PhysRevD.102.016007} {\bibfield  {journal} {\bibinfo
   {journal} {Phys.Rev.D}\ }\textbf {\bibinfo {volume} {102}},\ \bibinfo
  {pages} {016007} (\bibinfo {year} {2020})},\ \Eprint
  {http://arxiv.org/abs/1908.07051} {arXiv:1908.07051 [hep-th]} \BibitemShut
  {NoStop}%
\bibitem [{\citenamefont {Lamm}\ \emph {et~al.}(2020)\citenamefont {Lamm},
  \citenamefont {Lawrence},\ and\ \citenamefont {Yamauchi}}]{Lamm:2019uyc}%
  \BibitemOpen
  \bibfield  {author} {\bibinfo {author} {\bibfnamefont {H.}~\bibnamefont
  {Lamm}}, \bibinfo {author} {\bibfnamefont {S.}~\bibnamefont {Lawrence}}, \
  and\ \bibinfo {author} {\bibfnamefont {Y.}~\bibnamefont {Yamauchi}},\ }\href
  {\doibase 10.1103/PhysRevResearch.2.013272} {\bibfield  {journal} {\bibinfo
  {journal} {Phys.Rev.Res.}\ }\textbf {\bibinfo {volume} {2}},\ \bibinfo
  {pages} {013272} (\bibinfo {year} {2020})},\ \Eprint
  {http://arxiv.org/abs/1908.10439} {arXiv:1908.10439 [hep-lat]} \BibitemShut
  {NoStop}%
\bibitem [{\citenamefont {Kreshchuk}\ \emph {et~al.}(2022)\citenamefont
  {Kreshchuk}, \citenamefont {Kirby}, \citenamefont {Goldstein}, \citenamefont
  {Beauchemin},\ and\ \citenamefont {Love}}]{Kreshchuk:2020dla}%
  \BibitemOpen
  \bibfield  {author} {\bibinfo {author} {\bibfnamefont {M.}~\bibnamefont
  {Kreshchuk}}, \bibinfo {author} {\bibfnamefont {W.~M.}\ \bibnamefont
  {Kirby}}, \bibinfo {author} {\bibfnamefont {G.}~\bibnamefont {Goldstein}},
  \bibinfo {author} {\bibfnamefont {H.}~\bibnamefont {Beauchemin}}, \ and\
  \bibinfo {author} {\bibfnamefont {P.~J.}\ \bibnamefont {Love}},\ }\href
  {\doibase 10.1103/PhysRevA.105.032418} {\bibfield  {journal} {\bibinfo
  {journal} {Phys.Rev.A}\ }\textbf {\bibinfo {volume} {105}},\ \bibinfo {pages}
  {032418} (\bibinfo {year} {2022})},\ \Eprint
  {http://arxiv.org/abs/2002.04016} {arXiv:2002.04016 [quant-ph]} \BibitemShut
  {NoStop}%
\bibitem [{\citenamefont {Echevarria}\ \emph {et~al.}(2021)\citenamefont
  {Echevarria}, \citenamefont {Egusquiza}, \citenamefont {Rico},\ and\
  \citenamefont {Schnell}}]{Echevarria:2020wct}%
  \BibitemOpen
  \bibfield  {author} {\bibinfo {author} {\bibfnamefont {M.}~\bibnamefont
  {Echevarria}}, \bibinfo {author} {\bibfnamefont {I.}~\bibnamefont
  {Egusquiza}}, \bibinfo {author} {\bibfnamefont {E.}~\bibnamefont {Rico}}, \
  and\ \bibinfo {author} {\bibfnamefont {G.}~\bibnamefont {Schnell}},\ }\href
  {\doibase 10.1103/PhysRevD.104.014512} {\bibfield  {journal} {\bibinfo
  {journal} {Phys.Rev.D}\ }\textbf {\bibinfo {volume} {104}},\ \bibinfo {pages}
  {014512} (\bibinfo {year} {2021})},\ \Eprint
  {http://arxiv.org/abs/2011.01275} {arXiv:2011.01275 [quant-ph]} \BibitemShut
  {NoStop}%
\bibitem [{\citenamefont {Li}\ \emph {et~al.}(2022)\citenamefont {Li},
  \citenamefont {Guo}, \citenamefont {Lai}, \citenamefont {Liu}, \citenamefont
  {Wang}, \citenamefont {Xing}, \citenamefont {Zhang},\ and\ \citenamefont
  {Zhu}}]{Li:2021kcs}%
  \BibitemOpen
  \bibfield  {author} {\bibinfo {author} {\bibfnamefont {T.}~\bibnamefont
  {Li}}, \bibinfo {author} {\bibfnamefont {X.}~\bibnamefont {Guo}}, \bibinfo
  {author} {\bibfnamefont {W.~K.}\ \bibnamefont {Lai}}, \bibinfo {author}
  {\bibfnamefont {X.}~\bibnamefont {Liu}}, \bibinfo {author} {\bibfnamefont
  {E.}~\bibnamefont {Wang}}, \bibinfo {author} {\bibfnamefont {H.}~\bibnamefont
  {Xing}}, \bibinfo {author} {\bibfnamefont {D.-B.}\ \bibnamefont {Zhang}}, \
  and\ \bibinfo {author} {\bibfnamefont {S.-L.}\ \bibnamefont {Zhu}} (\bibinfo
  {collaboration} {QuNu}),\ }\href {\doibase 10.1103/PhysRevD.105.L111502}
  {\bibfield  {journal} {\bibinfo  {journal} {Phys.Rev.D}\ }\textbf {\bibinfo
  {volume} {105}},\ \bibinfo {pages} {L111502} (\bibinfo {year} {2022})},\
  \Eprint {http://arxiv.org/abs/2106.03865} {arXiv:2106.03865 [hep-ph]}
  \BibitemShut {NoStop}%
\bibitem [{\citenamefont {Qian}\ \emph {et~al.}(2022)\citenamefont {Qian},
  \citenamefont {Basili}, \citenamefont {Pal}, \citenamefont {Luecke},\ and\
  \citenamefont {Vary}}]{Qian:2021jxp}%
  \BibitemOpen
  \bibfield  {author} {\bibinfo {author} {\bibfnamefont {W.}~\bibnamefont
  {Qian}}, \bibinfo {author} {\bibfnamefont {R.}~\bibnamefont {Basili}},
  \bibinfo {author} {\bibfnamefont {S.}~\bibnamefont {Pal}}, \bibinfo {author}
  {\bibfnamefont {G.}~\bibnamefont {Luecke}}, \ and\ \bibinfo {author}
  {\bibfnamefont {J.~P.}\ \bibnamefont {Vary}},\ }\href {\doibase
  10.1103/PhysRevResearch.4.043193} {\bibfield  {journal} {\bibinfo  {journal}
  {Phys.Rev.Res.}\ }\textbf {\bibinfo {volume} {4}},\ \bibinfo {pages} {043193}
  (\bibinfo {year} {2022})},\ \Eprint {http://arxiv.org/abs/2112.01927}
  {arXiv:2112.01927 [quant-ph]} \BibitemShut {NoStop}%
\bibitem [{\citenamefont {Grieninger}\ \emph {et~al.}(2024)\citenamefont
  {Grieninger}, \citenamefont {Ikeda},\ and\ \citenamefont
  {Zahed}}]{Grieninger:2024cdl}%
  \BibitemOpen
  \bibfield  {author} {\bibinfo {author} {\bibfnamefont {S.}~\bibnamefont
  {Grieninger}}, \bibinfo {author} {\bibfnamefont {K.}~\bibnamefont {Ikeda}}, \
  and\ \bibinfo {author} {\bibfnamefont {I.}~\bibnamefont {Zahed}},\ }\href
  {\doibase 10.1103/PhysRevD.110.076008} {\bibfield  {journal} {\bibinfo
  {journal} {Phys.Rev.D}\ }\textbf {\bibinfo {volume} {110}},\ \bibinfo {pages}
  {076008} (\bibinfo {year} {2024})},\ \Eprint
  {http://arxiv.org/abs/2404.05112} {arXiv:2404.05112 [hep-ph]} \BibitemShut
  {NoStop}%
\bibitem [{\citenamefont {Chen}\ \emph {et~al.}(2025)\citenamefont {Chen},
  \citenamefont {Chen},\ and\ \citenamefont {Meher}}]{Chen:2025zeh}%
  \BibitemOpen
  \bibfield  {author} {\bibinfo {author} {\bibfnamefont {J.-W.}\ \bibnamefont
  {Chen}}, \bibinfo {author} {\bibfnamefont {Y.-T.}\ \bibnamefont {Chen}}, \
  and\ \bibinfo {author} {\bibfnamefont {G.}~\bibnamefont {Meher}},\
  }\href@noop {} {\enquote {\bibinfo {title} {{Parton Distributions on a
  Quantum Computer}},}\ } (\bibinfo {year} {2025}),\ \Eprint
  {http://arxiv.org/abs/2506.16829} {arXiv:2506.16829 [hep-lat]} \BibitemShut
  {NoStop}%
\bibitem [{\citenamefont {Zou}\ \emph {et~al.}(2026)\citenamefont {Zou},
  \citenamefont {Li}, \citenamefont {Liang}, \citenamefont {Wang},\ and\
  \citenamefont {Xing}}]{Zou:2026cfk}%
  \BibitemOpen
  \bibfield  {author} {\bibinfo {author} {\bibfnamefont {D.}~\bibnamefont
  {Zou}}, \bibinfo {author} {\bibfnamefont {T.}~\bibnamefont {Li}}, \bibinfo
  {author} {\bibfnamefont {J.}~\bibnamefont {Liang}}, \bibinfo {author}
  {\bibfnamefont {E.}~\bibnamefont {Wang}}, \ and\ \bibinfo {author}
  {\bibfnamefont {H.}~\bibnamefont {Xing}},\ }\href@noop {} {\enquote {\bibinfo
  {title} {{Hadronic tensor in lattice gauge theories by quantum computing}},}\
  } (\bibinfo {year} {2026}),\ \Eprint {http://arxiv.org/abs/2606.17003}
  {arXiv:2606.17003 [hep-ph]} \BibitemShut {NoStop}%
\bibitem [{\citenamefont {Ji}(1997)}]{Ji:1996nm}%
  \BibitemOpen
  \bibfield  {author} {\bibinfo {author} {\bibfnamefont {X.-D.}\ \bibnamefont
  {Ji}},\ }\href {\doibase 10.1103/PhysRevD.55.7114} {\bibfield  {journal}
  {\bibinfo  {journal} {Phys. Rev. D}\ }\textbf {\bibinfo {volume} {55}},\
  \bibinfo {pages} {7114} (\bibinfo {year} {1997})},\ \Eprint
  {http://arxiv.org/abs/hep-ph/9609381} {arXiv:hep-ph/9609381} \BibitemShut
  {NoStop}%
\bibitem [{\citenamefont {Byrnes}\ and\ \citenamefont
  {Yamamoto}(2006)}]{Byrnes:2005qx}%
  \BibitemOpen
  \bibfield  {author} {\bibinfo {author} {\bibfnamefont {T.}~\bibnamefont
  {Byrnes}}\ and\ \bibinfo {author} {\bibfnamefont {Y.}~\bibnamefont
  {Yamamoto}},\ }\href {\doibase 10.1103/PhysRevA.73.022328} {\bibfield
  {journal} {\bibinfo  {journal} {Phys.Rev.A}\ }\textbf {\bibinfo {volume}
  {73}},\ \bibinfo {pages} {022328} (\bibinfo {year} {2006})},\ \Eprint
  {http://arxiv.org/abs/quant-ph/0510027} {arXiv:quant-ph/0510027 [quant-ph]}
  \BibitemShut {NoStop}%
\bibitem [{\citenamefont {Alexandru}\ \emph {et~al.}(2019)\citenamefont
  {Alexandru}, \citenamefont {Bedaque}, \citenamefont {Harmalkar},
  \citenamefont {Lamm}, \citenamefont {Lawrence},\ and\ \citenamefont
  {Warrington}}]{Alexandru:2019nsa}%
  \BibitemOpen
  \bibfield  {author} {\bibinfo {author} {\bibfnamefont {A.}~\bibnamefont
  {Alexandru}}, \bibinfo {author} {\bibfnamefont {P.~F.}\ \bibnamefont
  {Bedaque}}, \bibinfo {author} {\bibfnamefont {S.}~\bibnamefont {Harmalkar}},
  \bibinfo {author} {\bibfnamefont {H.}~\bibnamefont {Lamm}}, \bibinfo {author}
  {\bibfnamefont {S.}~\bibnamefont {Lawrence}}, \ and\ \bibinfo {author}
  {\bibfnamefont {N.~C.}\ \bibnamefont {Warrington}},\ }\href {\doibase
  10.1103/PhysRevD.100.114501} {\bibfield  {journal} {\bibinfo  {journal}
  {Phys.Rev.D}\ }\textbf {\bibinfo {volume} {100}},\ \bibinfo {pages} {114501}
  (\bibinfo {year} {2019})},\ \Eprint {http://arxiv.org/abs/1906.11213}
  {arXiv:1906.11213 [hep-lat]} \BibitemShut {NoStop}%
\bibitem [{\citenamefont {Ji}\ \emph {et~al.}(2023)\citenamefont {Ji},
  \citenamefont {Lamm},\ and\ \citenamefont {Zhu}}]{Ji:2022qvr}%
  \BibitemOpen
  \bibfield  {author} {\bibinfo {author} {\bibfnamefont {Y.}~\bibnamefont
  {Ji}}, \bibinfo {author} {\bibfnamefont {H.}~\bibnamefont {Lamm}}, \ and\
  \bibinfo {author} {\bibfnamefont {S.}~\bibnamefont {Zhu}} (\bibinfo
  {collaboration} {NuQS}),\ }\href {\doibase 10.1103/PhysRevD.107.114503}
  {\bibfield  {journal} {\bibinfo  {journal} {Phys.Rev.D}\ }\textbf {\bibinfo
  {volume} {107}},\ \bibinfo {pages} {114503} (\bibinfo {year} {2023})},\
  \Eprint {http://arxiv.org/abs/2203.02330} {arXiv:2203.02330 [hep-lat]}
  \BibitemShut {NoStop}%
\bibitem [{\citenamefont {Gustafson}\ \emph {et~al.}(2024)\citenamefont
  {Gustafson}, \citenamefont {Ji}, \citenamefont {Lamm}, \citenamefont
  {Murairi}, \citenamefont {Perez},\ and\ \citenamefont
  {Zhu}}]{Gustafson:2024kym}%
  \BibitemOpen
  \bibfield  {author} {\bibinfo {author} {\bibfnamefont {E.~J.}\ \bibnamefont
  {Gustafson}}, \bibinfo {author} {\bibfnamefont {Y.}~\bibnamefont {Ji}},
  \bibinfo {author} {\bibfnamefont {H.}~\bibnamefont {Lamm}}, \bibinfo {author}
  {\bibfnamefont {E.~M.}\ \bibnamefont {Murairi}}, \bibinfo {author}
  {\bibfnamefont {S.~O.}\ \bibnamefont {Perez}}, \ and\ \bibinfo {author}
  {\bibfnamefont {S.}~\bibnamefont {Zhu}},\ }\href {\doibase
  10.1103/PhysRevD.110.034515} {\bibfield  {journal} {\bibinfo  {journal}
  {Phys.Rev.D}\ }\textbf {\bibinfo {volume} {110}},\ \bibinfo {pages} {034515}
  (\bibinfo {year} {2024})},\ \Eprint {http://arxiv.org/abs/2405.05973}
  {arXiv:2405.05973 [hep-lat]} \BibitemShut {NoStop}%
\bibitem [{\citenamefont {Assi}\ and\ \citenamefont
  {Lamm}(2024)}]{Assi:2024pdn}%
  \BibitemOpen
  \bibfield  {author} {\bibinfo {author} {\bibfnamefont {B.}~\bibnamefont
  {Assi}}\ and\ \bibinfo {author} {\bibfnamefont {H.}~\bibnamefont {Lamm}},\
  }\href {\doibase 10.1103/PhysRevD.110.074511} {\bibfield  {journal} {\bibinfo
   {journal} {Phys. Rev. D}\ }\textbf {\bibinfo {volume} {110}},\ \bibinfo
  {pages} {074511} (\bibinfo {year} {2024})},\ \Eprint
  {http://arxiv.org/abs/2405.12204} {arXiv:2405.12204 [hep-lat]} \BibitemShut
  {NoStop}%
\bibitem [{\citenamefont {Illa}\ \emph {et~al.}(2025)\citenamefont {Illa},
  \citenamefont {Savage},\ and\ \citenamefont {Yao}}]{Illa:2025dou}%
  \BibitemOpen
  \bibfield  {author} {\bibinfo {author} {\bibfnamefont {M.}~\bibnamefont
  {Illa}}, \bibinfo {author} {\bibfnamefont {M.~J.}\ \bibnamefont {Savage}}, \
  and\ \bibinfo {author} {\bibfnamefont {X.}~\bibnamefont {Yao}},\ }\href
  {\doibase 10.1103/3rwf-f844} {\bibfield  {journal} {\bibinfo  {journal}
  {Phys.Rev.D}\ }\textbf {\bibinfo {volume} {111}},\ \bibinfo {pages} {114520}
  (\bibinfo {year} {2025})},\ \Eprint {http://arxiv.org/abs/2503.09688}
  {arXiv:2503.09688 [hep-lat]} \BibitemShut {NoStop}%
\bibitem [{\citenamefont {Chen}\ \emph {et~al.}(2026)\citenamefont {Chen},
  \citenamefont {Müller},\ and\ \citenamefont {Yao}}]{Chen:2026hnh}%
  \BibitemOpen
  \bibfield  {author} {\bibinfo {author} {\bibfnamefont {V.}~\bibnamefont
  {Chen}}, \bibinfo {author} {\bibfnamefont {B.}~\bibnamefont {Müller}}, \
  and\ \bibinfo {author} {\bibfnamefont {X.}~\bibnamefont {Yao}},\ }\href@noop
  {} {\  (\bibinfo {year} {2026})},\ \Eprint {http://arxiv.org/abs/2601.10065}
  {arXiv:2601.10065 [hep-lat]} \BibitemShut {NoStop}%
\bibitem [{\citenamefont {Siew}\ \emph {et~al.}(2026)\citenamefont {Siew},
  \citenamefont {Chandrasekharan},\ and\ \citenamefont
  {Bhattacharya}}]{Siew:2026fax}%
  \BibitemOpen
  \bibfield  {author} {\bibinfo {author} {\bibfnamefont {R.~X.}\ \bibnamefont
  {Siew}}, \bibinfo {author} {\bibfnamefont {S.}~\bibnamefont
  {Chandrasekharan}}, \ and\ \bibinfo {author} {\bibfnamefont {T.}~\bibnamefont
  {Bhattacharya}},\ }\href@noop {} {\enquote {\bibinfo {title} {{Continuum
  limit of a qubit-regularized SU(3) lattice gauge theory with glueballs}},}\ }
  (\bibinfo {year} {2026}),\ \Eprint {http://arxiv.org/abs/2603.01215}
  {arXiv:2603.01215 [hep-lat]} \BibitemShut {NoStop}%
\bibitem [{\citenamefont {Yao}(2026)}]{Yao:2026rya}%
  \BibitemOpen
  \bibfield  {author} {\bibinfo {author} {\bibfnamefont {X.}~\bibnamefont
  {Yao}},\ }\href@noop {} {\enquote {\bibinfo {title} {{Quantum Simulation of
  QCD in Axial Gauge}},}\ } (\bibinfo {year} {2026}),\ \Eprint
  {http://arxiv.org/abs/2608.16783} {arXiv:2608.16783 [hep-lat]} \BibitemShut
  {NoStop}%
\bibitem [{\citenamefont {Atas}\ \emph {et~al.}(2021)\citenamefont {Atas},
  \citenamefont {Zhang}, \citenamefont {Lewis}, \citenamefont {Jahanpour},
  \citenamefont {Haase},\ and\ \citenamefont {Muschik}}]{Atas:2021ext}%
  \BibitemOpen
  \bibfield  {author} {\bibinfo {author} {\bibfnamefont {Y.~Y.}\ \bibnamefont
  {Atas}}, \bibinfo {author} {\bibfnamefont {J.}~\bibnamefont {Zhang}},
  \bibinfo {author} {\bibfnamefont {R.}~\bibnamefont {Lewis}}, \bibinfo
  {author} {\bibfnamefont {A.}~\bibnamefont {Jahanpour}}, \bibinfo {author}
  {\bibfnamefont {J.~F.}\ \bibnamefont {Haase}}, \ and\ \bibinfo {author}
  {\bibfnamefont {C.~A.}\ \bibnamefont {Muschik}},\ }\href {\doibase
  10.1038/s41467-021-26825-4} {\bibfield  {journal} {\bibinfo  {journal}
  {Nature Commun.}\ }\textbf {\bibinfo {volume} {12}},\ \bibinfo {pages} {6499}
  (\bibinfo {year} {2021})},\ \Eprint {http://arxiv.org/abs/2102.08920}
  {arXiv:2102.08920 [quant-ph]} \BibitemShut {NoStop}%
\bibitem [{\citenamefont {Ciavarella}\ and\ \citenamefont
  {Chernyshev}(2022)}]{Ciavarella:2021lel}%
  \BibitemOpen
  \bibfield  {author} {\bibinfo {author} {\bibfnamefont {A.~N.}\ \bibnamefont
  {Ciavarella}}\ and\ \bibinfo {author} {\bibfnamefont {I.~A.}\ \bibnamefont
  {Chernyshev}},\ }\href {\doibase 10.1103/PhysRevD.105.074504} {\bibfield
  {journal} {\bibinfo  {journal} {Phys.Rev.D}\ }\textbf {\bibinfo {volume}
  {105}},\ \bibinfo {pages} {074504} (\bibinfo {year} {2022})},\ \Eprint
  {http://arxiv.org/abs/2112.09083} {arXiv:2112.09083 [quant-ph]} \BibitemShut
  {NoStop}%
\bibitem [{\citenamefont {Ciavarella}(2025)}]{Ciavarella:2024lsp}%
  \BibitemOpen
  \bibfield  {author} {\bibinfo {author} {\bibfnamefont {A.~N.}\ \bibnamefont
  {Ciavarella}},\ }\href {\doibase 10.1103/PhysRevD.111.054501} {\bibfield
  {journal} {\bibinfo  {journal} {Phys.Rev.D}\ }\textbf {\bibinfo {volume}
  {111}},\ \bibinfo {pages} {054501} (\bibinfo {year} {2025})},\ \Eprint
  {http://arxiv.org/abs/2411.05915} {arXiv:2411.05915 [quant-ph]} \BibitemShut
  {NoStop}%
\bibitem [{\citenamefont {Ortiz}\ \emph {et~al.}(2001)\citenamefont {Ortiz},
  \citenamefont {Gubernatis}, \citenamefont {Knill},\ and\ \citenamefont
  {Laflamme}}]{Ortiz:2000gc}%
  \BibitemOpen
  \bibfield  {author} {\bibinfo {author} {\bibfnamefont {G.}~\bibnamefont
  {Ortiz}}, \bibinfo {author} {\bibfnamefont {J.}~\bibnamefont {Gubernatis}},
  \bibinfo {author} {\bibfnamefont {E.}~\bibnamefont {Knill}}, \ and\ \bibinfo
  {author} {\bibfnamefont {R.}~\bibnamefont {Laflamme}},\ }\href {\doibase
  10.1103/PhysRevA.64.022319} {\bibfield  {journal} {\bibinfo  {journal}
  {Phys.Rev.A}\ }\textbf {\bibinfo {volume} {64}},\ \bibinfo {pages} {022319}
  (\bibinfo {year} {2001})},\ \Eprint {http://arxiv.org/abs/cond-mat/0012334}
  {arXiv:cond-mat/0012334 [cond-mat]} \BibitemShut {NoStop}%
\bibitem [{\citenamefont {Somma}\ \emph {et~al.}(2002)\citenamefont {Somma},
  \citenamefont {Ortiz}, \citenamefont {Gubernatis}, \citenamefont {Knill},\
  and\ \citenamefont {Laflamme}}]{Somma:2001kjh}%
  \BibitemOpen
  \bibfield  {author} {\bibinfo {author} {\bibfnamefont {R.}~\bibnamefont
  {Somma}}, \bibinfo {author} {\bibfnamefont {G.}~\bibnamefont {Ortiz}},
  \bibinfo {author} {\bibfnamefont {J.}~\bibnamefont {Gubernatis}}, \bibinfo
  {author} {\bibfnamefont {E.}~\bibnamefont {Knill}}, \ and\ \bibinfo {author}
  {\bibfnamefont {R.}~\bibnamefont {Laflamme}},\ }\href {\doibase
  10.1103/PhysRevA.65.042323} {\bibfield  {journal} {\bibinfo  {journal}
  {Phys.Rev.A}\ }\textbf {\bibinfo {volume} {65}},\ \bibinfo {pages} {042323}
  (\bibinfo {year} {2002})},\ \Eprint {http://arxiv.org/abs/quant-ph/0108146}
  {arXiv:quant-ph/0108146 [quant-ph]} \BibitemShut {NoStop}%
\bibitem [{\citenamefont {Pedernales}\ \emph {et~al.}(2014)\citenamefont
  {Pedernales}, \citenamefont {Di~Candia}, \citenamefont {Egusquiza},
  \citenamefont {Casanova},\ and\ \citenamefont {Solano}}]{Pedernales:2014izf}%
  \BibitemOpen
  \bibfield  {author} {\bibinfo {author} {\bibfnamefont {J.}~\bibnamefont
  {Pedernales}}, \bibinfo {author} {\bibfnamefont {R.}~\bibnamefont
  {Di~Candia}}, \bibinfo {author} {\bibfnamefont {I.}~\bibnamefont
  {Egusquiza}}, \bibinfo {author} {\bibfnamefont {J.}~\bibnamefont {Casanova}},
  \ and\ \bibinfo {author} {\bibfnamefont {E.}~\bibnamefont {Solano}},\ }\href
  {\doibase 10.1103/PhysRevLett.113.020505} {\bibfield  {journal} {\bibinfo
  {journal} {Phys.Rev.Lett.}\ }\textbf {\bibinfo {volume} {113}},\ \bibinfo
  {pages} {020505} (\bibinfo {year} {2014})},\ \Eprint
  {http://arxiv.org/abs/1401.2430} {arXiv:1401.2430 [quant-ph]} \BibitemShut
  {NoStop}%
\bibitem [{\citenamefont {Roggero}\ and\ \citenamefont
  {Carlson}(2019)}]{Roggero:2018hrn}%
  \BibitemOpen
  \bibfield  {author} {\bibinfo {author} {\bibfnamefont {A.}~\bibnamefont
  {Roggero}}\ and\ \bibinfo {author} {\bibfnamefont {J.}~\bibnamefont
  {Carlson}},\ }\href {\doibase 10.1103/PhysRevC.100.034610} {\bibfield
  {journal} {\bibinfo  {journal} {Phys.Rev.C}\ }\textbf {\bibinfo {volume}
  {100}},\ \bibinfo {pages} {034610} (\bibinfo {year} {2019})},\ \Eprint
  {http://arxiv.org/abs/1804.01505} {arXiv:1804.01505 [quant-ph]} \BibitemShut
  {NoStop}%
\bibitem [{\citenamefont {Rall}(2020)}]{Rall:2020rsu}%
  \BibitemOpen
  \bibfield  {author} {\bibinfo {author} {\bibfnamefont {P.}~\bibnamefont
  {Rall}},\ }\href {\doibase 10.1103/PhysRevA.102.022408} {\bibfield  {journal}
  {\bibinfo  {journal} {Phys.Rev.A}\ }\textbf {\bibinfo {volume} {102}},\
  \bibinfo {pages} {022408} (\bibinfo {year} {2020})},\ \Eprint
  {http://arxiv.org/abs/2004.06832} {arXiv:2004.06832 [quant-ph]} \BibitemShut
  {NoStop}%
\bibitem [{\citenamefont {Kökcü}\ \emph {et~al.}(2024)\citenamefont
  {Kökcü}, \citenamefont {Labib}, \citenamefont {Freericks},\ and\
  \citenamefont {Kemper}}]{Kokcu:2023vwg}%
  \BibitemOpen
  \bibfield  {author} {\bibinfo {author} {\bibfnamefont {E.}~\bibnamefont
  {Kökcü}}, \bibinfo {author} {\bibfnamefont {H.~A.}\ \bibnamefont {Labib}},
  \bibinfo {author} {\bibfnamefont {J.}~\bibnamefont {Freericks}}, \ and\
  \bibinfo {author} {\bibfnamefont {A.~F.}\ \bibnamefont {Kemper}},\ }\href
  {\doibase 10.1038/s41467-024-47729-z} {\bibfield  {journal} {\bibinfo
  {journal} {Nature Commun.}\ }\textbf {\bibinfo {volume} {15}},\ \bibinfo
  {pages} {3881} (\bibinfo {year} {2024})},\ \Eprint
  {http://arxiv.org/abs/2302.10219} {arXiv:2302.10219 [quant-ph]} \BibitemShut
  {NoStop}%
\bibitem [{\citenamefont {Wang}\ \emph
  {et~al.}(2025{\natexlab{b}})\citenamefont {Wang}, \citenamefont {Xiong},
  \citenamefont {Cai},\ and\ \citenamefont {Yuan}}]{Wang:2025ojn}%
  \BibitemOpen
  \bibfield  {author} {\bibinfo {author} {\bibfnamefont {X.}~\bibnamefont
  {Wang}}, \bibinfo {author} {\bibfnamefont {L.}~\bibnamefont {Xiong}},
  \bibinfo {author} {\bibfnamefont {X.}~\bibnamefont {Cai}}, \ and\ \bibinfo
  {author} {\bibfnamefont {X.}~\bibnamefont {Yuan}},\ }\href {\doibase
  10.1103/z126-zdqj} {\bibfield  {journal} {\bibinfo  {journal}
  {Phys.Rev.Lett.}\ }\textbf {\bibinfo {volume} {135}},\ \bibinfo {pages}
  {230602} (\bibinfo {year} {2025}{\natexlab{b}})},\ \Eprint
  {http://arxiv.org/abs/2504.12975} {arXiv:2504.12975 [quant-ph]} \BibitemShut
  {NoStop}%
\end{thebibliography}%

\end{document}